\documentclass[aps,pra,showpacs,twoside,twocolumn,10pt,nofootinbib]{revtex4-1}
\usepackage[colorlinks=true, citecolor=red, urlcolor=blue ]{hyperref}
\usepackage{epsfig,newlfont,amssymb,amsmath,amsfonts,bm,graphicx,xcolor}
\usepackage{palatino,mathtools,amsthm,braket,soul,enumitem,color}
\usepackage[normalem]{ulem}

\begin{document}
 
\title{Scalable characterization  of localizable entanglement in noisy topological quantum codes}
\author{David Amaro$^1$, Markus M\"{u}ller$^{1,2,3}$, Amit Kumar Pal$^{1,4,5}$}
\affiliation{$^1$Department of Physics, Swansea University, Singleton Park, Swansea - SA2 8PP, United Kingdom  \\
$^2$Institute for Quantum Information, RWTH Aachen University, D-52056 Aachen, Germany\\
$^3$Peter Gr\"{u}nberg Institute, Theoretical Nanoelectronics, Forschungszentrum J\"{u}lich, D-52425 J\"{u}lich, Germany \\
$^4$Faculty of Physics, University of Warsaw, ul. Pasteura 5, PL-02-093 Warszawa, Poland\\
$^5$Department of Physics, Indian Institute of Technology Palakkad, Palakkad 678557, India
}

\begin{abstract}
Topological quantum error correcting codes have emerged as leading candidates towards the goal of achieving large-scale fault-tolerant quantum computers. However, quantifying entanglement in these systems of large size in the presence of noise is a challenging task. In this paper, we provide two different prescriptions to characterize noisy stabilizer states, including the surface and the color codes, in terms of localizable entanglement  over a subset of qubits. In one approach, we exploit  appropriately constructed entanglement witness operators to  estimate a witness-based lower bound of localizable entanglement, which is directly accessible in experiments. In the other recipe, we use graph states that are local unitary equivalent to the stabilizer state to determine a computable measurement-based lower bound of localizable entanglement. If used experimentally, this translates to a lower bound of localizable entanglement obtained from  single-qubit measurements in specific bases to be performed on the qubits outside the subsystem of interest. Towards computing these lower bounds, we discuss in detail the methodology of obtaining a local unitary equivalent graph state from a stabilizer state, which includes a new and scalable geometric recipe as well as an algebraic method that applies to general stabilizer states of arbitrary size.  Moreover, as a crucial step of the latter recipe, we develop a scalable graph-transformation  algorithm that creates a link between two specific nodes in a graph using a sequence of  local complementation operations.  We develop open-source Python packages for these transformations, and illustrate the methodology by applying it to a noisy  topological color code, and study how the witness and measurement-based lower bounds of localizable entanglement varies with the distance between the chosen qubits.   
\end{abstract}

\maketitle

\section{Introduction}
\label{sec:intro}

The advancements in the science of quantum information and computation~\cite{bennett2000} has put entanglement~\cite{horodecki2009} in a position of extreme importance for a number of quantum protocols including quantum teleportation~ \cite{horodecki2009,bennett1993,bouwmeester1997}, quantum dense coding~ \cite{bennett1992,mattle1996,sende2010}, quantum cryptography~\cite{ekert1991,jennewein2000}, and measurement-based quantum computation~\cite{raussendorf2001,raussendorf2003, briegel2009}. Besides, entanglement has also been identified as the  key ingredient in several other seemingly unrelated problems, such as the study of topological~\cite{pollmann2010,chen2010,jiang2012} and non-topological~\cite{osterloh2002,osborne2002,amico2008,chiara2017} quantum phases and corresponding quantum phase transitions in many-body systems, understanding transport properties in biological systems such as light-harvesting complexes~\cite{sarovar2010,zhu2012,lambert2013,chanda2014}, investigating the role of radical-pair mechanism in the navigability of animals in weak magnetic fields~\cite{cai2010},  and aspects of AdS/CFT correspondence in models of quantum gravity~\cite{hubeny2015,pastawski2015,almheiri2015,jahn2017}. Tremendous technological development has made the laboratory realization of entangled states, in both biparty and multiparty scenario, possible by using trapped ions~\cite{leibfried2003,leibfried2005,brown2016}, photonic systems~\cite{raimond2001,prevedel2009,barz2015},  nuclear magnetic resonance (NMR) molecules~\cite{negrevergne2006}, superconducting qubits~\cite{clarke2008,berkley2003}, cold atoms~\cite{hald1999,mandel2003,bloch2005,bloch2008}, solid-state systems~\cite{bernien2013}, and set-ups involving hybrids of these systems~\cite{togan2010, marinkovic2018}. This highlights the potential of realizing different quantum protocols that uses entanglement as resource in these systems. The theoretical aspect of this line of study involves quantum information processing using many-body systems realizable in the above substrates, which has particularly brought the importance of the study quantum many-body systems in a language consistent with the quantum information theory in focus. Being a key resource in quantum protocols,  entanglement has been the natural choice as the characterising feature of quantum many-body systems for this purpose. 

In the last decade, the extraordinary potential of  quantum computation~\cite{divincenzo2000,gottesman2010} in addressing otherwise intractable problems, such as simulating large quantum systems or decrypting codes efficiently, has led to major efforts  towards fabricating scalable systems, with a long-term goal of  fault-tolerance, using the available many-body substrates such as trapped ions~\cite{benhelm2008} and superconducting qubits~\cite{chow2012,barends2014}. There already exist noisy intermediate-scale quantum (NISQ) devices~\cite{preskill2018,paler2018,nash2019} made of 50-100 qubits, which are being viewed and investigated as  platforms to potentially achieve ``quantum supremacy"~\cite{preskill2012}, and as possible candidate systems to host logical qubits  as building blocks of envisioned large-scale quantum computers~\cite{fujii2015}.  To  date, the figures of merit for the usefulness of a quantum state, prepared in these systems, in a given quantum protocol are the different quantum correlations, such as entanglement,  that serves as resource in that protocol. However, characterizing such systems using entanglement proves to be difficult mainly due to exceeding resource requirements in traditional techniques such as complete quantum state or process tomography, even for moderately sized systems of a few qubits~\cite{ibort2009,lvovsky2009}. Another obstacle, from the computational point of view, is the presence of noise, which requires entanglement to be computed for a mixed state of a large system -- a long-standing problem of quantum information theory~\cite{horodecki2009,huang2014}.

\begin{figure}
\includegraphics[scale=0.4]{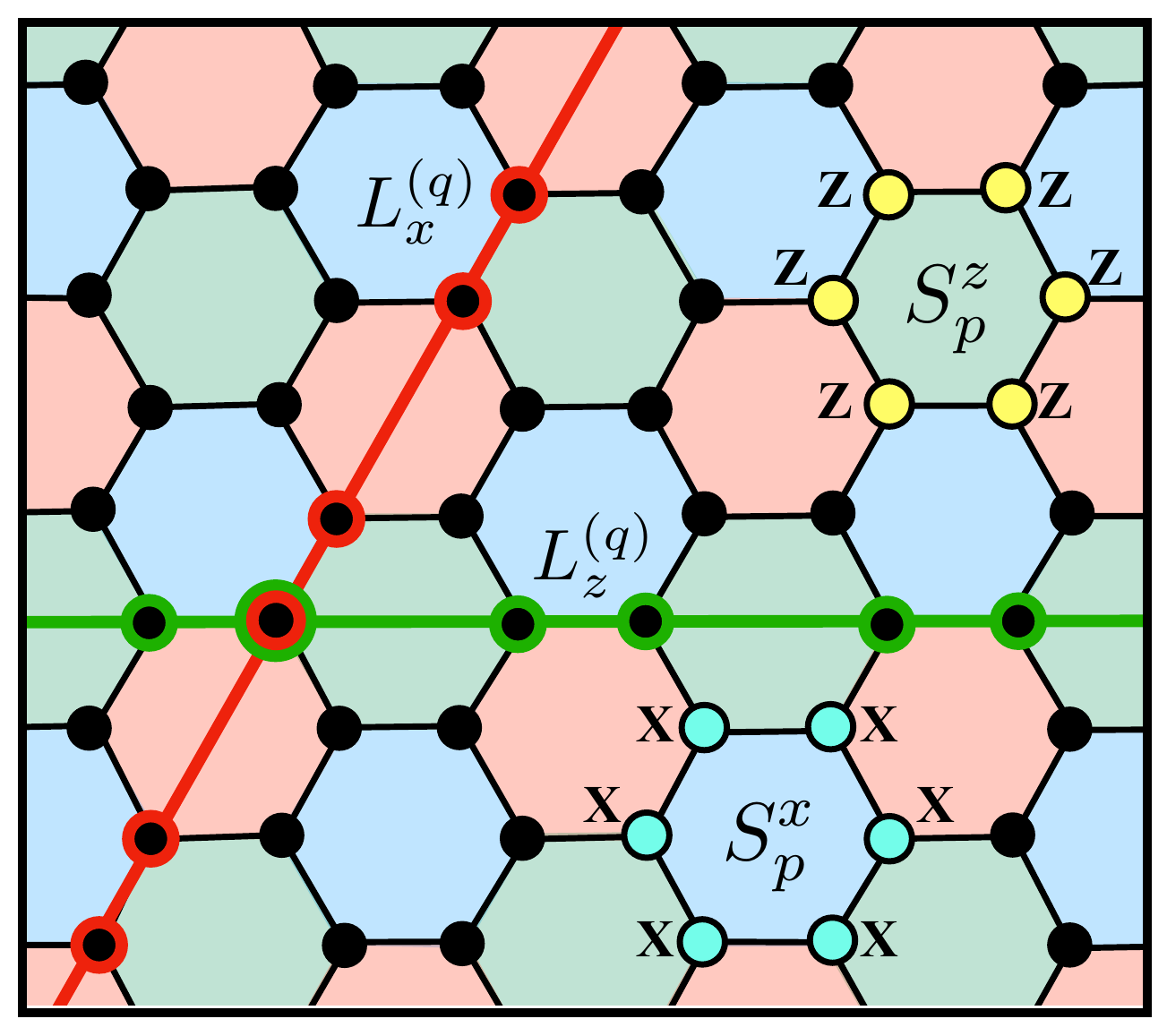}
\caption{(Color online.) \textbf{Topological color code.} A hexagonal lattice hosting a topological color code in which the $X$- and $Z$-type  plaquette operators are associated to plaquettes formed of  turquoise and yellow qubits respectively. The vertical (red) logical operator $L^{(q)}_x$ and the horizontal (green) logical operator $L^{(q)}_z$ are shown by the red and green lines respectively, intersecting each other at a single qubit. See Appendix~\ref{app:tcc} for a description of the salient features of a topological color code.}
\label{fig:tcc}
\end{figure}  

In the vision of building large-scale fault-tolerant quantum computers, topological quantum error correcting codes~\cite{nayak2008,pachos2014,lahtinen2017}, such as, eg., the surface codes~\cite{kitaev2001,kitaev2006} and the color codes~\cite{bombin2006,bombin2007} (see Fig.~\ref{fig:tcc} and Appendix~\ref{app:tcc} for a description), are being considered as  leading candidate systems. These systems are realized by arranging qubits on lattices of specific geometry, and are robust against external perturbation~\cite{dusuel2011,jahromi2013,zarei2015,jamadagni2018}, qubit loss~\cite{stace2009,stace2010,vodola2018}, as well as computational errors~\cite{katzgraber2009}. Attempts have recently been made to implement such systems  in the laboratory with, for example, trapped ions~\cite{nigg2014,linke2017,wright2019}, and superconducting qubits~\cite{kelly2015,gambetta2017}. In order to host a single logical qubit with low or moderate code distance, and to perform error correction protocols taking into account errors on multiple physical qubits, one needs to deal with systems with a large number of physical qubits in the presence of noise. This makes the characterization of these systems using entanglement measures difficult. Therefore, quantifying bipartite as well as multipartite entanglement in subsystems of a large-scale quantum many-body system like the topological quantum codes in the presence of noise has been an active field of research in recent times~\cite{castelnovo2007,castelnovo2008,schmitz2019}.

There exists a  plethora of approaches towards quantifying entanglement in subsystems of a quantum many-body system, which mainly follow two approaches -- (1) the partial trace-based~\cite{horodecki2009}, and (2) the local measurement-based~\cite{divincenzo1998,verstraete2004,popp2005,sadhukhan2017} approach. In the former, an entanglement measure is computed either between two subsystems denoted by $\Omega$ and $\overline{\Omega}$ of the full quantum many-body system, or over a subsystem $\Omega$ of a quantum many-body system by using the reduced state $\rho_\Omega=\text{Tr}_{\overline{\Omega}}(\rho)$ of $\Omega$. The reduced state is obtained by performing a partial trace operation on the rest of the system $\overline{\Omega}$, where $\Omega\cup\overline{\Omega}$ represents the entire system, and $\Omega\cap\overline{\Omega}=\emptyset$. This protocol is particularly useful in scenarios where the state of the entire system is pure, leading to concepts like topological entanglement entropy~\cite{kitaev2006a,levin2006,linke2018,brydges2019} between the subsystems $\Omega$ and $\overline{\Omega}$. This is also effective in situations where the partial trace operation results in a reduced density matrix $\rho_{\Omega}$ that faithfully provides a non-zero value of a chosen bipartite or multipartite  entanglement measure thereby quantifying entanglement over $\Omega$, when the subsystems constituting the region $\Omega$ are entangled accordingly. Apart from leading to the study of topological entanglement entropy as a function of the system parameters in topological quantum codes~\cite{kitaev2003,kargarian2008}, the partial trace-based prescription has also resulted in, for example,  investigations of the behaviour of entanglement over a collection of subsystems. These include pair of nearest-neighbor spins, in a large set of quantum many-body Hamiltonians, such as the transverse field Ising and the XY models, and the XXZ model in the presence of a magnetic field~\cite{osterloh2002,osborne2002,amico2008,chiara2017}.

\begin{figure*}
\includegraphics[width=\textwidth]{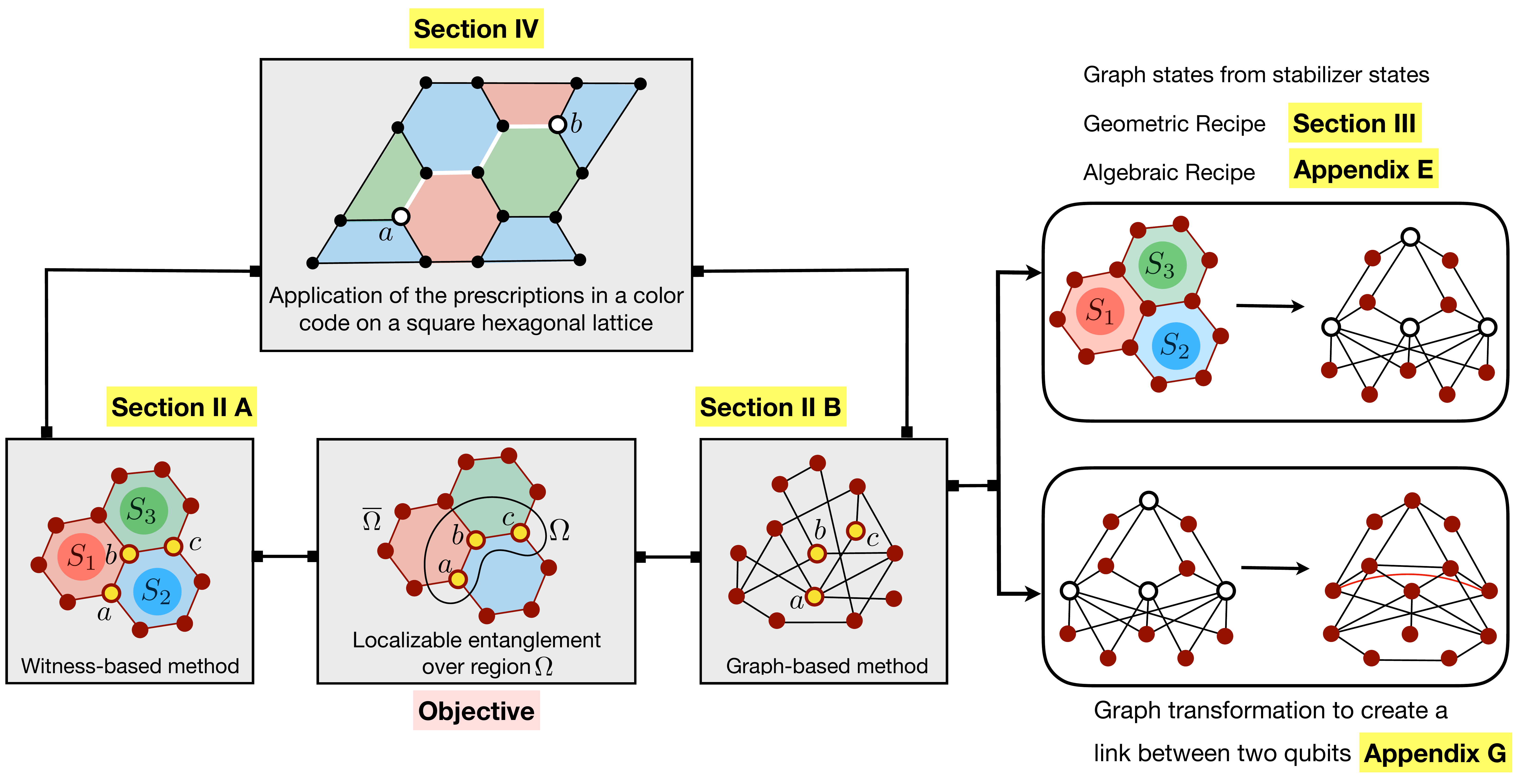}
\caption{(Color online.) \textbf{Schematic representation of the objectives of this paper.} This paper focuses on computing the localizable entanglement over a specific subsystem $\Omega$  of a noisy topological quantum error correcting code, via single-qubit projection measurements on the qubits in the rest of the system, designated by $\overline{\Omega}$.  Two prescriptions (Section~\ref{sec:le_bounds}), one witness-based (Section~\ref{subsec:wlb}) and the other graph-based (Section~\ref{subsec:mlb}),  are proposed. The latter approach requires a transformation of the stabilizer state to a graph state, where a geometric recipe for building the graph from the  stabilizer structure is discussed in Section~\ref{sec:graph-topo}, and an algebraic method to compute the adjacency matrices of the graphs is given in Appendix~\ref{app:algebraic_approach}. The graph-based method further requires a fail-safe recipe to create a link between  any two qubits if they do not already share a link. A graph-transformation algorithm is developed and is discussed in Appendix~\ref{app:graph-based} for this purpose. The corresponding methodologies are applied to a topological color code on a square-hexagonal lattice (Section~\ref{sec:apply}), and the dependence of the bound on the distance between two chosen qubits is investigated.}  
%The proposed methodology, and the corresponding developments of the technical aspects of the problem of computing lower-bounds of localizable entanglement in topological color codes in the presence of noise are shown in the form of a schematic diagram. The text-boxes contain descriptions of the highlights of this paper.
\label{fig:schematic}
\end{figure*} 

However, topological entanglement entropy fails to provide a faithful characterization of bipartite entanglement in a topological code when the state of the system is mixed -- for example, when there is noise in the system~\cite{horodecki2009,castelnovo2007,castelnovo2008}. Also, there exist logical states in topological error correcting codes for which the reduced state $\rho_{\Omega}$ is classical~\cite{nigg2014}, thereby providing zero entanglement as quantified by a chosen entanglement measure despite the existence of entanglement over $\Omega$. In these situations, the appropriate protocol is the local measurement-based approach~\cite{verstraete2004,popp2005,sadhukhan2017}, which relies on performing local projection measurements on the subsystems constituting $\overline{\Omega}$ in specific measurement bases in order to create \emph{entangled} post-measurement states over the region $\Omega$. This leads to non-zero average entanglement in the region $\Omega$ after the measurement, thereby appropriately quantifying the entanglement present in the subsystem $\Omega$. The maximum average post-measurement entanglement, maximized over all possible local projection measurements on the subsystems in $\overline{\Omega}$, is referred to as the \emph{localizable entanglement}~\cite{verstraete2004,popp2005,sadhukhan2017}, which is defined for both pure as well as mixed states.  Apart from being a good quantifier of local entanglement in stabilizer states~\cite{van-den-nest2004,hein2006,fujii2015,amaro2018}, including topological quantum codes, with or without noise, localizable entanglement is crucial also in other scenarios. For example, it has been used  in conceptualizing the correlation length in certain quantum many-body systems~\cite{verstraete2004,popp2005,verstraete2004a,jin2004}, for characterizing local entanglement in cluster-Ising~\cite{skrovseth2009,smacchia2011} and cluster-XY models~\cite{montes2012}, and in protocols including  measurement-based quantum computation~\cite{van-den-nest2004,hein2006,fujii2015} and entanglement percolation in quantum network~\cite{acin2007}.

Although the definition of localizable entanglement~\cite{verstraete2004,popp2005,sadhukhan2017} is straightforward to understand, the maximization involved in the definition often makes the computation of the quantity difficult if the size of the quantum many-body system is too large. Also, in the case of mixed states, which may originate in situations where there is noise in the system -- a practical scenario being topological quantum error correcting codes, the computation of the value of localizable entanglement requires calculation of a chosen entanglement measure for the states in the post-measurement ensemble over the region $\Omega$, which may turn out to be a challenging task. These obstacles highlight the need of a recipe to determine the localizable entanglement over a region of a large-scale topological quantum error correcting code in the presence of noise, which is the aim of this paper. 

In this paper, we calculate, if not the actual value, an effective lower bound of localizable entanglement in an efficient and experiment-friendly way via the use of entanglement witness operators. We illustrate our prescriptions for a region of two qubits in a color code under local uncorrelated Pauli noise. However, the recipe is fully applicable to the more general case of arbitrary stabilizer states, and it has the potential to be generalized to the case of arbitrary regions in arbitrary topological quantum error correcting stabilizer  codes. The major new results reported in this paper are as follows.
\begin{enumerate}
\item We propose two prescriptions for computing non-trivial lower bounds of localizable entanglement over a group of qubits in noisy topological quantum codes, and establish a connection between the two seemingly different prescriptions. The complete prescriptions are given in Section~\ref{sec:le_bounds}.
\begin{itemize} 
\item One of the prescriptions is a method based on entanglement witness operators~\cite{terhal2002,guhne2002,bourennane2004,guhne2009,guhne2005,alba2010,friis2018,amaro2019a}, which exploits appropriate construction of a local witness operator~\cite{alba2010,amaro2019a} from the topological code via a projection operator-based approach and the fact that a witness-based lower bound of localizable entanglement is computable from local witness operators~\cite{amaro2018}. We propose a specific construction of the local witness operator in the case of topological color codes, and explain the method in detail in Section~\ref{subsec:wlb}.
\item The second approach is  \emph{graph-based}. It exploits graph states~\cite{van-den-nest2004,hein2006} obtained via local unitary transformation from the stabilizer states representing topological codes, and subsequently transforming the graphs such that the chosen region $\Omega$ becomes connected in the transformed graph. The salient features of this method are given in Section~\ref{subsec:mlb}.
\end{itemize}  
\item The graph-based method requires obtaining the possible graph states from an aritrary stabilizer state in a topological quantum error correcting code.  A geometric prescription for the transformation of the surface code to the local unitary equivalent graph state has been proposed in Ref.~\cite{lang2012}. In this paper, we discuss the geometric recipe for transforming a general stabilizer state of  a topological color code to a graph state (Section~\ref{sec:graph-topo}). We also present, in connection to the geometric recipe, an algebraic methodology that transforms an arbitrary stabilizer state to a graph state, which is convenient for obtaining the adjacency matrix of  the graph from the structures of the stabilizer operators. This part is fairly technical, and to make the main text of the paper comprehensive and appropriate for a broad readership, we discuss the technicalities of this conversion in Appendix~\ref{app:algebraic_approach}. The method has also been given the form of an open source Python package. 
\item  Moreover, for the graph-based approach, we develop a fail-safe adaptive algorithm for connecting a region of two qubits in arbitrary connected graphs by local complementation operations, and implement the algorithm in the form of an open-source Python package. This involves technical details on graphs and their transformations under specific operations. Similar to the case of the algebraic approach for obtaining the adjacency matrices of graphs,  the more technical details of the methodology are discussed in  Appendix~\ref{app:graph-based}. 
\end{enumerate}
\noindent The detailed calculations of the decomposition of local entanglement witnesses, and a few examples and the necessary information on the topological color codes, graph states,  and noise models have been included in the Appendices~\ref{app:tcc}-\ref{app:stab_expt}. To demonstrate how the newly developed methodology can be used for topological quantum codes, we apply them to the specific example of the two-dimensional color code on the square hexagonal lattice, and discuss the results regarding the noise as well as distance dependence of localizable entanglement between a pair of qubits. The results are discussed in detail in Section~\ref{sec:apply}. Section~\ref{sec:conclude}  contains the concluding remarks.

%\textcolor{blue}{This paper is organized as follows.  In Section~\ref{sec:le_bounds}, we describe two different approaches to obtain computable lower bounds for localizable entanglement. Subsection~\ref{subsec:wlb} discusses the witness operator-based approach, which is advantageous from a experimental perspective. On the other hand,  Subsection~\ref{subsec:mlb} describes a methodology that relies on performing projection measurement in appropriate measurement bases, and which exploits the properties of graph states. The latter approach requires transformation of the stabilizer state of the topological quantum codes to graph states. A geometric recipe for transforming an arbitrary stabilizer state of a topological color code into a graph state is discussed in Section~\ref{sec:graph-topo}, and is also connected to an algebraic perspective, so that the adjacency matrix of the graph can be determined. As an illustration of the application of the methodology developed, in Section~\ref{sec:apply}, we discuss the computation of the witness-based lower bound of localizable entanglement over qubit-pairs situated in the bulk of a noisy color code. We also study the performance of the graph-based method applied to the same system.  We present the concluding remarks in Section~\ref{sec:conclude}. } 

 \begin{figure*}
\includegraphics[width=0.7\textwidth]{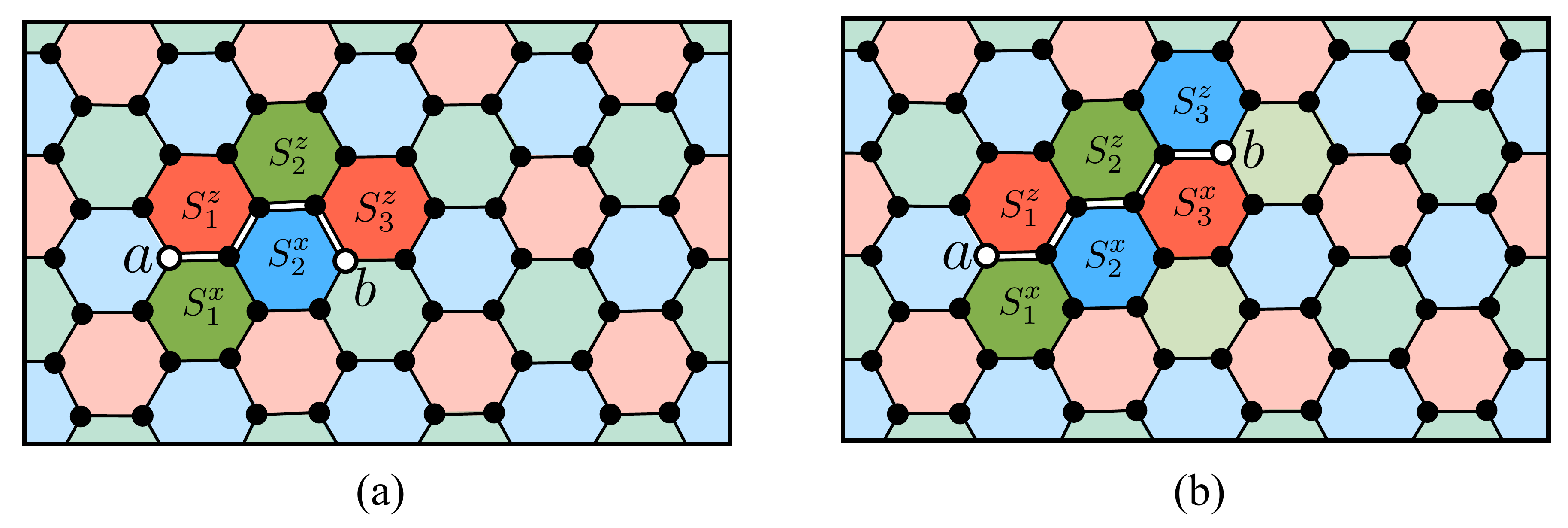}
\caption{(Color online.) \textbf{Construction of the local witness operator.} The local witness operator according to Eq.~(\ref{eq:witness_global}) for a two-qubit region $\Omega\equiv ab$ where the physical  distance between the two qubits is (a) $d=4$, and (b) $d=5$. The stabilizers $S^z$ and $S^x$ are obtained by multiplying respectively the $z$- and $x$-type stabilizers corresponding to the plaquettes on two adjacent paths of plaquettes, connecting the two qubits $a$ and $b$. In the present example, the two stabilizers $S_{j=1}$ and $S_{j=2}$ for constructing the local witness operator $\mathcal{W}_{\Omega}$ are constituted of the $x$- and $z$-type plaquette stabilizers as (a) ($d=4$) $S^z=\bigotimes_{p=1}^3S_p^z$, $S^x=\bigotimes_{p=1}^2S_p^x$, and (b) $(d=5)$ $S^z=\bigotimes_{p=1}^3S_p^z$, $S^x=\bigotimes_{p=1}^3S_p^x$, $p$ being the plaquette index. The distance between $a$ and $b$ is the length of the path constituted of the lattice links common to the two adjacent paths (shown by double continuous lines). 
%See Appendix~\ref{app:witness_decomposition} for the implications of the different colors of the qubits constructing the support of the stabilizers.
}  
\label{fig:local_witness}
\end{figure*}

\section{Bounds of localizable  entanglement}
\label{sec:le_bounds}

Localizable entanglement (LE)~\cite{verstraete2004,popp2005,sadhukhan2017,amaro2018} over a region $\Omega$ composed of the $N-m$ qubits in a $N$-qubit state $\rho$ is the maximum average entanglement that can be localized over $\Omega$, by performing local projection measurements on the $m$ qubits in $\overline{\Omega}$ -- the region  outside the chosen subsystem $\Omega$. Without any loss in generality, we assign the first $m$ qubits in the set of $N$ qubits labelled as  $1,2,\cdots,m,m+1,\cdots,N$   to the region $\overline{\Omega}$ and the rest in $\Omega$, with $\Omega\cup\overline{\Omega}$ representing the complete $N$-partite system with $\Omega\cap\overline{\Omega}=\emptyset$, and $1\leq m\leq N-2$.  Considering only rank-$1$ projection measurements on the qubits in $\overline{\Omega}$, LE over the region $\Omega$ is given by 
\begin{eqnarray}
E_{\Omega}(\rho)=\underset{\mathcal{M}}{\sup}\sum_{k=0}^{2^m-1}p_{k}\mathcal{E}(\rho_\Omega^{k}),
\label{eq:le}
\end{eqnarray}
where $\rho_{\Omega}^k=\text{Tr}_{\overline{\Omega}}\left[p_k^{-1}\mathcal{M}_k\rho\mathcal{M}_k\right]$ is the normalized  post-measurement state over the region $\Omega$, $p_k=\text{Tr}[\mathcal{M}_k\rho\mathcal{M}_k]$ is the probability of obtaining the measurement outcome $k$ with $\sum_{k=0}^{2^m-1}p_k=1$, and $\mathcal{M}_k=[\bigotimes_{i\in\overline{\Omega}}\ket{k_{i}}\bra{k_{i}}]\otimes I_{\Omega}$ forms the complete set $\mathcal{M}$ of rank-$1$ projection measurements over $\overline{\Omega}$. Here $\ket{k_{i}}$ $\in$ $\{\ket{\mathbf{0}}_{i},\ket{\mathbf{1}}_{i}\}$  are two arbitrary and mutually orthogonal single-qubit states in the Bloch sphere,  with $\ket{\mathbf{0}}_{i}=\cos (\theta_{i}/2)\ket{0}+e^{\text{i}\phi_{i}}\sin(\theta_{i}/2)\ket{1}$, $\ket{\mathbf{1}}_{i}=\sin (\theta_{i}/2)\ket{0}-e^{\text{i}\phi_{i}} \cos (\theta_{i}/2)\ket{1}$, $\{\ket{0},\ket{1}\}$ being the computational basis, and $\{\theta_{i},\phi_{i}\}$ are the real azimuthal and polar angles of the Bloch sphere ($0\leq\theta_{i}\leq \pi$, $0\leq \phi_{i}<2\pi$). The index $k$ here can be identified as  the multi-index $k\equiv k_{1}k_{2}\ldots k_{m}$, with $k_{i}=\mathbf{0},\mathbf{1}$. The optimization in Eq.~(\ref{eq:le}), therefore, reduces to an optimization over $2m$ real parameters, and is difficult to perform when $m$ is a large number. However, even  in cases where $m$ is small, analytical determination of LE is  possible only in the cases of a very limited number of multiqubit quantum states with special properties. 

In order to extract useful information about the properties of LE in situations where computing the exact value of LE proves difficult, a possible approach is to determine computable lower bounds of LE, which can provide insight of the behaviour of the actual quantity. In this spirit, one can perform the optimization over a restricted subset of the complete set of local projection measurements, for example, by allowing only local Pauli measurements over the qubits, thereby computing the \emph{restricted} LE (RLE), $E^R_{\Omega}(\rho)$,  of $\rho$~\cite{amaro2018}. However, computation of RLE is also non-trivial in the case of large $N$, where one has to consider a total of $3^{m}$ measurement settings, corresponding to three possible local Pauli measurements on each qubit, which is a large number when $m$ is large. A further lower bound can be found by considering a specific Pauli measurement setting where each qubit $i$ in $\overline{\Omega}$ is measured in the basis of a specific Pauli operator $\sigma_{i}$, where $\sigma_{i}=X,Y,Z$. The average entanglement in the region $\Omega$, corresponding to the chosen measurement setting $\mathcal{P}$, is given by 
\begin{eqnarray}
E_{\Omega}^\mathcal{P}(\rho)=\sum_{k=0}^{2^m-1}p_k\mathcal{E}(\rho_{\Omega}^k),
\label{eq:le_specific_pauli_setting}
\end{eqnarray}
where the superscript $\mathcal{P}$ represents an appropriately chosen specific Pauli measurement setting, and $E_{\Omega}(\rho)\geq E^R_{\Omega}(\rho) \geq E_{\Omega}^{\mathcal{P}}(\rho)$ by the definition of LE. 

%The challenge here is to choose the Pauli measurement setting $\mathcal{P}$ which would provide a non-trivial lower bound of LE. In this paper, we propose a graph-based methodology for obtaining computable MLBs in the case of an arbitrary stabilizer state under noise,  which is discussed in detail in Secs.~\ref{subsec:mlb} and \ref{sec:graph-based}. 

The challenge now is to choose an appropriate Pauli measurement setting $\mathcal{P}$, which would provide a non-trivial lower bound of LE.  In~\cite{amaro2018}, two avenues for constructing such lower bounds of the LE have been discussed -- (1) an experimentally accessible \emph{entanglement witness-based lower bound} (WLB) by using entanglement witness operators appropriate for the post-measurement states on the region $\Omega$, and (2) a \emph{graph-based approach} to determine an appropriate Pauli measurement set-up $\mathcal{P}$, which provides a non-trivial value of $E_{\Omega}^{\mathcal{P}}(\rho)$, denoted by the \emph{measurement-based lower bound} (MLB).  We consider the state $\rho$ to be a mixed one in general, originated from, for example, a stabilizer state $\rho_0$ due to application of noise $\rho_0\rightarrow\rho=\Lambda(\rho_0)$, $\Lambda(.)$ representing the noise model. In the following subsections, we present the underlying intricacies of computing the WLB and the MLB for LE in the case of arbitrary stabilizer states under noise. We use topological color codes~\cite{bombin2006,bombin2007} (see also Appendix~\ref{app:tcc} for definitions) for demonstration.

\subsection{Witness-based lower bound}
\label{subsec:wlb}

We begin our discussion with the WLB of LE, which employs an entanglement witness operator~\cite{terhal2002,guhne2002,bourennane2004,guhne2009,guhne2005,alba2010,friis2018,amaro2019a}, $\mathcal{W}$,  that indicates non-zero entanglement content in a quantum state $\rho$ via a negative expectation value, i.e., $\text{Tr}(\mathcal{W}\rho)<0$. A lower bound of the entanglement content in a quantum state $\rho$ can be determined using the set of expectation values of appropriately chosen entanglement witness operators as the solution of the optimization problem~\cite{brandao2005,brandao2006,eisert2007,guhne2007,guhne2008,amaro2018} 
\begin{eqnarray}
	\mathcal{E}_{\mbox{\scriptsize min\normalsize}}(\omega)=&&\inf \mathcal{E}(\rho),
	\label{eq:wlb_opt}
\end{eqnarray}
subject to $\text{Tr}\left(\rho\mathcal{W}\right) = \omega$, $\rho\geq 0$, and $\text{Tr}\left(\rho\right)=1$, where $\mathcal{E}$ is the chosen entanglement measure. This can be used in Eq.~(\ref{eq:le_specific_pauli_setting}) to write
\begin{eqnarray}
E_{\Omega}^\mathcal{P}(\rho)\geq  \sum_{k=0}^{2^m-1}p_k\mathcal{E}_{\mbox{\scriptsize min\normalsize}}(\omega_k) = E_{\Omega}^W(\rho),
\label{eq:le_wlb}
\end{eqnarray}
where $\omega_k$ is the expectation value of an appropriately chosen witness operator $W_{\Omega}^k$ for the post-measurement state $\rho_{\Omega}^k$ over the region $\Omega$ such that $\mathcal{E}_{\mbox{\scriptsize min\normalsize}}(\omega_k)\leq\mathcal{E}(\rho_{\Omega}^k)$, and $E_{\Omega}^W(\rho)$ is the WLB.

Note that to obtain $\omega_k$, one has to (1) perform local Pauli measurements on the qubits in the region $\overline{\Omega}$, (2) choose appropriate witness operators $W_{\Omega}^k$ for the post-measurement states $\rho_{\Omega}^k$ subject to the specific measurement outcome $k$ for the chosen Pauli measurement setting, and then (3) measure the expectation values $\omega_k=\text{Tr}\left[W_{\Omega}^k\rho_{\Omega}^k\right]$ in the post-measurements states. However, this protocol can prove difficult to carry out in experiments when $N$ and $m$ are large numbers. An alternative approach would be to look for a witness operator $\mathcal{W}_{\Omega}$, called the \emph{local} witness operator,  such that the expectation value of $\mathcal{W}_{\Omega}$, when determined with respect to the state $\rho$, detects entanglement in the region $\Omega$~\cite{alba2010,amaro2019a}, and a functional relation between $\text{Tr}(\mathcal{W}_{\Omega}\rho)$ and $E_{\Omega}^W(\rho)$ exists.  The challenge, however, in this approach is choosing an appropriate form of $\mathcal{W}_{\Omega}$ which can be connected to the specific measurement outcomes of the chosen Pauli measurement setting $\mathcal{P}$, such that the witnesses $W_{\Omega}^k$ corresponding to different values of $k$ can be constructed out of $\mathcal{W}_{\Omega}$.

For this purpose, we construct a local witness operator in terms of stabilizers describing the stabilizer state. A local witness operator $\mathcal{W}_{\Omega}$ that can detect genuine multiparty entanglement in the region $\Omega$ in an arbitrary stabilizer state $\ket{\psi_S}$, or a state $\rho_S$ close to it can be chosen  to be of the form~\cite{amaro2019a}
\begin{eqnarray}
\mathcal{W}_{\Omega}=\frac{1}{2}I-\prod_{S_j\in\mathcal{S}}\frac{I+S_j}{2},
\label{eq:witness_global}
\end{eqnarray}
where $\mathcal{S}=\{S_j\}$ is a subset of the complete set of stabilizers defining the state $\ket{\psi}$, given by $S_j=\bigotimes_i\tau_{u_{i,j}}$, $u_{i,j}=0,1,2,3$, $i$ is the qubit-index, $j$ indicates which stabilizer the qubit belongs to,   and $\tau_{u_{i,j}}=I$, $X$, $Y$, and $Z$, for $u_{i,j}=0,1,2$, and $3$ respectively (this is the same definitions of stabilizers  as given in Appendix~\ref{app:tcc}, with a new variable $u_{i,j}$ introduced in order to conveniently represent the Pauli matrices, which will be clear in subsequent discussions). One can write the stabilizers $S_j$ constructing $\mathcal{W}_{\Omega}$  by distinguishing the supporting qubits according to whether they belong to the region $\Omega$ or $\overline{\Omega}$,  as 
\begin{eqnarray}
S_j=\bigotimes_{i\in\overline{\Omega}}\tau_{u_{i,j}}^{\overline{\Omega}}\bigotimes_{l\in\Omega}\tau_{u_{l,j}}^{\Omega}=\left[\bigotimes_{i\in\overline{\Omega}}\tau_{u_{i,j}}^{\overline{\Omega}}\right]\otimes S_{j}^\Omega,
\label{eq:stab}
\end{eqnarray}
where $S_j^\Omega=\bigotimes_{l\in\Omega}\tau_{u_{l,j}}^{\Omega}$ is the part of the stabilizer $S_j$ with support on $\Omega$. For $\mathcal{W}_{\Omega}$  to detect entanglement in the region $\Omega$, the stabilizers $S_j$ in Eq.~(\ref{eq:witness_global}) have to be such that~\cite{amaro2019a}
\begin{enumerate}
\item[\text{(i)}]  $\left[\tau_{u_{i,j}}^{\overline{\Omega}},\tau_{u_{i^\prime,j^\prime}}^{\overline{\Omega}}\right]=0$ $\forall$ qubits $i,i^\prime$, and $\forall$ stabilizer pairs $j,j^\prime$, i.e., the supports of the stabilizers involved in constructing  $\mathcal{W}_{\Omega}$ must commute outside the region $\Omega$,   and
\item[\text{(ii)}] the set $\{S_j^\Omega\}$ obtained from the subset $\mathcal{S}$ of stabilizers $S_j$ of $\ket{\psi}$ is a complete set of stabilizer generators of a genuinely multipartite entangled state $\ket{\psi}_{\Omega}$ over $\Omega$.
\end{enumerate}
See Figs. \ref{fig:local_witness}(a) and (b) for demonstrations.

\begin{figure*}
\includegraphics[width=0.7\textwidth]{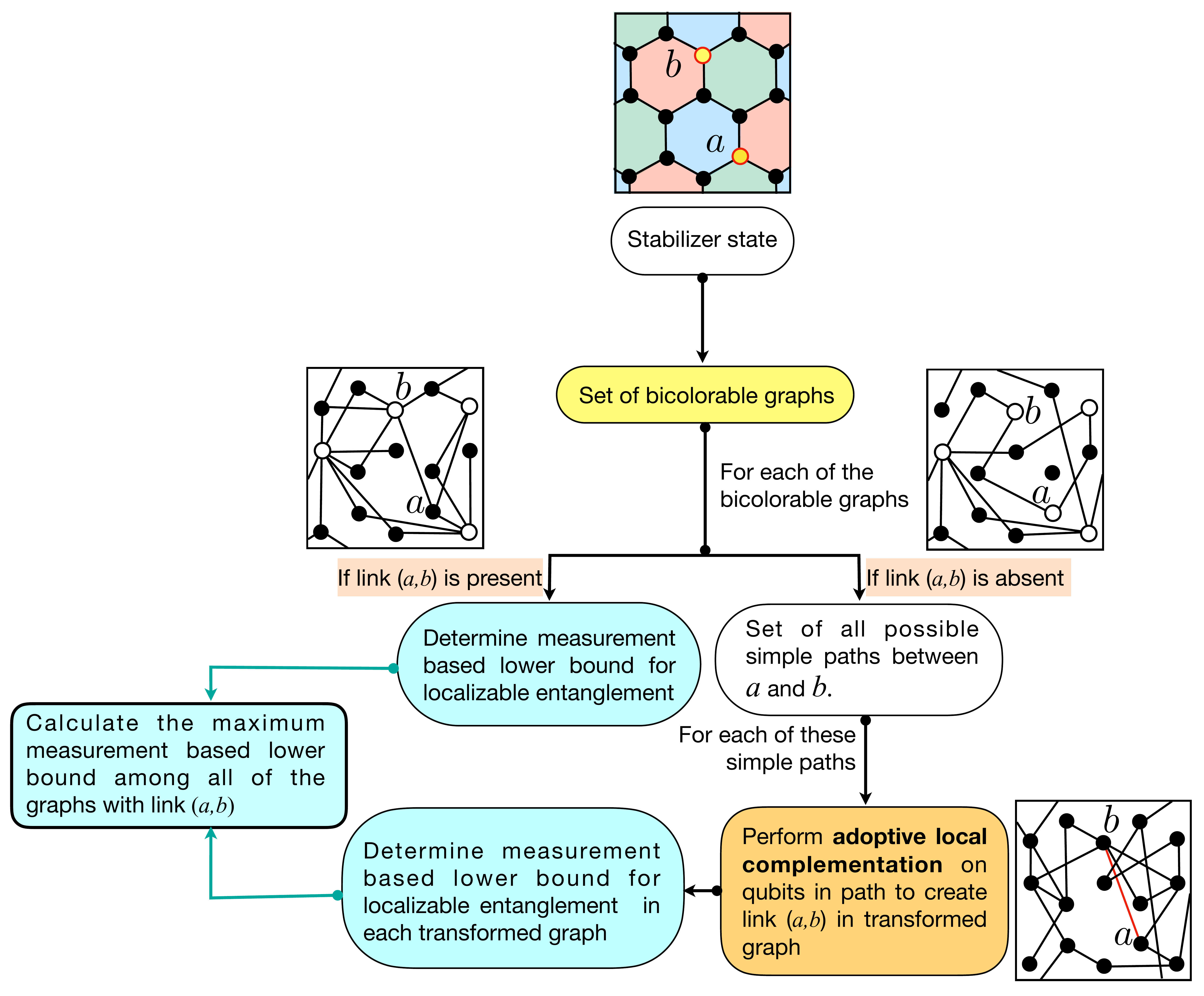}
\caption{(Color online.) \textbf{Graph-based method.} A schematic representation of the graph-based method to obtain non-zero measurement-based lower bounds of localizable entanglement, as presented in Sec.~\ref{subsec:mlb}. The input state is the  stabilizer state of the topological quantum error correcting code under specific noise, and a pair of chosen qubits on which entanglement is to be localized. The next step is to determine the set of bicolorable graphs from the stabilizer structure by using \textbf{Stabgraph} (see Sec.~\ref{sec:graph-topo} and Appendix~\ref{app:algebraic_approach} for the details on the procedure, and Ref.~\cite{amaro_github_2019} for the package). For each graph in the set, if the chosen qubits are connected, the methodology developed in Ref.~\cite{amaro2018} is to be used to compute the lower bound of localizable entanglement. On the other hand, if the qubits are not connected, the adaptive local complementation algorithm is to be used in the form of the \textbf{ALCPack} on the graph so that a link between the two qubits is generated via a graph transformation (see Appendix~\ref{app:graph-based} for details on the algorithm, and Ref.~\cite{pal_github_2019} for the package). Once the link is created,  the methodology developed in Ref.~\cite{amaro2018} can be used to compute the lower bound of localizable entanglement. The maximum of all of these values of lower bounds is  to be chosen to obtain the best measurement-based lower bound of localizable entanglement on the two chosen qubits.  
%(b) Transformation of a $7$-qubit topological color code to a graph $G^\prime$ by using the graph-based algorithm described in Sec.~\ref{subsec:mlb}, where the nodes $1$ and $5$ are connected by a link in $G^\prime$. A non-zero value of $E^{\mathcal{P}}_{15}(\rho^\prime)$ in $G^\prime$ requires local projection measurement in the $Z$-basis at all other qubits. This corresponds to $X$- and $Y$ measurement on qubits $7$  and $2$, respectively, and $Z$-measurement everywhere else. See Appendix~\ref{app:7qubit} for details.
}
\label{fig:graph_based_method}
\end{figure*}

The constructed local witness operator can be decomposed as a sum of the products of Pauli projection operators on $\overline{\Omega}$ and the witness operators $W^k_{\Omega}$, as (see Appendix~\ref{app:witness_decomposition} for the derivation)
\begin{eqnarray}
\mathcal{W}_{\Omega}=\sum_{k=0}^{2^{m}-1}P_{(k,v)}^{\overline{\Omega}}\otimes W_{\Omega}^{k},
\label{eq:decompose_2_main}
\end{eqnarray} 
where $m$ is the number of qubits in $\overline{\Omega}$, $P^{\overline{\Omega}}_{(k,v)}=2^{-m}\bigotimes_{i\in\overline{\Omega}}\left[I_i+(-1)^{k_i}\tau_{v_i}\right]$, and $k\equiv k_1k_2\cdots k_m$ and $v\equiv v_1v_2\cdots v_m$ are multi-indices with $v_i=1,2,3$, such that $k$ represents the outcome of the projection operation, and  $\tau_1=X$, $\tau_2=Y$, and $\tau_3=Z$. For a specific set of Pauli operators chosen for the qubits in $\overline{\Omega}$ (i.e., for a specific value of $v$), the expectation value of $\mathcal{W}_{\Omega}$ with respect to the state $\rho_S$ is 
\begin{eqnarray}
\omega=\text{Tr}(\mathcal{W}_{\Omega}\rho_S)=\sum_{k=0}^{2^m-1} p_{k}\text{Tr}(W_{\Omega}^{k}\rho_{\Omega}^{k})=\sum_k p_{k}\omega_{k},
\label{eq:average_witness}
\end{eqnarray} 
where $p_{k}=\text{Tr}\left[\left(P_{(k,v)}^{\overline{\Omega}} \otimes I_{\Omega}\right)\rho_S\right]$ is the probability of obtaining the measurement outcome $k$ if the projection operator $P_{(k,v)}^{\overline{\Omega}}$ is applied on the qubits in $\overline{\Omega}$, and to obtain this, we have used 
\begin{eqnarray}
\text{Tr}\left[P_{(k,v)}^{\overline{\Omega}}\otimes W_{\Omega}^{k}\rho_S\right]=\text{Tr}_{\Omega}\left[W_{\Omega}^{k}\text{Tr}_{\overline{\Omega}}\left(P_{(k,v)}^{\overline{\Omega}}\rho_S\right)\right],
\end{eqnarray}
with  $\rho_{\Omega}^{k}=\text{Tr}_{\overline{\Omega}}\left(P_{(k,v)}^{\overline{\Omega}}\rho_S\right)$. Note that we have now been able to write the expectation value of the witness operator as the sum of the product of a number of expectation values of witness operators confined to the region $\Omega$ and the corresponding probabilities of obtaining those witness operators in $\Omega$ via a projection operation on qubits in $\overline{\Omega}$  (Eq. (\ref{eq:average_witness})). This has direct resemblance with the definition of localizable entanglement, and therefore opens up a pathway to define a lower bound of LE according to  Eq. (\ref{eq:le_wlb}). However, choice of an appropriate entanglement measure $\mathcal{E}$ for the state $\rho_{\Omega}^k$ still remains a challenge. The required characteristics of the chosen entanglement measure would be as follows. 
\begin{enumerate}
\item In order to exploit Eq.~(\ref{eq:average_witness}), the functional relation $\mathcal{E}_{\text{min}}(\omega)$ should be such that 
\begin{eqnarray}
\mathcal{E}_{\text{\scriptsize min\normalsize}}\left(\omega\right)=\sum_{k=0}^{2^m-1}p_{k}\mathcal{E}_{\text{\scriptsize min\normalsize}}(\omega_k).
\end{eqnarray}
\item The forms of the witness operator $W^k_{\Omega}$  suggest that the chosen measure $\mathcal{E}$ should capture the genuine multiparty entanglement in $\rho^k_{\Omega}$.
\end{enumerate}

The first requirement indicates $\mathcal{E}_{\text{\scriptsize min\normalsize}}(\omega)$ to be a linear function of $\omega$, while the second requirement demands a computable genuine multiparty entanglement measure for mixed multiparty states. From here onward, we focus on regions $\Omega\equiv ab$ of size $2$ (as in the example in Fig.~\ref{fig:local_witness}), constituted of qubits, say, $a$ and $b$, and choose negativity as the entanglement measure (see Appendix~\ref{app:negativity} for definition) for mixed states. It has been shown in~\cite{amaro2018} that for a two-qubit state $\rho_{ab}$ and the corresponding expectation value $\omega$ of  witness operator $\mathcal{W}$ having the  form given in  Eq.~(\ref{eq:witness_global}), the lower bound of entanglement, $\mathcal{E}_{\text{min}}$, as measured by negativity~\cite{peres1996,horodecki1996} and as a function of $\omega$, can be obtained as $\mathcal{E}{\text{\scriptsize min\normalsize}}(\omega)=-2\omega$, which is linear in $\omega$. Also, negativity captures the genuine multiparty  entanglement over the two-qubit region $ab$. Therefore, from Eqs.~(\ref{eq:le_wlb}) and (\ref{eq:average_witness}), 
\begin{eqnarray}
E_{ab}^W(\rho_S)=-2\omega,
\label{eq:wlb_final}
\end{eqnarray}
which can be computed from the original state by measuring only the expectation value of the constructed local witness operator $\mathcal{W}_{\Omega}$.

\subsection{Measurement-based lower bound: Graph-based method}
\label{subsec:mlb}

Here we present a complete description of the methodology for determining the MLB of LE between any two qubits $a$ and $b$ in a large stabilizer state under noise. More specifically, we prescribe a logical choice for the specific Pauli measurements over the qubits in $\overline{\Omega}$ which guarantees a non-trivial MLB, by exploiting our previously reported results~\cite{amaro2018} on graph states (a primer on the graph states and the graph transformations required for our purpose is provided in Appendix~\ref{app:graphs}).  In order to make the paper comprehensive for a general audience, we postpone discussions on the technical details of the different steps involved for the Appendices.

\subsubsection{Graph-based method: Underlying mechanism}

Let us denote the stabilizer state of $N$ physical qubits, $\ket{\psi_S}$, under noise represented by the map $\Lambda(.):$
\begin{eqnarray}
\ket{\psi_S}\rightarrow\rho_S=\Lambda(\ket{\psi_S}\bra{\psi_S}).
\end{eqnarray}  
Any stabilizer state can be transformed to a graph state $\ket{\psi_G}$~\cite{van-den-nest2004} corresponding to a connected bicolorable graph $G$ by local unitary transformations, such that 
\begin{eqnarray}
\ket{\psi_G}=U_{S\rightarrow G}\ket{\psi_S}, 
\end{eqnarray}
where $U_{S\rightarrow G}=\bigotimes_i{U_i}$, $U_i$ being either a single-qubit Clifford unitary operation, or the identity operator in the qubit Hilbert space (a detailed discussion on a geometric approach towards this transformation is given in Sec.~\ref{sec:graph-topo}, and the algebraic details related to this methodology can be available at Appendix~\ref{app:algebraic_approach}). The graph is bicolorable since the qubits situated on the nodes can be divided into two disjoint sets -- a set of \emph{control} qubits and a set of \emph{target} qubits, where inter-set links are present in the graph, but intra-set links are not allowed. In the case of a graph state, we have provided a prescription for determining a non-trivial MLB of LE over a region $\Omega$ in~\cite{amaro2018} as long as $\Omega$ is connected, which, for a two-qubit region, implies the existence of a link between the two qubits constituting the region. However, in the present case, the underlying connected graph $G$  may or may not contain the link $(a,b)$ corresponding to the two-qubit region $\Omega\equiv ab$. In such scenario, a set of local complementation (LC) operations on a number of strategically chosen qubits situated on a selected simple path $\mathcal{L}_{ab}$ connecting the nodes $a$ and $b$ may result in a graph transformation $G\rightarrow G^\prime$, where the link $(a,b)$ exists in $G^\prime$~\cite{amaro2018}.   A sequence of LC  operations on a graph $G$ is equivalent to a local unitary operation $U_{G\rightarrow G^\prime}=\bigotimes_{i}U^\prime_i$ such that 
\begin{eqnarray}
\ket{\psi_{G^\prime}}=U_{G\rightarrow G^\prime}\ket{\psi_{G}},
\end{eqnarray} 
and $U^\prime_i$ is either a single-qubit Clifford unitary operation or the identity operator corresponding to the qubit $i$ situated on the chosen simple path $\mathcal{L}_{ab}$. Therefore,  the complete transition of the stabilizer state to the graph state $\ket{\psi_{G^\prime}}$ can  be described by a unitary operation (see Appendix~\ref{app:7qubit} for a demonstration with a $7$-qubit color code)
\begin{eqnarray}
U_{S\rightarrow G^\prime}=U_{G\rightarrow G^\prime}U_{S\rightarrow G}.
\end{eqnarray}

Let us proceed with the assumption that the graph state $\ket{\psi_{G^\prime}}$ corresponding to the graph $G^\prime$ containing the link $(a,b)$ has been created from $\ket{\psi_S}$ via local unitary transformation $U_{S\rightarrow G^\prime}=U_{G\rightarrow G^\prime}U_{S\rightarrow G}$. In presence of the noise represented by the map $\Lambda(.)$, 
\begin{eqnarray}
\rho_S &=&\Lambda(\ket{\psi_S}\bra{\psi_S})=U^{-1}_{S\rightarrow G^\prime}\Lambda^\prime(\ket{\psi_{G^\prime}}\bra{\psi_{G^\prime}})U_{S\rightarrow G^\prime}\nonumber\\ &=&U^{-1}_{S\rightarrow G^\prime}\rho^\prime U_{S\rightarrow G^\prime},
\end{eqnarray}
where $\rho^\prime=\Lambda^\prime(\ket{\psi_{G^\prime}}\bra{\psi_{G^\prime}})$, and $\Lambda(.)\rightarrow\Lambda^\prime(.)$ is the transformation of the noise due to the local unitary operation. Note that the LE and the RLE for $\rho_S$ are the same as respectively the LE and the RLE of  $\rho^\prime$ due to the local Clifford unitary connection between $\rho_S$ and $\rho^\prime$. But computation of the LE and the RLE over the region $ab$ in $\rho^\prime$ still remains difficult in the case of large $N$ which results in large $m=N-2$, and one has to look for an appropriate Pauli measurement set-up in $\overline{\Omega}$ which can provide a computable non-trivial MLB for $\rho^\prime$. The local unitary connection between $\rho_S$ and $\rho^\prime$ can then be exploited to connect the value of the MLB with a specific measurement setup in the case of $\rho_S$.  We have shown in~\cite{amaro2018} that in situations where noise in the system is low, and the link between qubits  $a$ and $b$ exists in the graph-state representation, an appropriate choice of measurement basis for a non-trivial MLB $E^\mathcal{P}_{ab}(\rho^\prime)$ of $E_{ab}(\rho^\prime)$ is local $Z$ measurements on all the qubits except qubits $a$ and $b$, which is an optimal basis for the LE over $ab$ in $\ket{\psi_{G^\prime}}$. The value of $E_{ab}^\mathcal{P}(\rho^\prime)$ $\left(=E_{ab}^\mathcal{P^\prime}(\rho_S)\right)$ represents the value of the MLB corresponding to a specific Pauli measurement setup $\mathcal{P}^\prime$ for the original state $\rho_S$, where  $\mathcal{P}^\prime$ can be obtained by transforming the $Z$ measurements on the qubits in $\overline{\Omega}\equiv\overline{ab}$ according to the over-all local unitary operation $U_{S\rightarrow G^\prime}$. 

The graph-based algorithm is summarized in Fig.~\ref{fig:graph_based_method} (a pseudo-code for the algorithm can be found in Appendix~\ref{app:pseudo_codes}). Evidently, the graph-based algorithm has two parts -- transforming the stabilizer state to the graph states, for which the graph adjacency matrix is to be determined, and transforming the graph via local complementation operations  to another graph in which a link exists between the chosen qubits. The first part and its different aspects have been discussed in Sec.~\ref{sec:graph-topo} and Appendix~\ref{app:algebraic_approach}, and the algorithm has been transformed into a Python open-source package,  namely, \textbf{StabGraph}~\cite{amaro_github_2019}, which generates the adjacency matrix of a graph corresponding to a graph state which is connected to a given stabilizer state via local unitary operators.  For the second part of the graph-based method, the key challenge is to ensure the certainty of creating a link between two chosen qubits by an optimal sequence of LC operations on a set of qubits in the graph. Towards this goal, we have developed an \emph{adaptive} LC  (ALC) algorithm, which, along with the corresponding graph transformations, is discussed in detail in Appendix~\ref{app:graph-based}. The crux of the algorithm depends on adapting itself according to the change in the graph after each local complementation operation on individual qubits, and subsequently choosing the qubit for the next local complementation operation according to the updated information. This algorithm has been made available as a package named  \textbf{ALCPack}~\cite{pal_github_2019}, which creates a link between any two given nodes in a simple, connected, and undirected graph via the adaptive local complementation method.

\subsubsection{Numerical aspects of the graph-based approach}

Note here that there exists a set of bicolorable graphs $\{G\}$ that can be obtained from a specific stabilizer state $\ket{\psi_S}$ by appropriately varying the local unitary operation $U_{S\rightarrow G}$. Moreover, for each such bicolorable graph $G$ where $a$ and $b$ are not connected, one can choose a set of simple paths  so that LC operations on each of these paths would provide a graph $G^\prime$ with the link $(a,b)$. Let us denote the complete set of all possible bicolorable graphs $G$ obtained from $\ket{\psi_S}$   by $\mathcal{S}_{G}$, having cardinality $N_{G}$.  In situations where the link $(a,b)$ $\notin$ $G$, let us denote the set of all possible simple paths $\mathcal{L}_{ab}$ connecting the qubits $a$ and $b$ be $\mathcal{S}_{\mathcal{L}_{ab}}$, having cardinality $N_{\mathcal{L}_{ab}}$. In the case of an arbitrary stabilizer state with large $N$, $N_G$ is usually a large number\footnote{For a color code with  $N_p$ plaquettes, there can be $N_p$ control qubits among $N$ qubits, and the number $N_G$ varies as $\binom{N}{N_p}$.}. Also, for a given graph and a given pair of nodes $\{a,b\}$, determination of all possible paths between $a$ and $b$ can be difficult when $N$ as well as the number of links in the graph is large. Therefore, for a large system $S$ described by a stabilizer state $\ket{\psi_{S}}$, it is difficult to obtain the optimal value of $E_{ab}^\mathcal{P}(\rho^\prime)$. However, since each graph $G^\prime$ results in a non-zero value of MLB, one can choose only a subset $\mathcal{S}^\prime_G$ of $\mathcal{S}_{G}$ with cardinality $n_G\leq N_G$, and a subset $\mathcal{S}^\prime_{\mathcal{L}_{ab}}$ of $\mathcal{S}_{\mathcal{L}_{ab}}$ with cardinality $n_{\mathcal{L}_{ab}}\leq N_{\mathcal{L}_{ab}}$ for each $G^\prime\in\overline{\mathcal{S}}_G$. This  leads to a set of  $n_Gn_{\mathcal{L}_{ab}}$  graphs $G^\prime$, for each of which a value of $E^\mathcal{P}_{ab}(\rho^\prime)$ can be obtained. The bound $\max E^\mathcal{P}_{ab}(\rho^\prime)$ can be tightened by increasing the value of $n_Gn_{\mathcal{L}_{ab}}$ according to the available numerical resources. Note  that the analytical computation of $E^\mathcal{P}_{ab}(\rho^\prime)$ depends also on the noise model $\Lambda$. As shown in~\cite{amaro2018}, in the case of local uncorrelated Pauli noise, $E^\mathcal{P}_{ab}(\rho^\prime)$ is analytically computable for arbitrary system size as long as the neighborhood of the region $\Omega\equiv ab$ in $G^\prime$ and the noise present in this region  is fully known.

%In Fig.~\ref{fig:graph_based_method_example}, a simple example of creating a graph $G^\prime$ from a color code and determining the transformation $\mathcal{P}\rightarrow \mathcal{P}^\prime$ due to the local unitary $U_{S\rightarrow G^\prime}$ is shown using the $7$-qubit color code. 

A word on the dependence of the run-time  of the graph-based method, in particular, the adaptive local-complementation algorithm and the algorithm used to get the graph from the stabilizer structure, on the system size $N$ is in order here. Since the ALC algorithm takes into account the transformed graphs at each of its steps, it is difficult to determine an exact dependence of the run-time of the algorithm on system size. However, one can determine a bound on how the run-time scales with $N$, by determining the maximum number of link operations during the ALC algorithm, which is $\leq N^3$  (see Appendix~\ref{app:graph-based} for an explicit derivation). On the other hand, the Stabgraph algorithm used Gauss elimination technique, and scales as $\sim N^3$. These indicate an overall polynomial scaling of the graph-based method with system size.

We point out here that one can also control the transformation $S\rightarrow G$ in such a way that the bicolorable graph $G$ directly contains the link $(a,b)$ (see Sec.~\ref{sec:graph-topo} for a discussion on how to ensure the creation of the link in $G$). Note that in the modified algorithm, which we refer to as the \emph{modified graph-based method}, the optimization of $E^{\mathcal{P}}_{ab}(\rho^\prime)$ is performed over a set of bicolorable graphs the corresponding states of which are connected to each other via local unitary operations. On the other hand, the former algorithm additionally uses graph states outside the set of bicolorable graphs. Therefore, the maximum value of $E^{\mathcal{P}}_{ab}(\rho^\prime)$ obtained from the former algorithm is higher than the same obtained from the latter. See Appendix~\ref{app:pseudo_codes} for a pseudo code of the modified graph-based method.

%In the next two sections, we discuss the technical issues related to the transformations $S\rightarrow G$ and $G\rightarrow G^\prime$ in detail. Readers interested in the application of the graph-based algorithm may skip ahead to Sec.~\ref{sec:apply}, where we use the methodology in a TCC embedded on a hexagonal lattice, and study the distance and noise dependence of the lower bound of LE.  

\section{Graphs from topological codes: geometric approach}
\label{sec:graph-topo}

An arbitrary stabilizer state $\ket{\psi_S}$ describing a system $S$ can be shown to be connected to a graph state $\ket{\psi_G}$ defined on a bicolorable  graph $G$ via a local unitary transformation~\cite{van-den-nest2004}. While this transformation has an established algebraic formulation, it has also been shown~\cite{lang2012} in the case of Kitaev's toric code that the graph $G$  can also be constructed from the structure of the stabilizers via a geometric recipe. In this Section, we introduce the geometric recipe for the topological color codes (see Appendix~\ref{app:tcc} for the details on the topological color codes) and explain the underlying idea which emerges from the  preparation protocols of the logical states of the code. 

%We also present the algebraic method for arbitrary stabilizer states, and propose an algorithm for explicitly obtaining the adjacency matrix of a bicolorable graph from the set of stabilizers.  

To discuss how a graph state can be obtained from a logical state of a topological color code via a geometric construction, we point out that the procedure for the creation of the logical states of a code on a lattice involves (1) initializing the qubits to  either $\ket{0}$ or $\ket{+}$ so that they collectively form a product state, and then (2) creating plaquette-wise GHZ state-type~\cite{greenberger1989} entanglement~\cite{nigg2014} via controlled entangling gates, which naturally labels one of the qubits as control (c), and the rest of the qubits as target (t) (see Appendix~\ref{app:ct} for an example with four qubits). The state on each of the plaquettes, however, can be further transformed to a graph state corresponding to a simple, connected, and undirected star-shapped graph with the control qubit $c$ as the central qubit, and the target qubits $t$ as the peripheral qubits, by applying local unitary transformations in the form of Hadamard operations on the target qubits. In terms of the stabilizer operators, the plaquette stabilizers corresponding to the plaquette are transformed to the graph-state generators via application of Hadamard operations on the target qubits. As an example, following this prescription, the $\ket{0}_L$ state of the $7$-qubit color code can be created by (1) choosing the qubits $1$, $5$, and $7$, located at the corners of the triangular code, as control qubits controlling the rest of the qubits in their respective plaquettes, (2) initializing the seven qubits to the state $\ket{+_10_20_30_4+_50_6+_7}$, and (3) applying controlled phase gate to the $(c,t_i)$ pairs. Subsequently, the local unitary connected graph state, which is obtained by applying Hadamard operations on all the target qubits in the $7$-qubit lattice, corresponds to a graph that is obtained by creating the three star graphs with central qubits $c=1,5,7$, and their respective target qubits (See Fig.~\ref{fig:circuit_color}(a)). 

\begin{figure}
\includegraphics[scale=0.25]{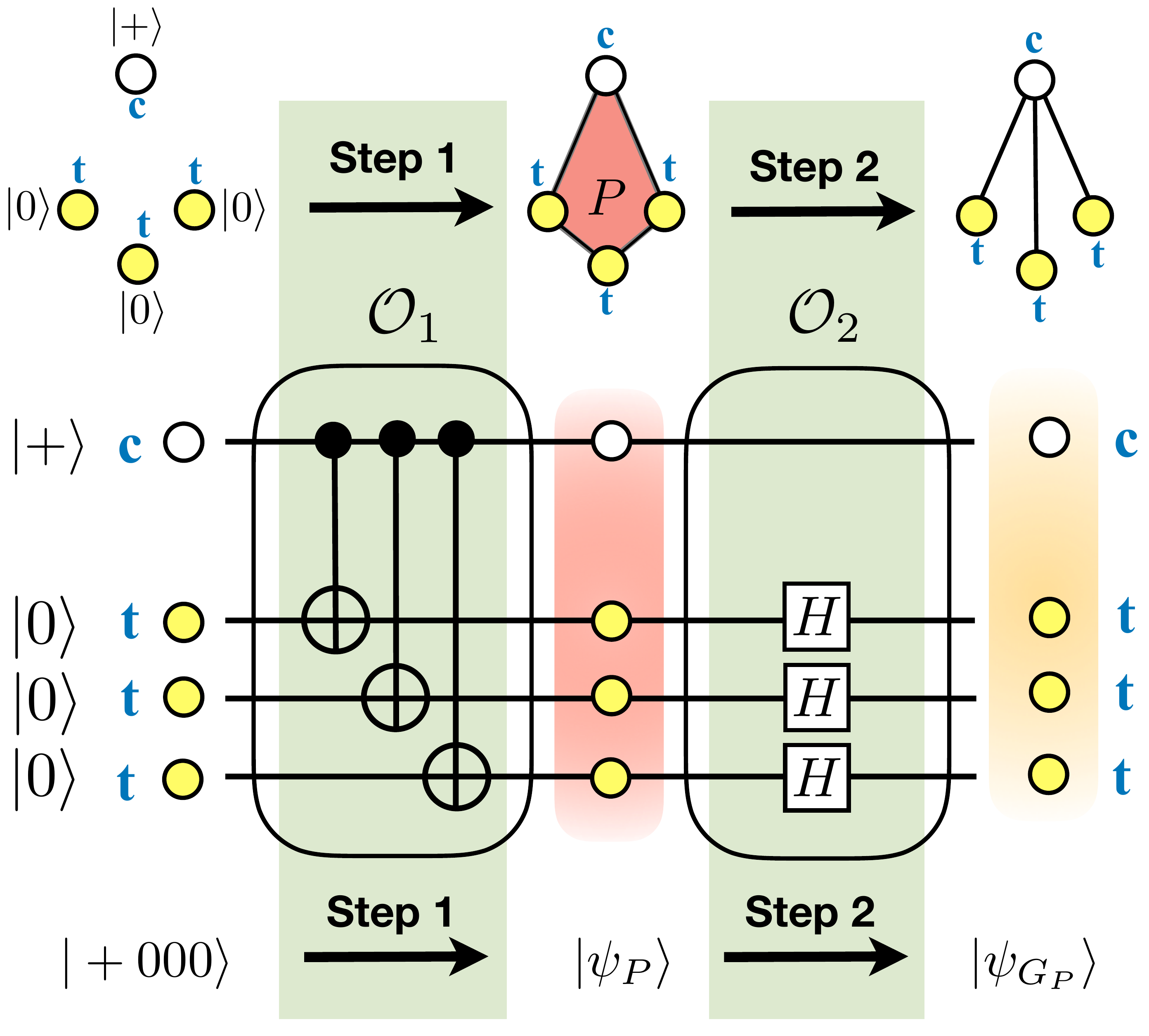}
\caption{(Color online.) \textbf{Quantum circuit creating GHZ-type entanglement.} The quantum circuit that takes the state $\ket{+000}$ to the four-qubit GHZ state $\ket{\psi_P}=\frac{1}{\sqrt{2}}(\ket{0000}+\ket{1111})$ (denoted by operation $\mathcal{O}_1$), and then further takes it to the four-qubit graph state $\ket{\psi_{G_P}}=\frac{1}{\sqrt{2}}(\ket{0+++}+\ket{1---})$ (denoted by operation $\mathcal{O}_2$). The implication of applying this circuit to four qubits arranged as a plaquette of a topological color code has also been shown at the top, where the graph $G_P$ corresponding to the state $\ket{\psi_{G_P}}$ is created on the four qubits, starting from four qubits in the product state $\ket{+000}$ and via the single-plaquette GHZ state $\ket{\psi_{P}}$. See Appendix~\ref{app:ct} for a detailed description.}
\label{fig:circuit}
\end{figure}

\begin{figure*}
\includegraphics[width=0.7\textwidth]{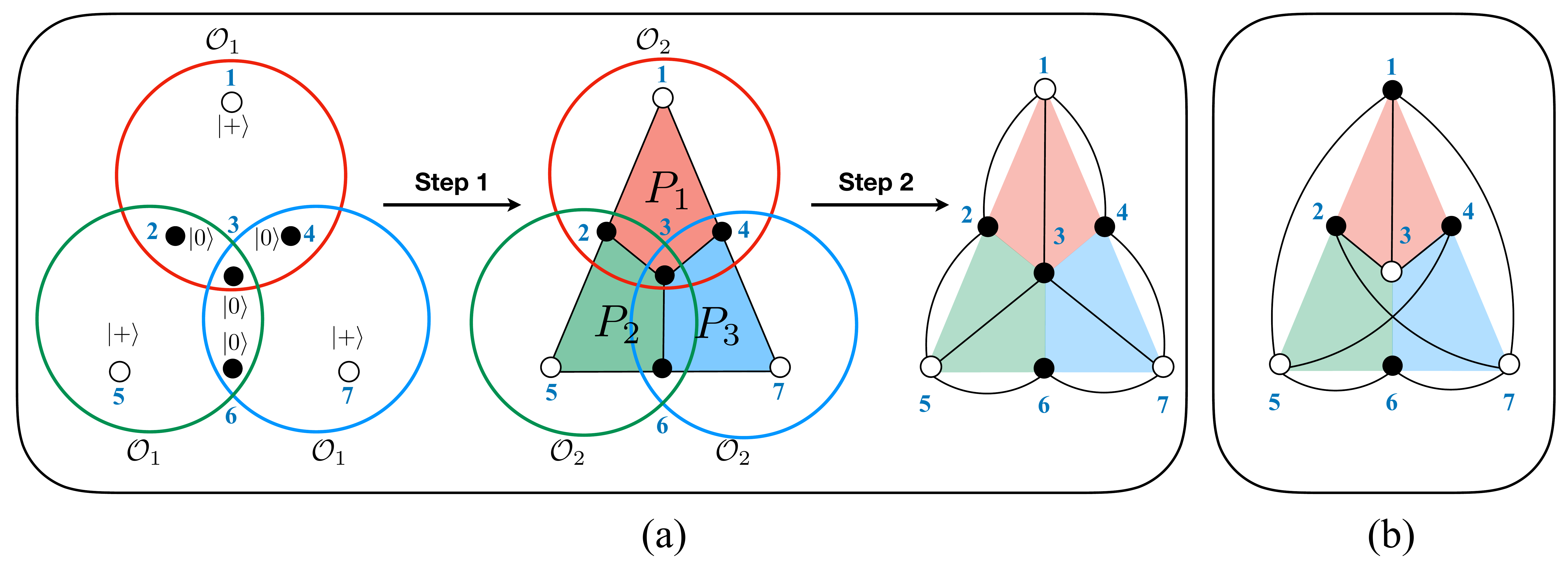}
\caption{(Color online.) \textbf{Graph states from the $7$-qubit color code.} (a) Application of $\mathcal{O}_1$ (Fig.~\ref{fig:circuit}) on groups of four qubits, given by $\{1,2,3,4\}$, $\{5,2,3,6\}$, and $\{7,3,4,6\}$, on $7$ qubits labelled as $\{1,2,\cdots,7\}$ and initialized at $\ket{+_10_20_30_4+_50_6+_7}$, leads to the $\ket{0}_L$ state of the $7$-qubit color code, where qubits $1$, $5$, and $7$ are chosen as control for the four groups of qubits respectively. Application of $\mathcal{O}_2$ on the same groups of qubits, i.e.,  application of Hadamard operation on the target qubits $2,3,4,6$ in the next step leads to a graph state corresponding to a graph obtained by creating three star graphs on the three groups of qubits, $\{1,2,3,4\}$, $\{5,2,3,6\}$, and $\{7,3,4,6\}$, where the control qubits $1$, $5$, and $7$ are used as central qubits respectively. (c) If one assumes the qubits $3$, $5$, and $7$ to be the control qubits, controlling the target quits in the plaquettes $P_1$, $P_1P_2$, and $P_1P_3$ respectively, then the resulting graph is obtained by combining the start graphs on the groups of qubits $\{1,2,3,4\}$, $\{1,4,5,6\}$, and $\{1,2,6,7\}$, using respectively qubits $3$, $5$, and $7$ as central qubits.}
\label{fig:circuit_color}
\end{figure*}

Note that in the case of topological color codes, the term \emph{plaquette} does not always represent the original plaquettes of the color code lattice.  It can be used in a broader sense, since a topological code can also be defined in terms of the products of its original plaquette stabilizers, defining the resultant plaquettes corresponding to the stabilizer operators obtained by multiplying two or more than two of the original plaquette stabilizer operators of the original lattice. Therefore, applying the above prescription for creating GHZ state-type entanglement needs to be suitably generalized for larger codes. Nevertheless, the above recipe can be applied to all plaquettes in a topological code in its logical state, where the challenge is  identifying the appropriate set of control qubits at correct positions of the code, and determining the correct sets of target qubits that are controlled by each of these control qubits. The graph corresponding to a graph state that is local unitary equivalent to the logical state of the topological code can then  simply be obtained by creating all the star-shaped  graphs that involve a control qubit and all its target qubits, with the control qubit as the central qubit. For example, the seven-qubit color code shown in Fig.~\ref{fig:circuit_color}(a) can also be expressed in terms of the plaquette stabilizer operators corresponding to the re-combined plaquettes $P_1$, $P_1P_2$, and $P_1P_3$. This can be understood from the example shown in Fig.~\ref{fig:circuit_color}(b), where the qubit $3$ controls the plaquette $P_1$, while the qubit $5$ controls the plaquette $P_1P_2$ constituted of qubits $1,2,5,6,3,4$. Although there are two control qubits, qubits $5$ and $3$, among these, the target qubits $\{1,6,4\}$ in  plaquette $P_1P_2$ are controlled by only the control qubit $5$, while the target qubit $2$ is controlled by none of the control qubits from the plaquette $P_1P_2$. The plaquette $P_1P_3$, on the other hand, is controlled by the qubit $7$. Note  that the $x$-type stabilizers associated to the plaquettes $\{P_1, P_1P_2, P3\}$ are an equivalent subset of the $x$-type generators associated to the plaquettes $\{P_1, P_2, P_3\}$, which ensures the validity of the geometric recipe.  It is also clear from the above discussion that the number of chosen control qubits has to be equal to the number of plaquettes in the topological color code.

\begin{figure*}
\includegraphics[width=0.7\textwidth]{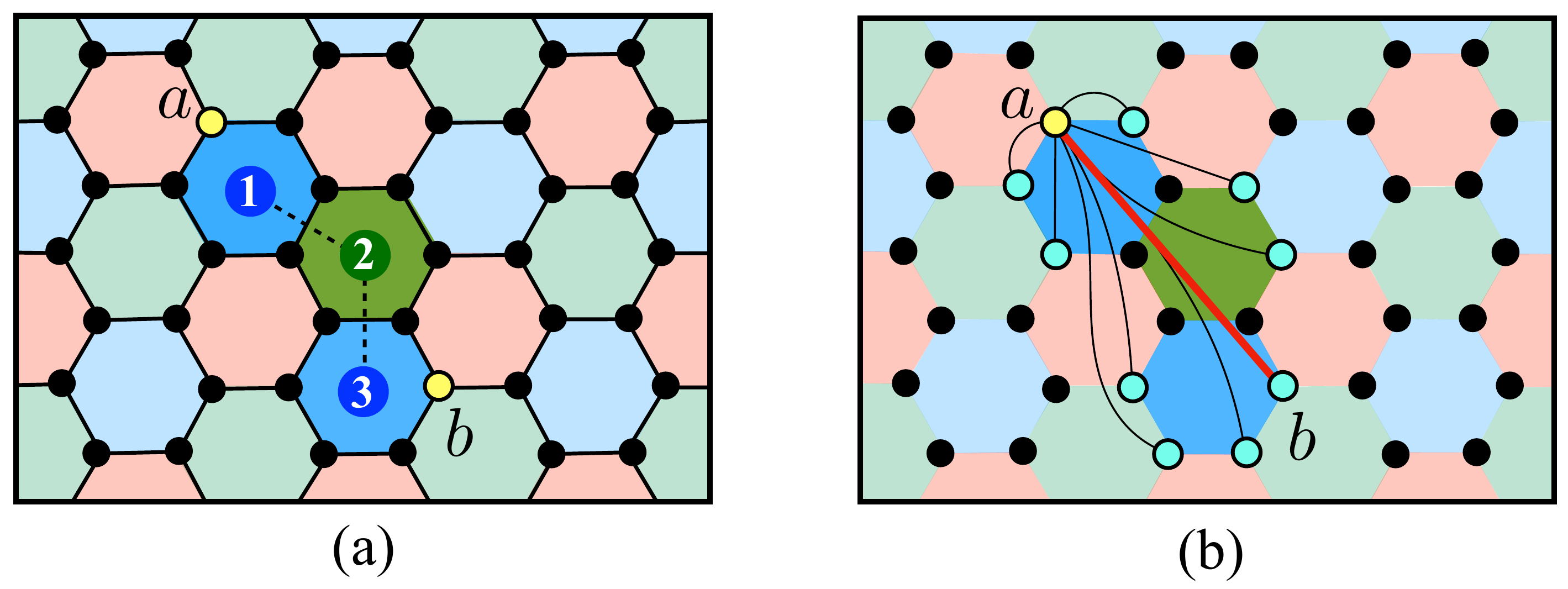}
\caption{(Color online.) \textbf{Modified graph-based method: Creation of a link without local complementation} (a) In order to create a link between two given qubits $a$ and $b$, a path of adjacent plaquettes, constituted of plaquettes $1$, $2$, and $3$ is chosen, so that the given qubits are contained by the composite larger plaquette constituted of the plaquettes from the path. (b) The next step of constructing the graph corresponding to the local unitary connected graph state would require creating a the star graph with the qubits on the larger plaquette, using qubit $a$ (yellow circle) as control, and qubit $b$ as one of the targets (turquoise circles).}
\label{fig:direct_link}
\end{figure*}

In view of all these aspects, for an arbitrary topological color code with an appropriate choice of the set of control qubits and the set of target qubits that are controlled by them,  one has to ensure the following conditions: 
\begin{enumerate}
\item a qubit, once chosen as a target (control) qubit, can not be chosen as a control (control or target) qubit any more, and 
\item the support of  all the plaquettes and the logical operators, as well as the products of them, must contain at least one control and one target qubits . 
\end{enumerate} 
Note here that the condition 1 ensures that the transformation from the logical state of a TCC to graph state ensures only $(c,t_i)$-type links, thereby dividing the qubits into two mutually disjoint sets of control and target qubits, and ensuring that the resulting graph is bicolorable, i.e., $(c,c)$ and $(t_i,t_j)$ links are not present. Note also that in the modified graph-based method discussed in Sec.~\ref{subsec:mlb}, one needs to ensure creation of a link between two given qubits. This can be done by choosing a path of adjacent plaquettes so that the given qubit-pair is contained by the large plaquette constituted by the plaquettes on the path, and then using one of the given qubit-pair as control qubit to create the graph by creating the $(c,t_i)$-type links (see Fig.~\ref{fig:direct_link} for a demonstration). 

The success of the geometric method depends explicitly on the correct choice of a set of control qubits, and the determination of the sets of target qubits that are controlled by the control qubits, ensuring the conditions (i) and (ii). While this is possible irrespective of the size of the code, the choice of a correct set of control qubits may prove difficult in the case of larger codes. This leads us to an algebraic approach for creating the graph, exploiting the binary picture of the code. Note that an algebraic treatment of the same problem was considered in~\cite{van-den-nest2004}, although the connection between the geometric recipe and the algebraic method was absent. We revisit the treatment, and provide a slightly modified version of the algebraic calculation in Appendix~\ref{app:algebraic_approach} for a general stabilizer state. The geometric recipe for the color code presented above is inherent in the algebraic approach for determining the adjacency matrix of the graph from the stabilizer structure of the color code. 
%Based on the algebraic calculation, one can develop an algorithm for obtaining the adjacency matrix $\Gamma$ corresponding to the graph $G$ representing the graph state that is local unitary equivalent to the stabilizer state, by utilizing the stabilizer structure of the state only. This algorithm has been realized in the form of a Python open-source  package called \textbf{StabGraph}~\cite{amaro_github_2019}, which generates the adjacency matrix corresponding to the graph underlying the graph state which is connected via local unitary operations with the stabilizer state (see Appendix~\ref{app:pseudo_codes} for a pseudo code of the algorithm).} 

\begin{figure*}
\includegraphics[width=0.7\textwidth]{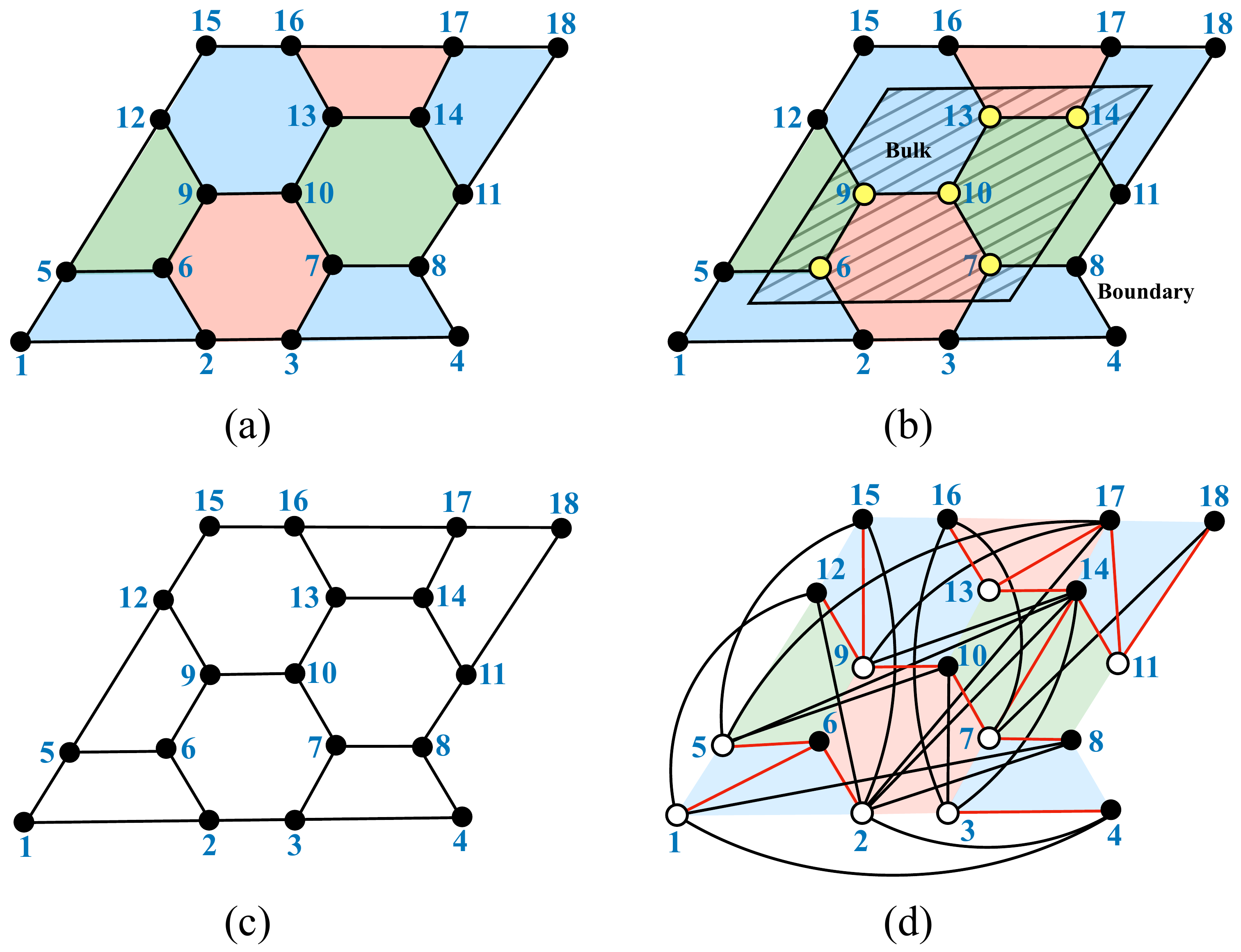}
\caption{(Color online.) \textbf{Local and non-local graphs.} (a) A color code of distance $D=4$ hosted in a square hexagonal lattice with $N=18$ qubits, and $N_p=8$ plaquettes, containing $k=N-2N_p=2$ logical qubits. (b) The bulk qubits (yellow circles) are obtained by removing $D/4$ qubits in the direction from the boundary to the center on all sides of the square hexagonal lattice. (c) A local graph equivalent to the square hexagonal color code of distance $D=4$, where all the links are local links.  (d) A non-local graph obtained by the geometric approach presented in Sec.~\ref{sec:graph-topo}, which can also be found using the algebraic method in Appendix~\ref{app:algebraic_approach}. Here, the local (non-local) links refer to the links connecting qubits belonging to the same (different) plaquettes, and are marked by red (black) continuous lines. The chosen control qubits are represented by the white circles, while the target qubits are marked by black circles.}
\label{fig:non_local_links}
\end{figure*}

\section{Application: Color code on a square hexagonal lattice}
\label{sec:apply}

In this Section, we apply the methodology developed through Secs.~\ref{sec:le_bounds}-\ref{sec:graph-topo} in the case of a color code hosted in a square hexagonanl lattice with open boundary condition on a plane, where the `square' indicates the shape of the lattice (our methodology works irrespective of whether the lattice is square or triangular; for an example of the `triangular' lattice, see the $7$-qubit code shown in Fig.~\ref{fig:graph_based_method}(b)). The number of qubits, $N$, in the code is represented by the distance, $D$, of the code, where, for the square hexagonal  lattice, $N$ increases quadratically with increasing $D$ (see Appendix~\ref{app:tcc}).
An example of the square hexagonal lattice hosting a color code with $D=4$ and containing two logical qubits is given in  Fig.~\ref{fig:non_local_links}(a), where $N=18$. We will be computing the lower bound of LE over a qubit pair $\{a,b\}$ in the bulk, which is constructed from the lattice by removing $D/4$ qubits in the direction towards the center from each boundary (see Fig.~\ref{fig:non_local_links}(b)). Note that the choice of the control and the target qubits depends explicitly on the choice of the stabilizer state. In the present case, we  consider the color code to be in $\ket{+}_L$ for all our discussions, and the local unitary connected graph state is obtained by applying Hadamard operations over the control qubits. This choice is justified as the logical Pauli-eigenstates in 2D color codes are connected by local unitary operators due to the transversality of logical Clifford gate operations~\cite{bombin2006,bombin2007}. 

\begin{figure*}
\includegraphics[width=0.7\textwidth]{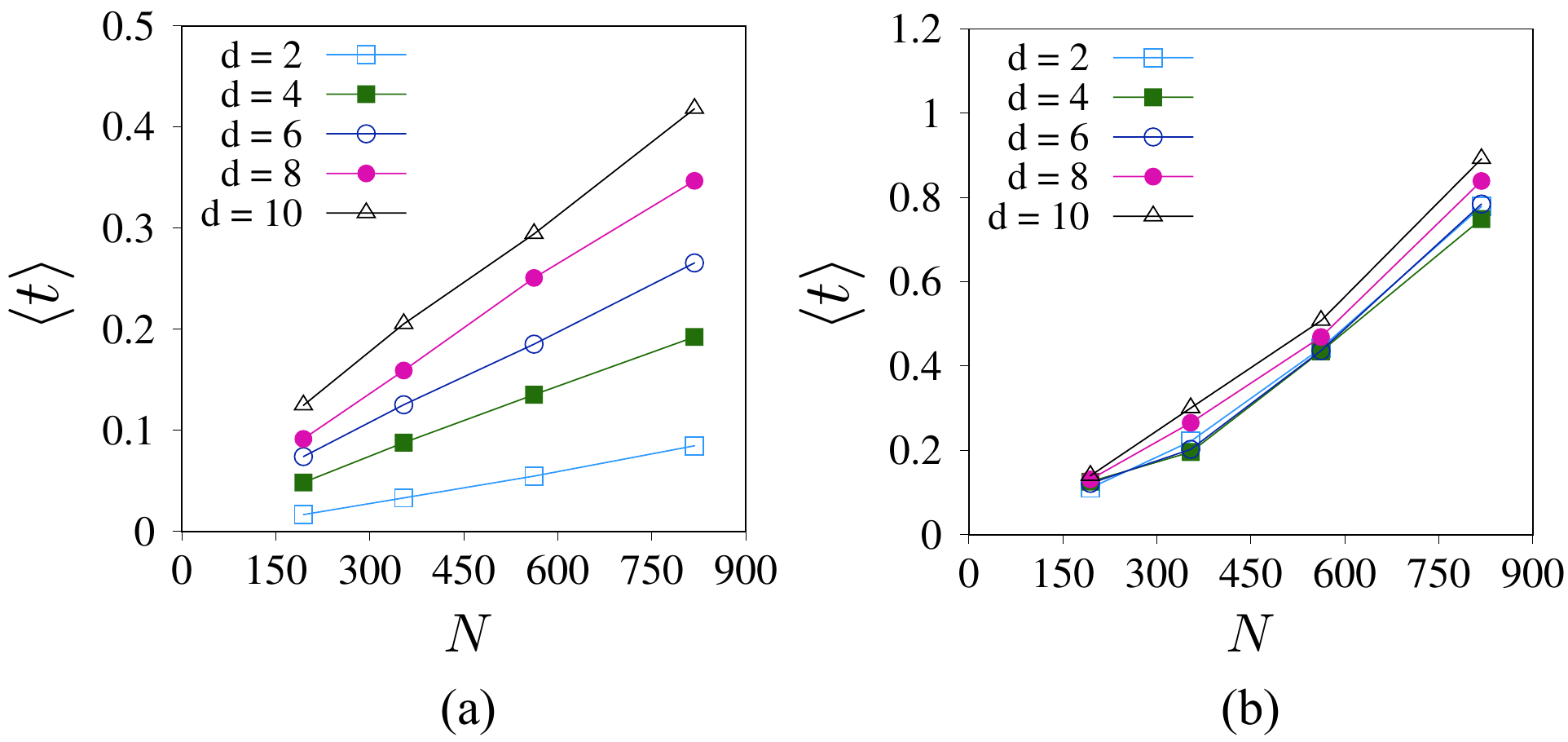}
\caption{(Color online.) \textbf{Variation of $\left\langle t\right\rangle$ with $N$.} The average time taken by the adaptive local complementation algorithm to create a link between two chosen qubits with a specific distance $d$ from each other in the bulk of (a) a local graph mimicking the square hexagonal lattice hosting a color code of  code-distance $D$, and (b) a non-local graph obtained by the method discussed in Sec.~\ref{sec:graph-topo},  as a function of the system size $N$. The value of $\langle t\rangle$ increases with increasing $d$ for a fixed system size. The unit of the $\langle t\rangle$ axis is in seconds, and the $N$ axis is dimensionless. The codes are run in a standard desktop computer. }
\label{fig:tlc}
\end{figure*}

\subsection{Computation of witness-based lower bound}
\label{subsec:appl_wlb}

Here, we explicitly compute the expectation value $\omega$ of the local witness operator $\mathcal{W}_{ab}$ of the form given in Eq.~(\ref{eq:witness_global}), constructed for a pair $\{a,b\}$ of qubits in the bulk. As discussed in Sec.~\ref{subsec:wlb} and demonstrated in Fig.~\ref{fig:local_witness}, we build these witnesses from two stabilizers denoted by $S^x$ and $S^z$, one being  $x$-type and the other  $z$-type, obtained by multiplying respectively the $x$-  and $z$-type stabilizers  corresponding to the plaquettes on two adjacent  paths of plaquettes connecting the qubits $a$ and $b$, so that the necessary conditions for the construction of $\mathcal{W}_{ab}$ (see Sec.~\ref{subsec:wlb}) are satisfied. The  path constituted of the common lattice-links between the adjacent paths of plaquettes provide the path connecting the two chosen qubits, and the length of this path is the distance $d$ between the chosen qubits. In all our discussions, we  compute the distance between the chosen qubits with respect to the square hexagonal color code lattice. 

Expanding the form of $\mathcal{W}_{ab}$ from  Eq.~(\ref{eq:witness_global}) using $S^x$ and $S^z$, the WLB $-2\omega$ is calculated as
\begin{eqnarray}
E_{ab}^W(\rho_S)=-2\omega=\frac{1}{2}\left[\omega_x+\omega_z+\omega_{xz}-1\right],
\label{eq:omega_expanded} 
\end{eqnarray}
where $\omega_{x(z)}=\text{Tr}\left[\rho_S S^{x(z)}\right]$, and $\omega_{xz}=\text{Tr}\left[\rho_S S^{x}S^{z}\right]$. For demonstration, we  consider the state $\rho_S$ originating from the application of single-qubit uncorrelated Pauli noise channels~\cite{nielsen2010,holevo2012} to the stabilizer state $\ket{\psi_S}$ (see Appendix~\ref{app:noise} for a brief description), including the bit-flip (BF), phase-flip (PF), bit-phase-flip (BPF), and depolarizing (DP) noise. For the BF (PF) noise, one can show that $\omega_x=1$ $(\omega_z=1)$ and $\omega_z=\omega_{xz}=(1-q)^{n_z}$ ($\omega_x=\omega_{xz}=(1-q)^{n_x}$), where $q$ is the strength of the noise $(0\leq q\leq 1)$ which we asssume to be the same for all qubits, and $n_x$ ($n_z$) is the number of qubits in the support $R_x$ ($R_z$) of the stabilizer $S^x$ ($S^z$). See Appendix~\ref{app:stab_expt} for the calculation in the case of PF noise.  The calculation in the case of the DP channel is similar to the same in the case of the PF channel, where the expectation values are $\omega_x=(1-q)^{n_x}$, $\omega_z=(1-q)^{n_z}$, and $\omega_{xz}=(1-q)^{n_x+n_z-2}$, where $n_x+n_z-2$ is the number of qubits in the support of $S^xS^z$.   Using these, Eq.~(\ref{eq:omega_expanded}) becomes 
\begin{eqnarray}
E_{ab}^W(\rho_S)&=&(1-q)^{n_x} \text{ for PF noise,}\nonumber \\
E_{ab}^W(\rho_S)&=&(1-q)^{n_z} \text{ for BF noise, and} \nonumber \\
E_{ab}^W(\rho_S)&=&\frac{1}{2}\big[(1-q)^{n_x}+(1-q)^{n_z}\nonumber\\ &&+(1-q)^{n_x+n_z-2}-1\big], \text{ for DP noise}.
\end{eqnarray}   
Note from the design of the local witness operator (see Fig.~\ref{fig:local_witness}) that the types of the plaquette stabilizer operators corresponding to each of the two adjacent paths of plaquettes -- one above and the other below the path made of lattice-links connecting qubits $a$ and $b$ -- are different from each other, one being $z$-type while the other $x$-type.  One obtains a valid local witness operator even when $x$- and $z$-types of the stabilizers above and below the path connecting qubits $a$ and $b$,  and contributing to $\mathcal{W}_{ab}$, are interchanged.  Note also that the values of $n_x$ and $n_z$ depends on the distance between the chosen qubits, implying that the dependence of WLB on $d$ is decided by how $n_x$ and $n_z$ grow with increasing $d$. The exact dependence of $n_x$ and $n_z$ on $d$ depends explicitly on the  construction of $\mathcal{W}_{ab}$, and the layout of the path made of lattice-links connecting the qubits $a$ and $b$ on the lattice. In general, the number of plaquettes involved in the local witness operators constructed in this way, and therefore the support of the stabilizers grows as $\sim a+bd$, with some constants $a$ and $b$, implying an exponential dependence of $\omega$ on $d$.  As an example, let us consider the layout of the path of lattice-links connecting the qubits $a$ and $b$ in the bulk as shown in Fig.~\ref{fig:local_witness}(b), for which
\begin{eqnarray}
n_x&=&6+2\left\lfloor\frac{d-1}{2}\right\rfloor,\nonumber \\
n_z&=&6+2\left\lceil\frac{d-1}{2}\right\rceil.
\end{eqnarray}

\begin{figure*}
\includegraphics[width=0.7\textwidth]{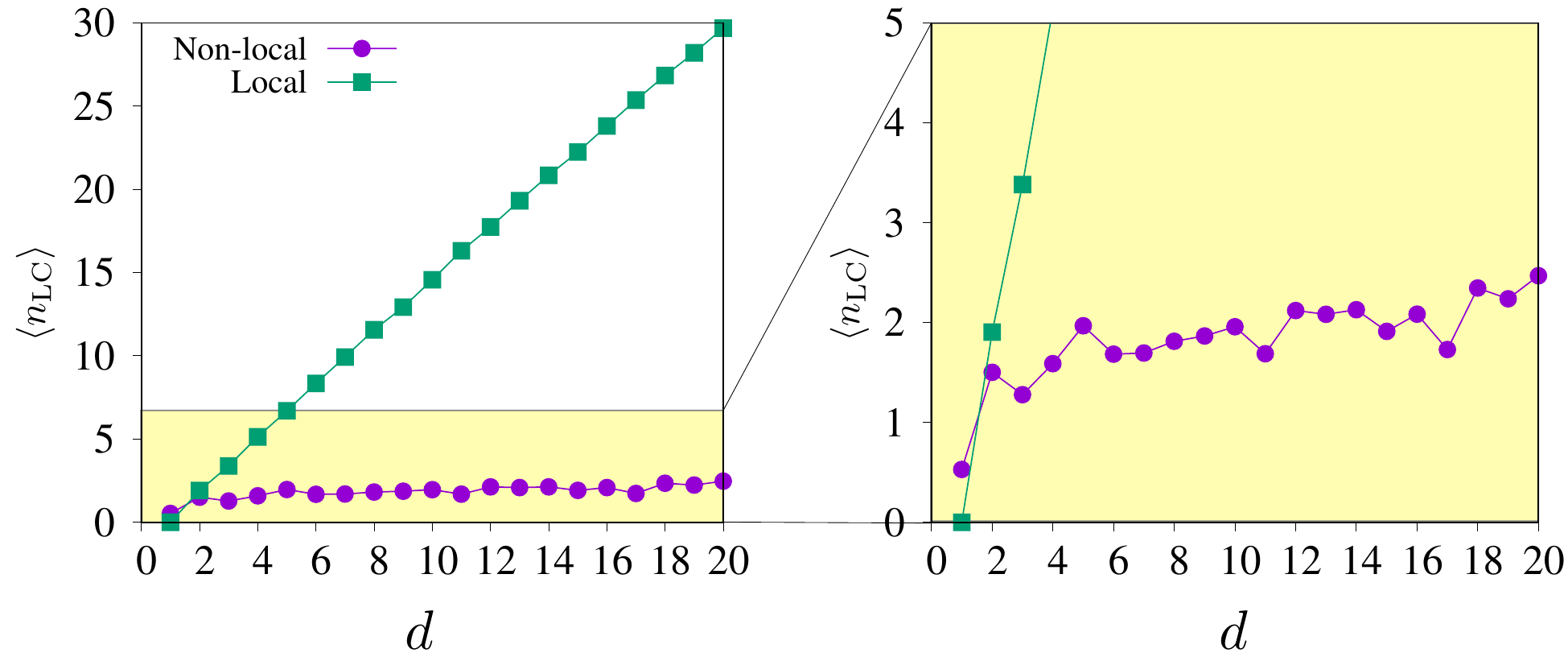}
\caption{(Color online.)\textbf{Variation of $\left\langle n_{\text{\scriptsize LC\normalsize}}\right\rangle$ with $d$. } Average number of local complementation operations required to create a link between two given qubits $a$ and $b$ in the bulk, as a function of the distance between the two qubits in the case of a local and a non-local graph corresponding to the square hexagonal code with $D=20$. The average value of $n_{\text{\scriptsize LC\normalsize}}$ is calculated over a sample of $n_{\mathcal{L}_{ab}}=10^4$ graphs. The range $0\leq \left\langle n_{\text{\scriptsize LC\normalsize}}\right\rangle \leq 5$ is enlarged. }
\label{fig:nlc}
\end{figure*}

\subsection{Computation of measurement-based lower bound}
\label{subsec:appl_mlb}

We now investigate the variation of MLB with the distance between the chosen qubits in the bulk of the square hexagonal lattice of code-distance $D$. Before moving on to the analysis of the numerical results, a word on the notion of the existence of local and non-local links in the graph, and its relation with the topological properties of the system is in order here.  In the color code lattice, the links are all \emph{local} links since they connect qubits  belonging to a specific plaquette. This notion of locality comes from the fact that the plaquette stabilizer operators are intrinsically \emph{local} in the sense that they operate on the qubits belonging to the same plaquette. Therefore, a graph constructed following the lattice of the color code (where one  considers lattice sites as nodes in the graph where the qubits are situated, and introduces links in the graph  according to the links in the color-code lattice -- see Fig.~\ref{fig:non_local_links}(c) for an example) contains only local links. However, the graph obtained from the code  using  the methodology presented in Sec.~\ref{sec:graph-topo} may contain a number of \emph{non-local} links connecting a control and a target qubit belonging to two different and distant plaquettes (see Fig.~\ref{fig:non_local_links}(d)), which is in contrast with the characteristics of a local graph. It has been shown that small and simple setups diminishes the effect of these non-local links, and a critical size is to be achieved in order to observe the effect of the topological properties in terms of the existence of the non-local links in the graph~\cite{lang2012}. However, one can also take a different perspective, and ask whether a differentiation can be made in terms of entanglement. Our graph-based algorithms are appropriate for such investigations.  

We use the developed packages \textbf{StabGraph} and \textbf{ALCPack} to apply the graph-based algorithms I and II to determine the MLB for LE over qubit-pairs situated in the bulk of the lattice of the square hexagonal code. In the case of the graph-based algorithm I, for a pair of chosen qubits separated from each other by distance $d$, we set $n_G=1$, and optimize the MLB of LE over a set of graphs $G^\prime$ generated via LC operations on the qubits situated on randomly chosen paths $\mathcal{L}_{ab}$ connecting the qubits $a$ and $b$, where the size of the set of graphs $G^\prime$ equals to the number of random paths $n_{\mathcal{L}_{ab}}$.   We first test how the ALC algorithm scales with the system size by looking at the average time $\langle t\rangle$ taken by the ALC algorithm to create a link between the chosen qubits $a$ and $b$. We  vary the code-distance $D$ of a square hexagonal code as $D=12,16,20,24$, such that the number of qubits in the system are $N=194,354,562,818$ respectively. In Fig.~\ref{fig:tlc}, we present the variation of $\langle t\rangle$ as a function of $N$ for qubit pairs with different distances $d=2,4,6,8,10$, where the average value $\langle t\rangle$ is determined over a sample of size $n_{\mathcal{L}_{ab}}=10^4$ for each value of $d$. We separately consider a local graph that follows the structure of the square hexagonal lattice, and a non-local graph obtained from the TCC defined on the square hexagonal lattice by using the methodology described in Sec.~\ref{sec:graph-topo}.  From Fig.~\ref{fig:tlc}, it is evident that $\langle t\rangle$ increases with $N$ for a fixed $d$, and increases with $d$ for a fixed $N$ when the graph is local, which is in contrast with the variation of $\langle t\rangle$ with $N$ in the case of a non-local graph obtained from the square hexagonal code. In the latter case, $\langle t\rangle$ increases only negligibly with $d$ for a fixed value of $N$. This can be understood from the fact that there exists considerable number of non-local links in the case of the local unitary equivalent graph obtained from the TCC, which results in comparable lengths of the paths connecting the chosen qubits in the graph irrespective of the actual distance $d$ between the qubits. This  leads to a similar number of required LC operations, which results in slowly increasing values of $\langle t\rangle$ with $d$ for a fixed $N$.  On the other hand, in the local graph, the typical length of a path $\mathcal{L}_{ab}$ connecting $a$ and $b$ increases with increasing distance $d$ between qubits $a$ and $b$ in the TCC lattice, subsequently increasing the required number of LC operations, and therefore the average value of $\langle t\rangle$.  Also, the variation of $\langle t\rangle$ with increasing $N$ for a fixed $d$ clearly validates the polynomial scaling of the ALC algorithm as discussed in Sec.~\ref{subsec:mlb}.

\begin{figure*}
\includegraphics[width=0.7\textwidth]{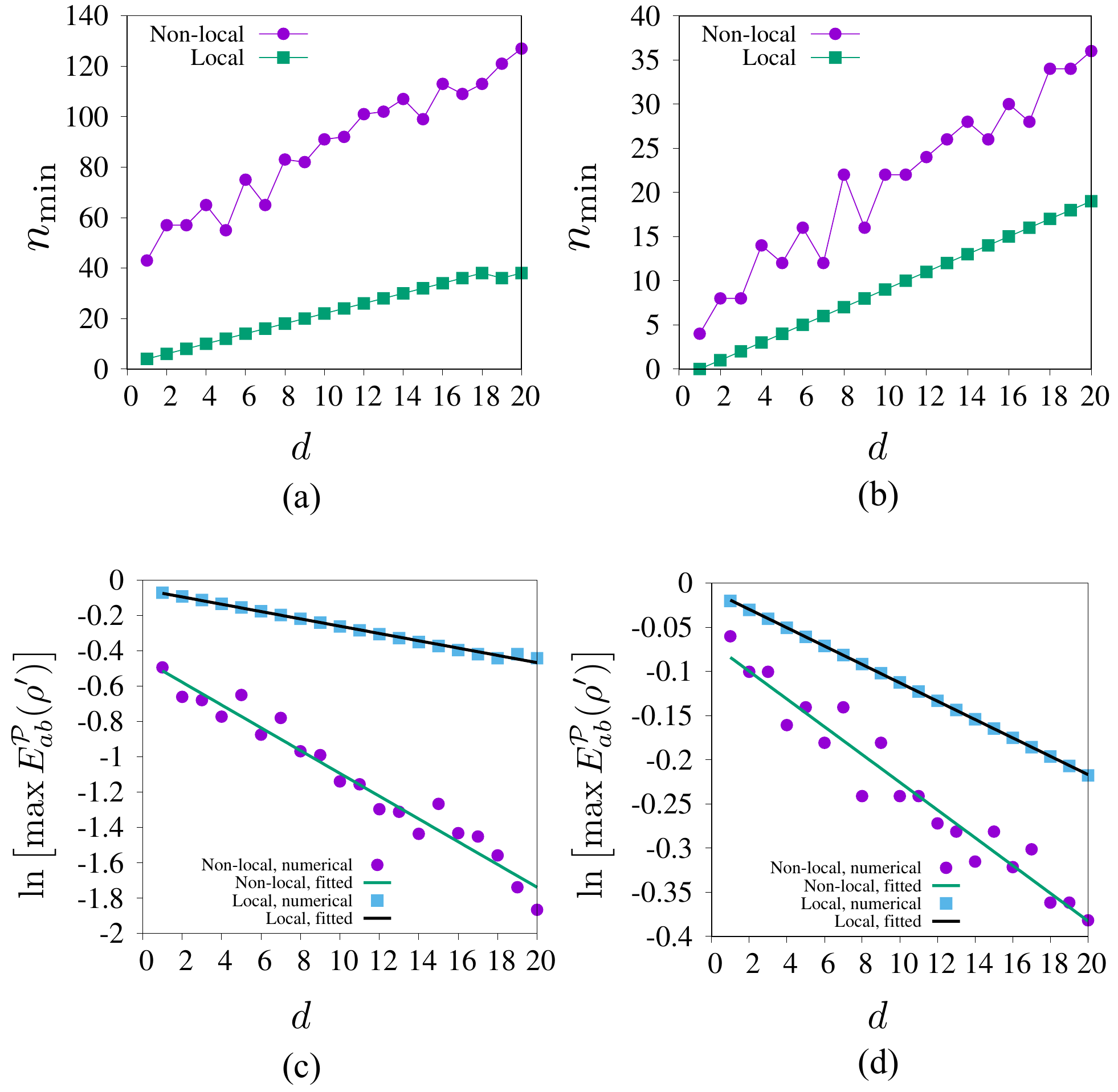}
\caption{(Color online.) \textbf{Variation of $n_{\text{\scriptsize min\normalsize}}$ with $d$.} Minimum size of the neighborhood against $d$, of the qubit-pair $\{a,b\}$ in the bulk of the local and the non-local graphs corresponding to the square hexagonal color code of code distance $20$ where all the qubits are exposed to (a) depolarizing or (b) phase-flip noise.  \textbf{Variation of $\max E_{ab}^{\mathcal{P}}(\rho^\prime)$ with $d$. } Maximum value of MLB against $d$, over the qubit-pair $\{a,b\}$ in the bulk of the local and the non-local graphs corresponding to the square hexagonal color code of code distance $20$ where all the qubits are under (a) depolarizing and (b) phase-flip noise. The fitted curves are represented by the continuous lines, where the values of the fitting parameters are obtained as (a) Non-local graph: $a^\prime=-0.064(3)$, $b=-0.44(4)$, Local graph: $a^\prime=-0.0207(4)$, $b=-0.053(3)$, (b) Non-local graph: $a^\prime=-0.015(1)$, $b=-0.06(1)$, Local: $a^\prime=-0.0104(2)$, $b=-0.0089(2)$.  }
\label{fig:ngh_ent}
\end{figure*}

Given the above discussion, it is interesting to investigate whether the average number of LC operation, $\left\langle n_{\text{\scriptsize LC\normalsize}}\right\rangle$, required to create a link between two chosen qubits in the bulk varies with the distance between the qubits, when the system-size is fixed. Fig.~\ref{fig:nlc} depicts the variation of $\left\langle n_{\text{\scriptsize LC\normalsize}}\right\rangle$ with $d$ in the case of the local and non-local graphs corresponding to the square hexagonal color code lattice of $D=20$, where the averaging has been done over a sample size of $n_{\mathcal{L}_{ab}}=10^4$ for each value of $d$. In the case of the local graph, $\left\langle n_{\text{\scriptsize LC\normalsize}}\right\rangle$ rapidly increases with increasing $d$, while for the non-local graph, the increasing trend of $\left\langle n_{\text{\scriptsize LC\normalsize}}\right\rangle$ slows down considerably when $d$ increases. These results are in agreement with the variations of $\langle t\rangle$ against $d$ for a fixed value of  $N$. 

Since both $U_{S\rightarrow G}$ and $U_{G\rightarrow G^\prime}$ are constituted of local Clifford unitary operators (see Sec.~\ref{subsec:mlb}), the transformed noise $\Lambda^\prime$ is also local uncorrelated Pauli noise similar to $\Lambda$, although the individual bases in which the noise processes take place corresponding to $\Lambda^\prime$ on each qubit may differ from that in $\Lambda$.  As shown in~\cite{amaro2018}, for uncorrelated single-qubit Pauli noise applied to a graph $G^\prime$ in which a link between the two chosen qubits is present, a high value of MLB is favourable when the size `$n$' of the neighborhood of the qubit-pair $\{a,b\}$, for which the noise does not commute with the $Z$-measurement, is low. In Figs.~\ref{fig:ngh_ent}(a)-(b), we plot the variation of the minimum value of $n$, represented by $n_{\text{\scriptsize min\normalsize}}$, as a function of $d$ for the local as well as non-local graphs corresponding to the square hexagonal code with $D=20$, where the minimization of $n$ is achieved over a sample size of $n_{\mathcal{L}_{ab}}=10^4$ for all values of $d$, in the case of (a) the DP and (b) the PF noise. Note that in the former case, the value of $n$ equals the number of qubits in the full neighborhood, while in the latter, $n$ is the number of qubits in the neighborhood with BF or BPF noise. This implies a higher value of $n$ in the former case than the latter, which is clearly demonstrated also in the values of $n_{\text{\scriptsize min\normalsize}}$ in the Fig.~\ref{fig:ngh_ent}. In both cases of the local and the non-local graphs corresponding to the TCC, and for both types of noise, the value of $n_{\text{\scriptsize min\normalsize}}$ increases monotonically with $d$.

Next, we plot the natural logarithm of the maximum value of MLB, denoted by $\max E_{ab}^{\mathcal{P}}(\rho^\prime)$ $\left(=\max E_{ab}^{\mathcal{P}^\prime}(\rho_S)\right)$ and computed following the methodology developed in~\cite{amaro2018} as a function of $d$. The plots are shown in Figs.~\ref{fig:ngh_ent}(c) (for the DP noise) and (d) (for the PF noise), where we choose the noise strength $q=10^{-2}$, and the maximization is achieved over the same set of $n_{\mathcal{L}_{ab}}$ graphs as in the cases of $\langle n_{\text{\scriptsize LC\normalsize}}\rangle$ and $n_{\text{\scriptsize min\normalsize}}$. We fit the variation of the value of the natural logarithm of $\max E_{ab}^{\mathcal{P}}(\rho^\prime)$ with $d$ using the equation $\ln\left[\max E_{ab}^{\mathcal{P}}(\rho^\prime)\right]=a^\prime+bd$, such that the MLB decays exponentially with $d$ according to the equation $\max E_{ab}^{\mathcal{P}}(\rho^\prime)=a\text{e}^{bd}$ with $a^\prime=\ln a$. Here, $a^\prime$ and $b$ are fitting parameter, which are expected to be functions of the noise strength $q$.  See Fig.~\ref{fig:ngh_ent} for the values of the fitting parameters $a^\prime$ and $b$, obtained in the example.

\begin{figure*}
\includegraphics[width=0.7\textwidth]{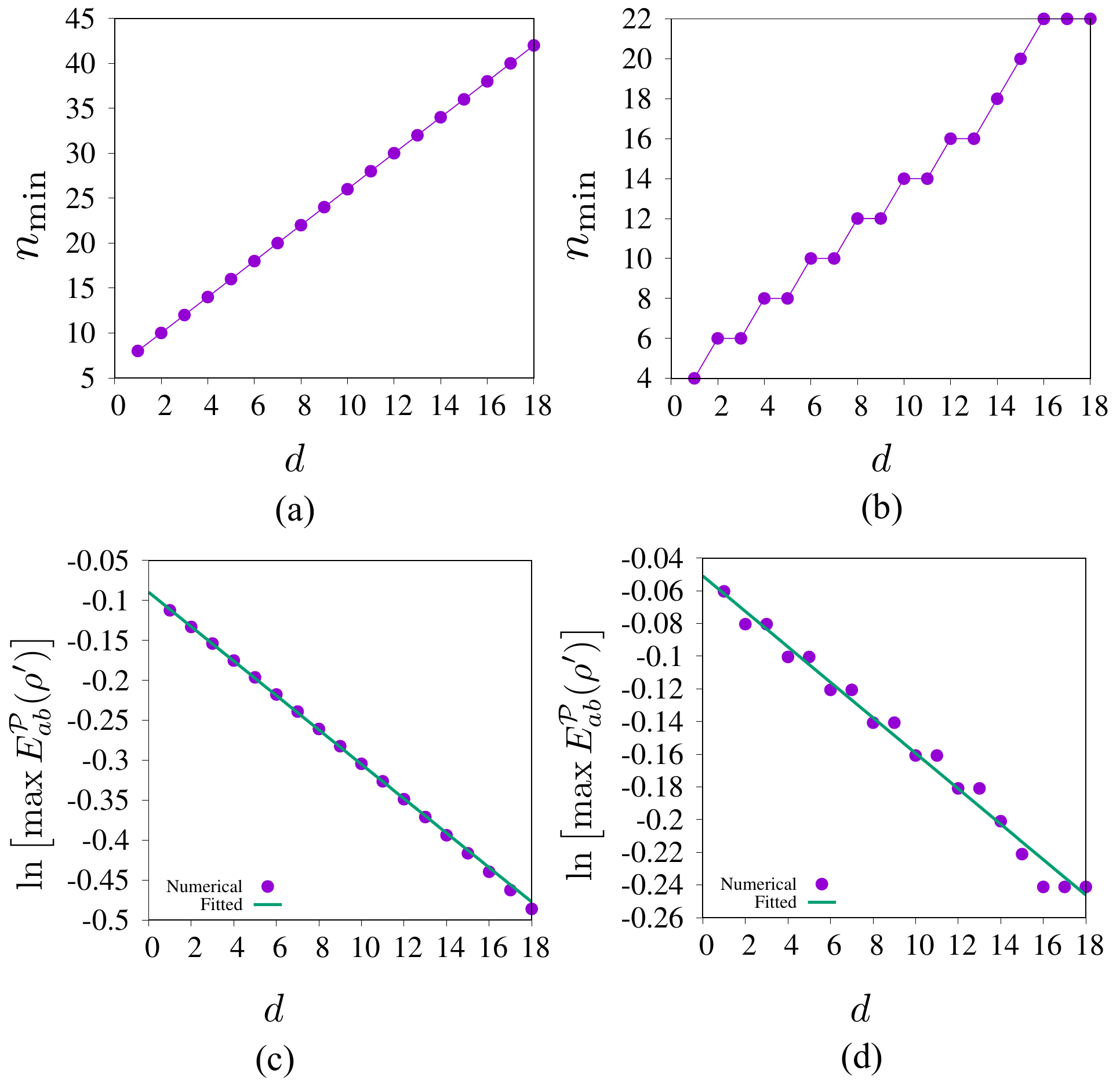}
\caption{(Color online.) \textbf{Variation of $n_{\text{\scriptsize min\normalsize}}$ and $\max E_{ab}^{\mathcal{P}}(\rho^\prime)$ with $d$.} The minimum size of the neighborhood of the qubit-pair $\{a,b\}$ under local uncorrelated Pauli noise that do not commute with $Z$-measurement, and the natural logarithms of the maximum value of MLB as functions of $d$, over the qubit-pair $\{a,b\}$ in the bulk of the non-local graphs corresponding to the square hexagonal color code of code distance $20$, where all the qubits are under (a) depolarizing and (b) phase-flip noise. The fitted curves in the case of MLB are represented by the continuous lines, where the values of the fitting parameters are obtained as (a) $a^\prime=-0.0215(3)$, $b^\prime=-0.089(4)$, (b) $a^\prime=-0.0108(4)$, $b^\prime=-0.05(1)$. }
\label{fig:dent}
\end{figure*}

So far, we have considered a variant of the graph-based algorithm by setting $n_G=1$, $n_{\mathcal{L}_{ab}}=10^4$. We now apply the \emph{modified} graph-based algorithm in order to investigate the features of MLB where a link between $a$ and $b$ is created every time that one transforms the stabilizer state into a graph state. Therefore, the optimization this time is over a large number $n_G$ of bicolorable graphs $G$, obtained directly from $S$ (see Sec.~\ref{subsec:mlb}). The data obtained for $n_{\text{\scriptsize min\normalsize}}$ and $\ln\left[\max E_{ab}^{\mathcal{P}}(\rho^\prime)\right]$ as functions of $d$ by using graph-based algorithm II are presented in Fig.~\ref{fig:dent}. 

The numerical results presented in this Section are illustrations of the applicability of the witness- and graph-based methodologies  developed and discussed in Sec.~\ref{sec:le_bounds} for determining non-trivial lower bounds of the localizable entanglement over bulk qubit-pairs in the case of arbitrary TCC lattices, such as the square hexagonal lattice with open boundary condition. The lower bounds decrease with increasing $d$, and the trend agrees with our understanding of the growth of the size of the neighborhood with noise having bases that do not commute with $Z$-measurements around the link connecting the qubits in the chosen pair, with the increasing distance between them. However, in order to infer the exact dependence of the MLB on $d$, one could consider the optimality of the algorithms for obtaining the maximum value of the MLB. This  issue could  be thoroughly investigated, which is beyond the scope of the present work.

\section{Conclusions and outlook}
\label{sec:conclude}

In this paper, we provide two specific pathways to estimate the lower bound of localizable entanglement over a subset of qubits in a large system of noisy topological codes, including the surface and the color codes. In one approach, we use an appropriately constructed local entanglement witness operator constituted of the stabilizer generators of the code, and estimate a lower bound of localizable entanglement over the chosen region of qubits via the expectation value of the witness operator.  We also propose a specific construction of the the local witness operator for this purpose.  On the other hand, we also propose a methodology for estimating a lower bound of localizable entanglement corresponding to a specific projection measurement setup on the qubits outside the specified subset of qubits in a noisy stabilizer state. This uses the fact that an arbitrary stabilizer state can be connected to a graph state via local unitary transformations. We discuss in detail a scalable geometric recipe for determining the graph underlying the local unitary connected graph state from the stabilizer state describing a color code. We also present an algebraic methodology to determine the adjacency matrix of the graph corresponding to a graph state obtained via local unitary transformation from a stabilizer state of arbitrary size. Moreover, we develop an algorithm which creates a link between any two chosen nodes in a simple, connected, and undirected graph by using a sequence of local complementation.  We also develop appropriate numerical packages~\cite{amaro_github_2019,pal_github_2019} for these purposes. We predict that the runtime of the graph-based algorithm scales polynomially with the system size, which is  supported by our numerical findings corresponding to a topological color code on a square hexagonal lattice. 

We also determine the witness- and measurement-based lower bounds of localizable entanglement in the case of qubit pairs situated in the bulk of a topological color code described on a  square hexagonal lattice. We explicitly compute the expectation value of the local entanglement witness operator constructed according to our prescription in the case of a stabilizer state of the system under local uncorrelated Pauli noise. Our calculations show that the bound obtained from the proposed construction of the local witness operator exponentially decreases with increasing support of the witness operator, and therefore with increasing  distance between the chosen qubits.   In the case of the measurement-based method, along with computing the bounds of localizable entanglement over a qubit-pair in the bulk of a topological color code via the graph-based methods I and II, we also determine the bounds of localizable entanglement corresponding to a qubit pair in the case of a local graph that follows the square hexagonal lattice. 

Note that the numerical data presented in this paper corresponding to the measurement-based lower bound is obtained as a proof of  the functionality of the graph-based method and the modified graph-based method developed in this paper. It could be interesting to study the optimality of the algorithm in order to optimize the bound.  The algorithms proposed in this paper could also be generalized for regions beyond two qubits
leading to the notion of multiparty localized entanglement~\cite{sadhukhan2017}  in topological quantum codes, which could allow to reveal long-range multiparty quantum correlations.

\acknowledgements

We acknowledge support by the EU Quantum Technology Flagship grant AQTION 820495, the ERC Starting Grant QNets 804247, and by U.S. A.R.O. through Grant No. W911NF-14-1-010. The research is also based upon work supported by the Office of the Director of National Intelligence (ODNI), Intelligence Advanced Research Projects Activity (IARPA), via the U.S. Army Research Office Grant No. W911NF-16-1-0070. The views and conclusions contained herein are those of the authors and should not be interpreted as necessarily representing the official policies or endorsements, either expressed or implied, of the ODNI, IARPA, or the U.S. Government. The U.S. Government is authorized to reproduce and distribute reprints for Governmental purposes notwithstanding any copyright annotation thereon. Any opinions, findings, and conclusions or recommendations expressed in this material are those of the author(s) and do not necessarily reflect the view of the U.S. Army Research Office. AKP acknowledges support from the National Science Center (Poland) Grant No. 2016/22/E/ST2/00559. Numerical simulations have been performed on the Swansea SUNBIRD system. The Swansea SUNBIRD system is part of the Supercomputing Wales project, which is part-funded by the European Regional Development Fund (ERDF) via the Welsh Government. We thank Davide Vodola for letting us use the numerical tools developed by him to construct the square hexagonal lattice. We also thank Ciarán Ryan-Anderson for useful discussions.

\appendix

\section{Topological color codes}
\label{app:tcc}

We use topological color codes (TCC)~\cite{bombin2006,bombin2007}  as the testing ground for our results, and the defining  features of a TCC are briefly discussed in this appendix. A TCC model is constructed on a two-dimensional (2D), three-colorable, and trivalent lattice, where each vertex of the lattice contains a physical qubit, and the lattice can be embedded on a compact surface having arbitrary topology of genus $g$ (for example, a torus with $g=1$). The three-colorability of the lattice implies that the faces of the lattice, also known as the plaquettes, can be painted with three different colors, where neighbouring plaquettes always have different colors. The trivalency of the lattice implies that each vertex is connected to three links. The lattice can also be characterized by coloring the three links connected to each vertex with the same set of three different colors as the colors of the plaquettes, such that the neighboring plaquettes share a link having a color different than both colors of the plaquettes sharing the link.  There is a number of such regular lattices available, such as the hexagonal (honeycomb) lattice, the square-octagonal lattice, and the square-hexagon-dodecahedron lattice.  The number of logical qubits in a TCC is  given by $k=4-2\chi$, where $\chi$ is the Euler characteristic of the surface, thereby ensuring that $k$ depends on the topology of the surface . The methodology developed in this paper are aimed for arbitrary stabilizer states, and therefore applies to arbitrary TCC. For the purpose of testing our prescriptions, in this paper, we shall focus on the 2D honeycomb lattice with open boundary conditions. To keep the figures uncluttered, we shall use the three-colorability of only the plaquettes for demonstration (see Fig. \ref{fig:tcc} for a schematic of a honeycomb lattice, where the three colors corresponding to the TCC are red $(R)$, green $(G)$, and blue $(B)$).

We now set up the terminology for the stabilizer description of a TCC, which requires the definition of two key concepts, (1) the \emph{stabilizer subspace}, and (2) the \emph{logical operators}. The \emph{stabilizer subspace} of the TCC is determined by the \emph{stabilizer group} of operators, which is generated by a set of plaquette operators denoted by $S_p^\alpha$. There are two types of plaquette operators, corresponding to $\alpha=x$ and $\alpha=z$ for each plaquette $P_p$, called the $x$-type and the $z$-type operators, given by 
\begin{eqnarray}
S_p^x=\bigotimes_{i\in P_p}X_i,\;\; S_p^z=\bigotimes_{i\in P_p}Z_i,
\label{eq:plaquette_operators}
\end{eqnarray} 
where $X$ and $Z$ are respectively the $x$ and $z$ components of Pauli matrices.  Each of the plaquette operators squares to the  identity, i.e., $\left(S_p^\alpha\right)^2=I$ $\forall$ $p$, $\alpha=x,z$, and they mutually commute, $[S_p^\alpha,S_{p^\prime}^{\alpha^\prime}]=0$, since all plaquettes have an even number of vertices in a trivalent three-colorable lattice, and  since adjacent plaquettes share even number of vertices. The stabilizer subspace $\mathcal{H}_S$ in the full Hilbert space $\mathcal{H}$ of the TCC is given by 
\begin{eqnarray}
\mathcal{H}_S=\{\ket{\psi}:S_p^\alpha\ket{\psi}=\ket{\psi} \forall p, \alpha=x,z\}.
\label{eq:stabilizer_subspace}
\end{eqnarray}
Note that the plaquette stabilizer operators are intrinsically \emph{local} in the sense that they operate on the qubits belonging to the same plaquette. This notion of locality will be crucial for the discussions presented in this paper. In all our considerations, we define the physical distance between any two lattice points  by the length of the shortest path constituted of the lattice links and  connecting the two lattice points.  

A TCC hosts a total of $2k$ independent logical generators $L^{(q)}_{\alpha}$, where $q=1,2,...,k$ is the number of logical qubit, and $\alpha=x,z$ indicates the local Pauli matrices corresponding to the lattice sites that constitute the logical operators. They are defined on homologically non-trivial (i.e., non-contractible) strings across the lattice, and they commute with all stabilizers:
\begin{eqnarray}
[L_{\alpha}^{(q)},S_p^{\alpha^\prime}]=0.
\end{eqnarray}
The $x$- and $z$-type logical operators define the computational basis $\{\ket{0}_L,\ket{1}_L\}$ for the logical qubits, such that 
\begin{eqnarray}
L_{z}^{(q)}\ket{0}_L&=&\ket{0}_L,\; L_{z}^{(q)}\ket{1}_L=-\ket{1}_L,
\end{eqnarray}
where the subscript `$L$' denotes the logical states. The eigenbasis of $L_x^{(q)}$ and $L_y^{(q)}$ in terms of $\{\ket{0}_L,\ket{1}_L\}$ are  $\ket{\pm}_L=(\ket{0}_L\pm\ket{1}_L)/\sqrt{2}$ and $\ket{\pm\text{i}}_L=(\ket{0}_L\pm\text{i}\ket{1}_L)/\sqrt{2}$ respectively.

In this paper, we focus on a 2D color code defined on a square-hexagonal lattice, as described in Sec.~\ref{sec:apply}. The number of  physical qubits, $N$, in a 2D topological color code depends on the code distance $D$. In the case of the topological color code defined on  the square-hexagonal lattice with code distance $D$, $N$ varies with $D$ as 
\begin{eqnarray}
N= \frac{3D^2}{2}-2(D-1),
\end{eqnarray}   
where  $D=4l$ $(l=1,2,3,4,\cdots)$.

\section{Decomposition of local witness operators}
\label{app:witness_decomposition}

\begin{figure}
\includegraphics[scale=0.4]{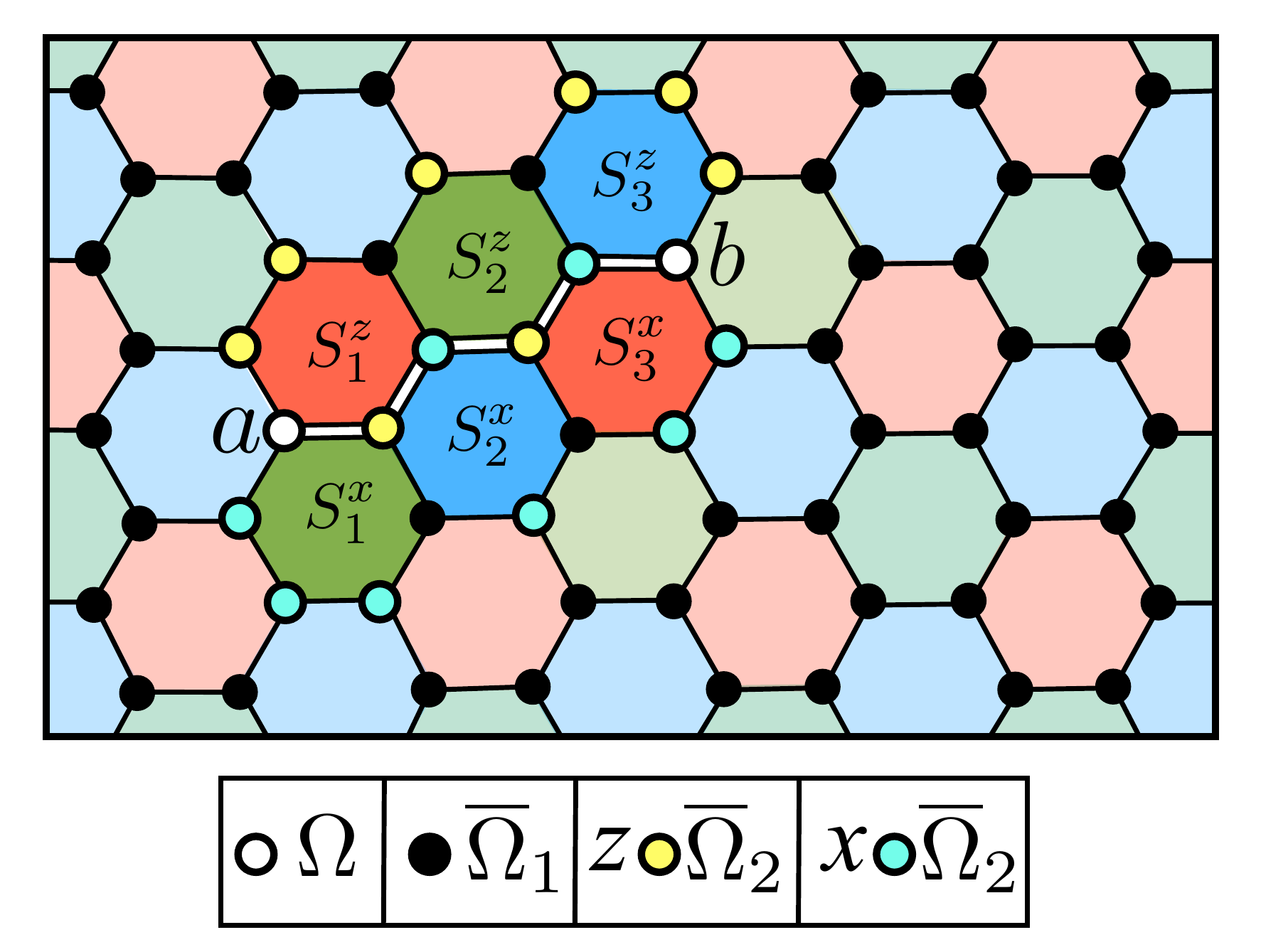}
\caption{(Color online.) \textbf{The supports of the witness operator in Fig. \ref{fig:local_witness}(b)}. The two-qubit region is marked by the white nodes, while the black, turquoise, and yellow nodes mark the region $\overline{\Omega}$. The black nodes represent the sub-region $\overline{\Omega}_1$, while the turquoise and yellow nodes stand for the sub-region $\overline{\Omega}_2$. The two different colors, turquoise and yellow on a qubit $i\in\overline{\Omega}_2$ represent $\tau_{v_i}^{\overline{\Omega}_2}=X$ and $\tau_{v_i}^{\overline{\Omega}_2}=Z$ respectively.} 
\label{fig:lwfine}
\end{figure}

In this appendix,  we present the detailed  calculation for decomposing a local entanglement witness operator of  the form in Eq.~(\ref{eq:witness_global}) into the form given in  Eq.~(\ref{eq:decompose_2_main}). In order to decompose $\mathcal{W}_{\Omega}$ in terms of local projection operators on $\overline{\Omega}$ and witness operators $W^k_{\Omega}$ on $\Omega$, notice that the commutation property in (i)  splits $\overline{\Omega}$ into two regions, (1) $\overline{\Omega}_{1}$ constituted of qubits for which $u_{i,j}=0$ $\forall$ $j$, and (2) $\overline{\Omega}_{2}$ consisting of qubits such that there exists at least one stabilizer $S_j\in\mathcal{S}$  in which $u_{i,j}\neq 0$. Therefore, for qubits $i\in\overline{\Omega}_1$, $\bigotimes_{i\in\overline{\Omega}_1}\tau_{u_{i,j}}^{{\overline{\Omega}_1}}=I_{\overline{\Omega}_1}$, $I_{\overline{\Omega}_1}$ being the identity operator in the Hilbert space of the qubits in $\overline{\Omega}_1$. On the other hand,  in the case of a qubit $i$ in $\overline{\Omega}_2$, if $u_{i,j}\neq 0$ for more than one stabilizers $S_j\in\mathcal{S}$, then the values of $u_{i,j}$ are identical for all $j$ for which $u_{i,j}\neq 0$. This assigns a specific Pauli operator $\tau_{v_i}^{\overline{\Omega}_2}$ ($v_i=1,2,$ or $3$) for the qubits $i\in\overline{\Omega}_2$. The corresponding projection operators can be written as  $P_{(k_{i},v_i)}^{\overline{\Omega}_2}=\left[I_i+(-1)^{k_i}\tau_{v_i}^{\overline{\Omega}_2}\right]/2$, $I_i$ being the identity operator in the Hilbert space of the qubit $i\in\overline{\Omega}_2$ in stabilizer $S_j$, and $k_i$ $(=0,1)$ can be interpreted as the outcome of the projection measurement via $P_{(k_i,v_i)}^{\overline{\Omega}_2}$.  The constructions of the local witness operator has been demonstrated in Figs.~\ref{fig:local_witness}(a) and (b). The supports of the stabilizers $S^z$ and $S^x$, denoted by $R_z$ and $R_x$ respectively, are the sets of nodes corresponding to which the Pauli operator contributing to the stabilizer is not an identity. For example, the sizes of $R_z$ and $R_x$, denoted respectively by $n_z$ and $n_x$, are $n_z=8$, $n_x=6$, for the figure (a), while for (b), $n_x=n_z=8$. See Fig.~\ref{fig:lwfine} for an illustration of the supports is the case depicted in Fig.~\ref{fig:local_witness}(b).

Before breaking the mathematics any further, let us consider the effect of the application of a projection operator $P_{(k_i,v_i)}^{\overline{\Omega}_2}$ on each of the qubits in $\overline{\Omega}_2$, which results in 
\begin{eqnarray}
P_{(k_i,v_i)}^{\overline{\Omega}_2} S_j=\eta_{i,j}P_{(k_i,v_i)}^{\overline{\Omega}_2}\otimes \left[\bigotimes_{\underset{l\neq i}{l\in\overline{\Omega}_2}}\tau_{u_{l,j}}^{\overline{\Omega}_2}\right]\otimes I_{\overline{\Omega}_1}\otimes S_j^{\Omega}, \nonumber\\ 
\end{eqnarray}
where 
\begin{eqnarray}
\eta_{i,j}=\left\{
 \begin{array}{cc}
 (-1)^{k_i}, & \text{for } u_{i,j}\neq 0,  \\
 1, &  \text{for } u_{i,j}=0.
\end{array}\right.
\end{eqnarray}                                             
Therefore, application of a projection operator of the form $P_{(k_i,v_i)}^{\overline{\Omega}_2}$ on each qubit $i\in\overline{\Omega}_2$ yields 
 \begin{eqnarray}
P_{(k,v)}^{\overline{\Omega}_2}S_j=\eta_j P_{(k,v)}^{\overline{\Omega}_2}\otimes I_{\overline{\Omega}_1}\otimes S_j^\Omega,
\label{eq:projector_on_stab}
\end{eqnarray}
with $\eta_j=\prod_{i\in\overline{\Omega}_2}\eta_{i,j}$, and
\begin{eqnarray}
P_{(k,v)}^{\overline{\Omega}_2}=\bigotimes_{i\in\overline{\Omega}_2}P_{(k_i,v_i)}^{\overline{\Omega}_2},
\label{eq:big_projector}
\end{eqnarray}
where $k\equiv k_{1}k_{2}\cdots k_{{m^\prime}}$ and $v\equiv v_{1}v_{2}\cdots v_{{m^\prime}}$ are multi-indices\footnote{Note here that the allowed values of $v_i$ are $1,2,$ and $3$ instead of $\{0,1,2,3\}$ which forms a complete base $4$ of decimal numbers. Therefore, the allowed values of $u_{i,j}$ form only a subset of $0,1,\cdots,4^{m^\prime}-1$}, and we denote  the size of $\overline{\Omega}_2$ by $m^\prime$ ($\leq m$). From Eq. (\ref{eq:projector_on_stab}), it is clear that the application of local projection operations in the basis of Pauli operators fixed by the witness $\mathcal{W}_{\Omega}$ outside the region $\Omega$ results in local stabilizers $S_j^\Omega$. These stabilizers  correspond to the genuine multiparty entangled state $\ket{\psi}_{\Omega}$ -- the same state that is obtained over the region $\Omega$ by performing projection operation $P_{(k,v)}^{\overline{\Omega}_2}$ on the stabilizer state $\ket{\psi}$ (see condition (ii)).  

We now point out that the completeness of the basis formed by the eigenstates of the  Pauli matrices in the Hilbert space $\mathcal{H}_i$ of a qubit allows one to write $I_{i}$ as
\begin{eqnarray}
I_{i}=\sum_{k_{i}=0}^1 P_{(k_{i},v_i)}^{\overline{\Omega}_2}.
\label{eq:identity_expand}
\end{eqnarray}
This can be extended to the identity operator in the Hilbert space $\mathcal{H}=\bigotimes_{i=1}^n\mathcal{H}_i$ of the full system, which, when multiplied with $\mathcal{W}_{\Omega}$ from the left and expanded in terms of the projectors $P_{(k,v)}^{\overline{\Omega}_2}$ yields 
\begin{eqnarray}
\mathcal{W}_{\Omega}=\left[\sum_{k=0}^{2^{m^\prime}-1}P_{(k,v)}^{\overline{\Omega}_2}\otimes W_{\Omega}^{k}\right]\otimes I_{\overline{\Omega}_1},
\label{eq:decompose_1}
\end{eqnarray}
with 
\begin{eqnarray}
W_{\Omega}^{k}=\frac{1}{2}I_{\Omega}-\prod_{S_j\in\mathcal{S}}\frac{I_{\Omega}+\eta_jS_j^\Omega}{2},
\end{eqnarray} 
where   $I_{\Omega}$ is the identity operator in the Hilbert space of the qubits in $\Omega$. Note here that given the condition (ii) for the construction of witness operators, $W_{\Omega}^{k}$ detects genuine multiparty entanglement in the region $\Omega$.   Note also that one can use Eq. (\ref{eq:identity_expand}) for the identity operators in the Hilbert space of the qubits in $\overline{\Omega}_1$ as well, by choosing any one of the three Pauli operators, $X$, $Y$, or $Z$ to define the projectors, since the form of $W^{k}_{\Omega}$ remains independent of this choice. Each set of Pauli operators defining the projectors over  the region $\overline{\Omega}_1$ leads to a specific Pauli measurement setup over the qubits in $\overline{\Omega}$. For each of these setups, Eq. (\ref{eq:decompose_1}) takes the form
\begin{eqnarray}
\mathcal{W}_{\Omega}=\sum_{k=0}^{2^{m}-1}P_{(k,v)}^{\overline{\Omega}}\otimes W_{\Omega}^{k}.
\label{eq:decompose_2}
\end{eqnarray}

\section{Negativity of a two-qubit state}
\label{app:negativity}
We formally define negativity~\cite{peres1996,horodecki1996} as a bipartite entanglement measure. Negativity of an arbitrary bipartite (e.g. two-qubit) state $\varrho_{ab}$ is defined as
\begin{eqnarray}
N_g(\varrho_{ab})=||\varrho_{ab}^{\text{T}_a}||_1-1.
\end{eqnarray}
Here, $\varrho_{ab}^{\text{T}_a}$ is the partial transposition of $\varrho_{ab}$ w.r.t. qubit $a$, and $||\mathbf{x}||_1=\text{Tr}\left[\sqrt{\mathbf{x}^\dagger\mathbf{x}}\right]$ denotes the trace norm. In terms of the eigenvalues $\{\lambda_i\}$ of $\varrho_{ab}^{\text{T}_a}$, $N_g$ is given by 
\begin{eqnarray}
N_g(\varrho_{ab})=2\sum_{\lambda_i<0}|\lambda_i|. 
\end{eqnarray}

\section{Graphs and graph states}
\label{app:graphs}

Here we discuss some basic concepts regarding graphs and graph states~\cite{hein2006}.

\noindent\textbf{Graph.} A graph $G$ is a collection of \emph{nodes} connected to each other by \emph{links}. We denote the number of nodes by $N$, and label them as $1,2,\cdots,N-1,N$, while a link connecting the nodes $i$ and $j$ is designated as $(i,j)$ ($i\neq j$, $i,j\in G$). A graph is represented by an $ N\times N $ binary matrix $\Gamma$, called the adjacency matrix, given by 
\begin{eqnarray}
\Gamma_{ij}=\begin{cases}
                         1, & \text{for } (i,j) \in G,  \\
                         0, & \text{for } (i,j) \notin G.
                         \end{cases}
                         \label{eq:adjacency}
\end{eqnarray} 
We shall focus on \emph{simple}, \emph{undirected}, and \emph{connected} graphs only. A simple graph is one without any loop (a node connected to itself) or multiple links connecting a pair of nodes.  Such a graph  is connected iff for each pair of sites $ \{i,j\}\in V$, there exists at least one path $\mathcal{L}_{ij}$ between $i$ and $j$, constituted of a set of links $\{(k,l)\}\in G$ with $k,l\in V$. In an undirected graph, the links $(i,j)$ and $(j,i)$ are equivalent.  The neighbourhood of a node $i$ in $G$ is denoted by by $N_{i}\subset V$, which is the set of nodes $\{j\}$ that are directly connected to $i$ by  links, i.e., $(i,j)\in G$ $\forall$ $j\in N_i$.

\noindent\textbf{Graph state.} A graph state $\ket{\psi_G}$ corresponding to an underlying graph $G$ can be created by (i) considering a qubit located at every node, initialized in the state $\ket{+}$ such that the state of the $N$-qubit system is $\ket{+}^{\otimes N}$, and then (ii) applying a controlled phase gate, $U^z_{(i,j)}$ (see Eq.~(\ref{eq:czgate})), on each pair of qubits $\{i,j\}$ if $(i,j)\in G$. The resulting graph state is 
\begin{eqnarray}
\ket{\psi_G}=\left[\prod_{(i,j)\in G}U^z_{(i,j)}\right]\ket{+}^{\otimes N}.
\end{eqnarray}
Note that further application of the same controlled phase unitaries on the qubits in the graph state completely disentangles the graph state to $\ket{+}^{\otimes N}$, since $\left[U^z_{(i,j)}\right]^2=I$.

\noindent\textbf{Simple paths.} A \emph{simple} path $\mathcal{L}_{ab}$ connecting the two nodes $\{a,b\}$ by a sequence of nodes is  given by 
\begin{eqnarray}
\mathcal{L}_{ab}=[a\equiv m_1,m_2,\cdots ,m_n\equiv b],
\label{eq:simple_path}
\end{eqnarray}
with $\{m_2,m_3,\cdots ,m_{n-1}\}$ being the nodes that are visited while traveling from the \emph{source} $a$ to the \emph{target} $b$ along $\mathcal{L}_{ab}$, where none of the nodes is repeated in the sequence, and the link $(m_i,m_{i+1})\in G$, $i=1,\cdots,n-1$.  The number of links traversed while going from $a$ to $b$ is the length $l=n-1$ of the path. We denote a simple path between $a$ and $b$ having length $l$ as $\mathcal{L}^{(l)}$. To keep our notations uncluttered,  we discard the subscript ``$ab$" and the superscript ``$(l)$", and denote a simple path of the form given in Eq.~(\ref{eq:simple_path}) by $\mathcal{L}$, unless we need to distinguish between two paths of different source and/or target nodes, or of different length.   

\noindent\textbf{Shortest paths.} There can be more than one simple paths of different or same lengths between two nodes  $a$ and $b$ in a graph. The shortest path is the simple path between $a$ and $b$ having the minimal length $l=l_{min}$, where the minimization is taken over all possible simple paths between $a$ and $b$. There can be more than one shortest paths between a specific pair of nodes.

\noindent\textbf{Local complementation.} The local complementation (LC) operation with respect to a node $i$, denoted by $\tau_i(.)$, on a graph $G$ deletes all the links $\{(j,k)\}$ if $j,k\in N_i$, and $(j,k)\in G$, and creates all the links $\{(j,k)\}$ if $j,k\in N_i$, and $(j,k)\notin G$.  A sequence of LC operations on $n$ nodes denoted by $\mathbf{m}\equiv\{m_1,m_2,\cdots,m_n\}$ of a graph results in a graph transformation
\begin{eqnarray}
\tau_{\mathbf{m}}&=&\tau_{m_n/m_{n-1}/\cdots/m_1}(.)\nonumber \\
&=&\tau_{m_n}\circ\tau_{m_{n-1}}\circ\cdots\circ\tau_{m_1}(.),
\label{eq:lc_sequence}
\end{eqnarray}
where the LC operation is performed on the node $m_1$ first, and then according to the sequence $\{m_1,m_2,\cdots,m_n\}$. Note that this specific sequence is important, as LC operations, in general, do not commute.

\noindent\textbf{Stabilizer formalism.} The stabilizer description of a graph state $\ket{\psi_G}$ corresponding to a graph $G$ uses $N$ generators $g_i$ of the form
\begin{eqnarray}
g_i=X_i\bigotimes_{j\in\mathcal{N}_i}Z_j,
\end{eqnarray}
such that $g_i\ket{\psi_G}=\ket{\psi_G}$ $\forall$ $i$ $\in$ $V$, which forms a subset of the Pauli group.  The generators $\{g_i\}$ mutually commute, thereby sharing a common eigenbasis, and the graph state $\ket{\psi_G}$ represents the common eigenstate with eigenvalue $+1$. The rest of the eigenstates of the generators can be represented as $\{\ket{\psi_G}_\nu=Z_\nu\ket{\psi_G}\}$, where $\nu$ is a multi-index representing the sequence $\nu_1\nu_2\cdots\nu_N$ of indices $\{\nu_i\}$ corresponding to the qubit $i\in G$, where $\nu_i=0,1$, and $Z_\nu=\bigotimes_{i\in V}Z_i^{\nu_i}$.

\noindent\textbf{Local complementation as local unitary operation.} An LC operation on a graph $G$ with respect to the node $i$ is equivalent to a local unitary transformation of the corresponding graph state $\ket{\psi_G}$, where the unitary operator is given by 
\begin{eqnarray}
U_i=\exp\left(-\text{i}\frac{\pi}{4}X_i\right)\bigotimes_{j\in\mathcal{N}_i}\exp\left(\text{i}\frac{\pi}{4}Z_j\right). 
\label{eq:lc_unitary}
\end{eqnarray}
A sequence of LC operations on a number of chosen nodes in a graph $G$, denoted by Eq.~(\ref{eq:lc_sequence}), is given by 
\begin{eqnarray}
U=\bigotimes_{i\in\mathbf{m}} U_i,
\end{eqnarray}
where $U_i$ is as in Eq.~(\ref{eq:lc_unitary}).

\section{Algebraic approach}
\label{app:algebraic_approach}

In this appendix, we put the graphical recipe discussed in Sec.~\ref{sec:graph-topo} in a mathematical footing, and show how the adjacency matrix of the graph can be obtained from the structure of the stabilizers, revisiting the results obtained in~\cite{van-den-nest2004}. We use the binary picture~\cite{van-den-nest2004,hein2006} for the description of the stabilizer state of the $N$-qubit system, which is a $2N\times N$ binary matrix composed of two $N\times N$ blocks corresponding to the $Z$- and $X$-type stabilizers, given by 
\begin{equation}
	A=\left( \begin{array}{c}
	\mathcal{Z}  \\ \hline
	\mathcal{X}
	\end{array}\right). 
	\label{eq:binary_stab}
\end{equation}
Here each column of  $A$, marked by the index $j$, represents one stabilizer $S_j$, in which the Pauli matrix corresponding to the qubit $i$ is represented by $\left( \begin{array}{c}
	\mathcal{Z}_{i,j}  \\ \hline
	\mathcal{X}_{i,j}
	\end{array}\right)$, where $\mathcal{Z}_{i,j}$, $\mathcal{X}_{i,j}$ $=0,1$ (see Appendix~\ref{app:binary} for details).  Since the stabilizers, or more precisely, the generators are independent, the matrix $A$ is of $\text{rank}(A)=N$. Assuming that the rank of $\mathcal{X}$ is $n$, the stabilizer state can be represented by a new set of stabilizers obtained by re-combining the original stabilizers through Gaussian elimination among the columns in such a way that $\mathcal{X}_{ij}=0$ for $j=n+1\cdots,N$, where we arrange the columns as $j=1,2,3,\cdots,n,n+1,\cdots,N$ without any loss in generality. We point out here that all the operations in the binary picture are performed modulo $2$. Denoting ``left" by $l$ and ``right" by $r$, the stabilizer state now becomes  
\begin{equation}
	A^\prime=\left( \begin{array}{cc}
	\mathcal{Z}_l 	& \mathcal{Z}_r \\ \hline
	\mathcal{X}_l	& 0
	\end{array}\right),
	\label{eq:binary_stab_recombined}
\end{equation}
where $\mathcal{X}_l$ is an $N\times n$ matrix of full rank. Note here that in the case of a CSS code~\cite{nielsen2010}, $\mathcal{Z}_l$ can be considered to be $0$, ensuring that the left block of columns in $A^\prime$ corresponds to only $x$-type stabilizers, while the stabilizers represented by the  right block of columns are only $z$-type.   

Since the matrix $\mathcal{X}_l$ has rank $n$, the same number of linearly independent rows can be chosen from it, which will then form $n\times n$ invertible matrix $\mathcal{X}_{l,c}$ corresponding to the $n$ \emph{control} qubits, the subscript $c$ denoting \emph{control}. As a consequence, $\mathcal{X}_{l,c}$ is an invertible matrix. The rest of the rows in $\mathcal{X}_{l}$  stand for the target qubits, denoted by the subscript $t$. We always label the rows as $i=1,2,\cdots,n,n+1,\cdots,N$ from the top, and consider the first $n$ rows to  be corresponding to the control qubits, while the rest $N-n$ rows are for target qubits. Therefore the matrix $A^\prime$ takes the form 
\begin{eqnarray}
	A^\prime=\left( \begin{array}{ccc}
	\mathcal{Z}_{l,c} 	& \mathcal{Z}_{r,c} \\ 
	\mathcal{Z}_{l,t}	& \mathcal{Z}_{r,t} \\ \hline
	\mathcal{X}_{l,c}	& 0 \\
    \mathcal{X}_{l,t}	    & 0 	
	\end{array}\right). 
	\label{eq:binary_stab-ct}
\end{eqnarray}
Our approach towards extracting the adjacency matrix $\Gamma$ underlying  a local unitary connected graph state from a stabilizer state represented by  $A^\prime$  can be summarized via the following equation:
\begin{eqnarray}
QA^\prime R=\left( \begin{array}{c}
	\Gamma\\ \hline
	I
	\end{array}\right),
\label{eq:main}
\end{eqnarray} 
where $R$ is an invertible binary matrix, and  $Q$ is constituted of local Clifford unitary operations, having the form $Q=U_ZH_{\text{target}}$. Here, $U_Z$ is a local $\frac{\pi}{2}$ rotation w.r.t. the $z$-axis on a subset of the control qubits that we are going to specify later, and $H_{\text{target}}$ represents Hadamard operations on all the target qubits. In the subsequent discussion, we explicitly calculate the L.H.S of Eq.~(\ref{eq:main}), and demonstrate the extraction of $\Gamma$.

The first step is to apply Hadamard operations on all the target qubits, which, in  the binary picture, implies the interchange of the elements above and below the horizontal line in Eq.~(\ref{eq:binary_stab_recombined}), i.e., 
\begin{eqnarray}
     H_{\text{target}}A^\prime=\left( \begin{array}{ccc}
	\mathcal{Z}_{l,c} 	& \mathcal{Z}_{r,c} \\ 
	\mathcal{X}_{l,t}	    & 0 \\ \hline
	\mathcal{X}_{l,c}	& 0 \\	
    \mathcal{Z}_{l,t}	& \mathcal{Z}_{r,t}
	\end{array}\right). 
	\label{eq:binary_stab-hadamard}
\end{eqnarray}
It can be proved that the lower block of $ H_{\text{target}}A' $ is invertible, and its inverse, $R$, can be determined. To do this, we take a different approach that the one used in \cite{van-den-nest2004}. We apply to $ H_{\text{target}}A' $ a Clifford operation represented by $ U $ that makes the upper block vanish and leaves the lower block unchanged. The unitary that we construct for the purpose of the proof is represented by
\begin{equation}
	U=\left( \begin{array}{c|c}
	I &  \begin{array}{cc}
	C &	B^{\text{T}}  \\
	B	& 0									
	\end{array}\\
	\hline
	0 &	I										
	\end{array}\right),
\end{equation}
where explicit forms of $ B,\,C $ will be given  in Eqs.~(\ref{eq:b}) and (\ref{eq:c}) respectively. The matrix $ UH_{\text{target}}A' $ is given by 
\begin{equation}
\left( \begin{array}{cc}
\mathcal{Z}_{l,c}+C\mathcal{X}_{l,c}+B^{\text{T}}\mathcal{Z}_{l,t} &  \mathcal{Z}_{r,c}+B^{\text{T}}\mathcal{Z}_{r,t} \\
\mathcal{X}_{l,t}+B\mathcal{X}_{l,c} &	0  \\\hline
\mathcal{X}_{l,c}	& 0	\\
\mathcal{Z}_{l,t} &	\mathcal{Z}_{r,t}										
\end{array}\right).
\end{equation}
From the form of the matrix $ C $ in Eq.~(\ref{eq:b}), the non-zero diagonal terms of the upper block vanish, while the off-diagonal terms vanish due to Eq.~(\ref{eq:e_2}). Hence the proof of the invertibility of the lower block of $ H_{\text{target}}A' $. Note that Clifford operations are represented by full-rank matrices, and therefore $ U $ has preserved the rank  $N$ of $ H_{\text{target}}A' $. Consequently,  the matrix $ U H_{\text{target}}A' $, which contains only the lower block, is of  full rank, and therefore, the lower block is invertible.

In order to determine $ R $, we note that the lower block of $ H_{\text{target}}A' $ has the form of a lower triangular matrix, and  the diagonal terms $ \mathcal{X}_{l,c} $ and $ \mathcal{Z}_{r,t} $ of the lower block of $ H_{\text{target}}A' $ are also invertible due to the invertibility of the lower block of $ H_{\text{target}}A' $. Therefore,  $ R $ is another lower triangular matrix by blocks having the form 
\begin{equation}
	R=\left( \begin{array}{cc}
	\mathcal{X}_{l,c}^{-1} & 0 \\ V & \mathcal{Z}_{r,t}^{-1}
	\end{array}\right) 
\end{equation}
where $ V $ must satisfy that $ \mathcal{Z}_{l,t}\mathcal{X}_{l,c}^{-1}+\mathcal{Z}_{r,t}V=0 $, leading to $ V=\mathcal{Z}_{r,t}^{-1}\mathcal{Z}_{l,t}\mathcal{X}_{l,c}^{-1} $. 
%Hence, we arrive at the form of $ R $ in Eq.~(\ref{eq:r}).

%The next step is to multiply $H_{\text{target}}A^\prime$ by $R$ from the right. We choose $R$ to be the inverse of the lower block of $H_{\text{target}}A^\prime$ (see Appendix~\ref{app:symmetry} for a proof of invertibility of $R$). Since the lower block  of $H_{\text{target}}A^\prime$ has a lower-triangular form,  and $\mathcal{X}_{l,c}$ and $\mathcal{Z}_{r,t}$ are both invertible,  $R$ can be written as 
%\begin{eqnarray}
%R=\left( \begin{array}{cc}
%\mathcal{X}_{l,c}^{-1} & 0  \\
%\mathcal{Z}_{r,t}^{-1}\mathcal{Z}_{l,t}\mathcal{X}_{l,c}^{-1} &	\mathcal{Z}_{r,t}^{-1}										
%\end{array}\right).
%\label{eq:r}
%\end{eqnarray}
The lower block of $H_{\text{target}}A^\prime$, when multiplied by $R$ from the right, becomes the identity, while the upper block becomes 
\begin{eqnarray}
\left( \begin{array}{cc}
\mathcal{Z}_{l,c}\mathcal{X}_{l,c}^{-1}+\mathcal{Z}_{r,c}\mathcal{Z}_{r,t}^{-1}\mathcal{Z}_{l,t}\mathcal{X}_{l,c}^{-1} & \mathcal{Z}_{r,c}\mathcal{Z}_{r,t}^{-1}  \\
\mathcal{X}_{l,t}\mathcal{X}_{l,c}^{-1} &	 0										
\end{array}\right).
\end{eqnarray} 
We now show that the upper block of  $H_{\text{target}}A^\prime R$ is symmetric, for which we exploit the fact that every stabilizer state, say,  $A^\prime$, due to the communitativity between all pairs of stabilizers, has to satisfy $\left(A^\prime\right)^{\text{T}}DA^\prime=0$, where $D$ is a $2N\times 2N$ binary matrix with zeros in the two $N\times N$ diagonal blocks, and $N\times N$ identity matrices in the off-diagonal blocks~\cite{hein2006} (also see Appendix~\ref{app:binary}). This leads to 
\begin{eqnarray}\label{eq:e_1}
\mathcal{X}_{l,c}^\text{T}\mathcal{Z}_{l,c}+\mathcal{X}_{l,t}^\text{T}\mathcal{Z}_{l,t}&=&\mathcal{Z}_{l,c}^\text{T}\mathcal{X}_{l,c}+\mathcal{Z}_{l,t}^\text{T}\mathcal{X}_{l,t},\\
\mathcal{X}_{l,c}^\text{T}\mathcal{Z}_{r,c}&=&\mathcal{X}_{l,t}^\text{T}\mathcal{Z}_{r,t}.
\label{eq:e_2}
\end{eqnarray}
Multiplying Eq.~(\ref{eq:e_2}) from the left hand side by $\left(\mathcal{X}_{l,c}^\text{T}\right)^{-1}$, and from the right hand side by $\mathcal{Z}_{r,t}^{-1}$,  we obtain 
\begin{eqnarray}
\mathcal{Z}_{r,c}\mathcal{Z}_{r,t}^{-1}=\left[\mathcal{X}_{l,t}\mathcal{X}_{l,c}^{-1}\right]^\text{T}.
\label{eq:e_3}
\end{eqnarray}
Next, we multiply Eq.~(\ref{eq:e_1}) from the left by $\left(\mathcal{X}_{l,c}^\text{T}\right)^{-1}$ and from the right by $\mathcal{X}_{l,c}^{-1}$ to obtain 
\begin{eqnarray}
&& \left[\mathcal{Z}_{l,c}+\left(\mathcal{X}_{l,c}^\text{T}\right)^{-1}\mathcal{X}_{l,t}^\text{T}\mathcal{Z}_{l,t}\right]\mathcal{X}_{l,c}^{-1}\nonumber \\ &&=\left(\mathcal{X}_{l,c}^\text{T}\right)^{-1} \left[\mathcal{Z}_{l,c}^\text{T}+\mathcal{Z}_{l,t}^\text{T}\mathcal{X}_{l,t}\mathcal{X}_{l,c}^{-1}\right].
\label{eq:e_4}
\end{eqnarray}  
Use of  Eqs.~(\ref{eq:e_3}) leads to the modified form of  the upper block of $H_{\text{target}}A^\prime R$ as
\begin{eqnarray} 
\Gamma^\prime=  \left( \begin{array}{cc}
C& B^{\text{T}}   \\
B &	 0										
\end{array}\right),
\end{eqnarray}
where 
\begin{eqnarray}\label{eq:b}
B&=& \mathcal{X}_{l,t}\mathcal{X}_{l,c}^{-1},\\ 
C&=&  \left[\mathcal{Z}_{l,c}+\left(\mathcal{X}_{l,c}^\text{T}\right)^{-1}\mathcal{X}_{l,t}^\text{T}\mathcal{Z}_{l,t}\right]\mathcal{X}_{l,c}^{-1} ,
\label{eq:c}
\end{eqnarray}
with $C$ being symmetric, given Eq.~(\ref{eq:e_4}). Therefore, in view of our goal to explicitly calculate $\Gamma$ from Eq.~(\ref{eq:main}), we have now  obtained 
\begin{eqnarray}
H_{\text{target}}A^\prime R=\left( \begin{array}{c}
	\Gamma^\prime \\ \hline
	I
	\end{array}\right). 
\end{eqnarray}

For the above matrix $\Gamma^\prime$ to be the adjacency matrix $\Gamma$ of a graph, the diagonal elements of $C$ must vanish, which is achieved by multiplying $H_{\text{target}}A^\prime R$  by $U_Z$ from the left, given by (see Appendix~\ref{app:binary})
\begin{eqnarray}
U_Z=\left( \begin{array}{c|c}
I &  \begin{array}{cc}
\text{diag}(C) &	0  \\
0	& 0									
\end{array}\\
\hline
0 &	I										
\end{array}\right),
\end{eqnarray} 
where $U_Z$ represents  the single-qubit $\frac{\pi}{2}$ rotations with respect to the $z$ axis on the control qubits for which $C_{ii}=1$.  This leads to the adjacency matrix $\Gamma$ of the form 
\begin{eqnarray}
\Gamma=\left( \begin{array}{c|c}
C+\text{diag}(C) &	B^{\text{T}}  \\
\hline
B	& 0		
\label{eq:adjacency_gamma}							
\end{array}\right).
\end{eqnarray}

Based on the above discussion, one can develop an algorithm for obtaining the adjacency matrix $\Gamma$ corresponding to the graph $G$ representing the graph state that is local unitary equivalent to the stabilizer state, by utilizing the stabilizer structure of the state only. This algorithm has been realized in the form of a Python open-source  package called \textbf{StabGraph}~\cite{amaro_github_2019}, which generates the adjacency matrix corresponding to the graph underlying the graph state which is connected via local unitary operations with the stabilizer state (see Appendix~\ref{app:pseudo_codes} for a pseudo code of the algorithm). Our  recipe using the Gaussian elimination technique scales as $\sim N^3$ with the system size, $N$. Note here that in the case of CSS codes~\cite{nielsen2010}, $\mathcal{Z}_{l,c}$ and $\mathcal{Z}_{l,t}$ in $A^\prime$ can be set to zero by stabilizer recombination, implying $C=0$, thereby ensuring that the adjacency matrix 
\begin{eqnarray}
\Gamma=\left( \begin{array}{c|c}
0 &	B^{\text{T}}  \\
\hline
B	& 0									
\end{array}\right)
\end{eqnarray} 
has vanishing diagonal blocks. This also implies that links between a pair of control qubits and a pair of target qubits are prohibited, thereby ensuring that the resulting graph is bicolorable, although in the general stabilizer state, the existence of $(c,c)$-type links is allowed.  Moreover, transformation of the stabilizer state to the graph state does not require the application of the local unitary operation $U_Z$, if $C=0$.

\begin{figure*}
\includegraphics[width=0.7\textwidth]{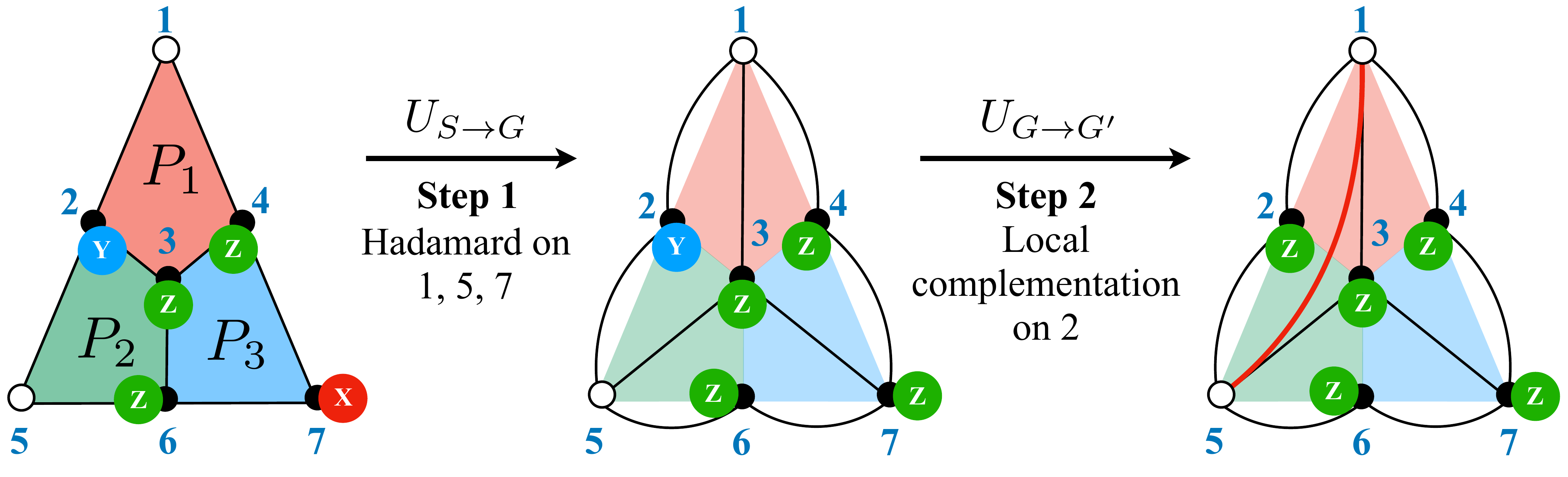}
\caption{(Color online.) Transformation of a $7$-qubit topological color code to a graph $G^\prime$ by using the graph-based algorithm described in Sec.~\ref{subsec:mlb}, where the nodes $1$ and $5$ are connected by a link in $G^\prime$. A non-zero value of $E^{\mathcal{P}}_{15}(\rho^\prime)$ in $G^\prime$ requires local projection measurement in the $Z$-basis at all other qubits. This corresponds to $X$- and $Y$ measurement on qubits $7$  and $2$, respectively, and $Z$-measurement everywhere else. See Appendix~\ref{app:7qubit} for details.}
\label{fig:graph_based_method_example}
\end{figure*}

\section{Graph-based algorithm on a 7-qubit color code}
\label{app:7qubit}

The graph-based method for computing the measurement-based lower bound of LE has been discussed in Sec.~\ref{subsec:mlb}. In this appendix, for illustration, we apply the graph-based method described in Sec.~\ref{subsec:mlb} on the smallest 2D color code, constructed over a triangular lattice of seven qubits arranged in three adjoined plaquettes, where one physical qubit is placed at each vertex. The stabilizer operators corresponding to the 7-qubit color code are constituted of 4-qubit $X$- and $Z$-type operators for each plaquette, given by
\begin{eqnarray}
S_1^x&=&X_1X_2X_3X_4,\;S_1^z=Z_1Z_2Z_3Z_4,\nonumber \\
S_2^x&=&X_2X_3X_5X_6,\;S_2^z=Z_2Z_3Z_5Z_6,\nonumber\\
S_3^x&=&X_3X_4X_6X_7,\;S_3^z=Z_3Z_4Z_6Z_7.
\end{eqnarray}
Let us assume that the two-qubit region is given by the qubits $a\equiv1$ and $b\equiv 5$ (see Fig.~\ref{fig:graph_based_method_example}), where $\overline{\Omega}=\{2,3,4,6,7\}$.  The logical state $\ket{+}_L$ of this TCC can be transformed into a graph state $\ket{\psi_G}$ via the unitary transformation $U_{S\rightarrow G}=H_1\otimes H_5 \otimes H_7$, where the set of control qubits here is composed of the qubits $\{1,5,7\}$, and the set of targets holds the qubits $\{2,3,4,6\}$.  The link $(a,b)$ is absent in the graph $\mathcal{G}$. Therefore, a subsequent LC operation on the qubit $2$ is performed to create the graph $\mathcal{G}^\prime$, where the link $(a,b)$ exists. This LC transformation is equivalent to $U_{G\rightarrow G^\prime}=\exp (\text{i}\pi Z_{1}/4)\otimes\exp (-\text{i}\pi X_{2}/4)\otimes\exp (\text{i}\pi Z_{5}/4)$ (see Appenndix~\ref{app:graphs}).  Therefore, the unitary operator $U_{S\rightarrow G^\prime}$ consists of the local unitary operators $U_2=\exp(-\text{i}\pi X_2/4)$, $U_3=U_4=U_6=I$, and $U_7=H_7$ in the region $\overline{\Omega}\equiv\overline{ab}$. Consequently, $E^{\mathcal{P}}_{ab}(\rho^\prime)$ as a MLB with $\mathcal{P}$ $\equiv$ $Z$ measurement on all qubits $\in$ $\overline{ab}$ in $\rho^\prime$ is equivalent to $Y$ measurement on qubit $2$, $X$ measurement on qubit $7$, and $Z$ measurements on qubits $3$, $4$, and $6$, denoted altogether by $\mathcal{P}^\prime$ in $\rho_S$, so that $E^{\mathcal{P}}_{ab}(\rho^\prime)=E^{\mathcal{P}^\prime}_{ab}(\rho_S)$.

\section{Graph transformations via Adaptive Local Complementation}
\label{app:graph-based}

As mentioned in Sec.~\ref{subsec:mlb}, for the successful implementation of the graph-based protocol, one needs to develop a graph trannsformation algorithm that creates a link between any two chosen nodes in a simple, connected, undirected graph via a sequence of  local complementation operations over a set of nodes in the graph. We have developed an adaptive local complementation technique that updates itself at every step of the graph transformation via local complementation over one qubit, and chooses the next node for local complementation using the updated information. Here, we discuss the technical details of the adaptive LC technique developed for the graph transformation $G\rightarrow G^\prime$ used in Sec.~\ref{subsec:mlb}.  The prescription works for a generic simple, connected, and undirected graph $G$, irrespective of whether it is bicolorable or not. 

We shall denote a link connecting two nodes $i$ and $j$ by $(i,j)$.  We shall also represent a simple path in the graph, connecting the two nodes $a$ and $b$ and having length $l$, by $\mathcal{L}_{ab}^{(l)}$, which, without any loss in generality, is given by (see Appendix~\ref{app:graphs})
\begin{eqnarray}
\mathcal{L}_{ab}=[a\equiv 1,2,\cdots ,n\equiv b].
\label{eq:simple_path_main_text}
\end{eqnarray}
Here, $\{2,3,\cdots ,n-1\}$ are the nodes that are visited while traveling from $a$ to $b$ along $\mathcal{L}_{ab}^{(l)}$,  such that $(i,i+1)\in G$, $i=1,\cdots,n-1$, and none of the nodes is repeated in the sequence.  The number of links traversed while going from $a$ to $b$ is the length $l=n-1$ of the path. To keep the notation uncluttered, we will drop the subscript $ab$ and the superscript $(l)$ from the notation of the simple path unless it is explicitly required for comparison.  For the purpose of this discussion, one needs a specific classification of the simple paths connecting two nodes, and a definition of the distance between them, as follows.  

\begin{figure}
\includegraphics[scale=0.3]{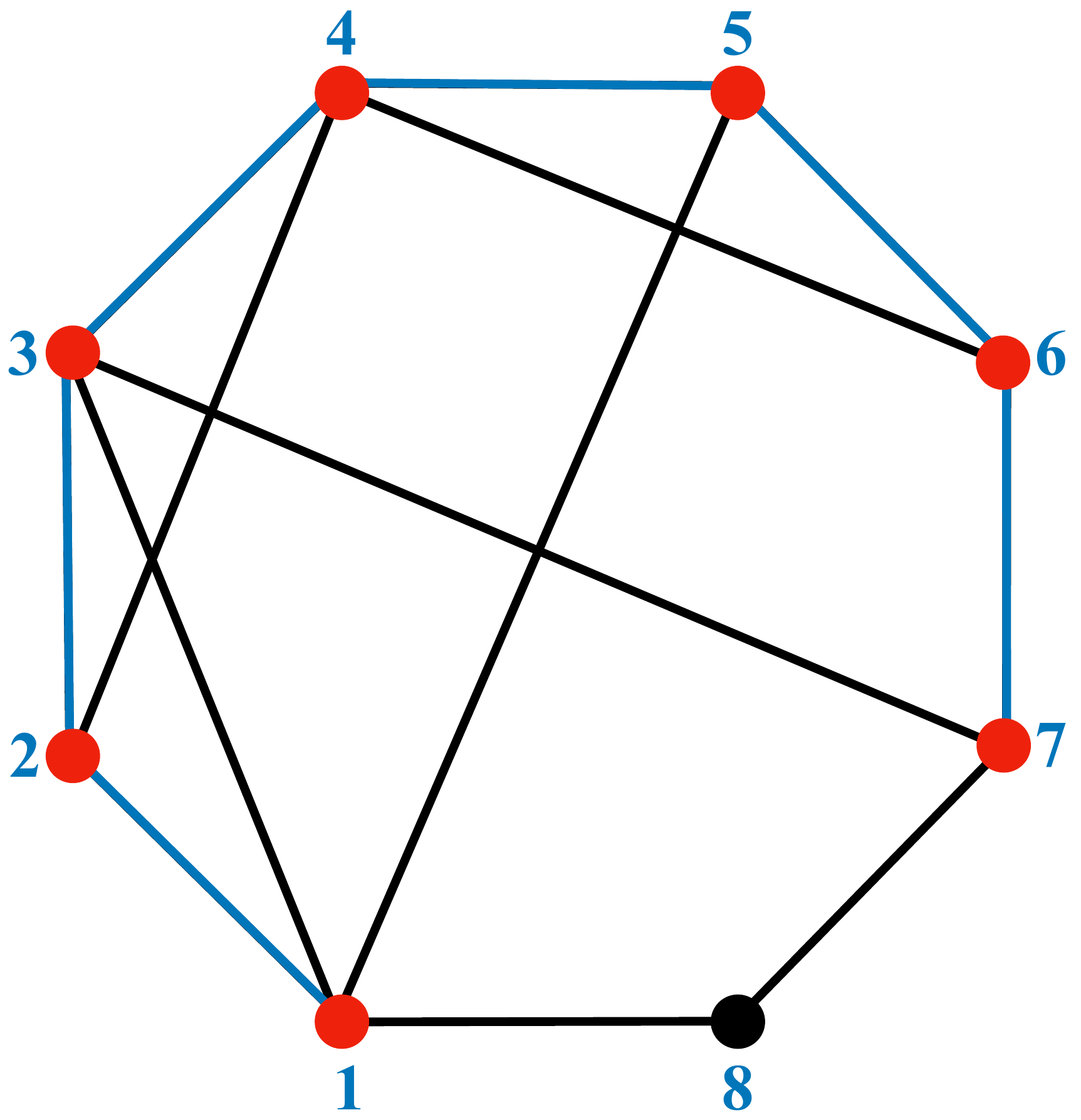}
\caption{(Color online.) \textbf{Examples of $\mathcal{C}_1$ and $\mathcal{C}_2$ paths.} A simple, connected, and undirected graph of $8$ nodes. For demonstration, we consider the path $\mathcal{L}=[a\equiv 1,2,3,4,5,6,7\equiv b]$, which is a $\mathcal{C}_2$ path due to the existence of the links $(1,3)$, $(1,5)$, $(2,4)$, $(3,7)$, and $(4,6)$. From these links and the links constructing the path $\mathcal{L}$, three $\mathcal{C}_1$ paths of different lengths $l=2,3,4$ can be distilled, given by $\mathcal{L}_{1}^{(4)}=[1,2,4,6,7]$, $\mathcal{L}_{1}^{(3)}=[1,5,6,7]$,  and $\mathcal{L}_3^{(2)}=[1,3,7]$. The path $\mathcal{L}_2^{(3)}$ can be determined via the $\mathcal{C}_2\rightarrow\mathcal{C}_1$ algorithm. }
\label{fig:graph}
\end{figure}

\noindent\textbf{Category of simple paths.} We first divide the set of all possible simple paths between two given nodes $a$ and $b$ into two categories, $\mathcal{C}_1$ and $\mathcal{C}_2$. A simple path $\mathcal{L}$ of the form given in Eq. (\ref{eq:simple_path_main_text}) belongs to the category  $\mathcal{C}_1$ if a qubit on $\mathcal{L}$ is not connected to another qubit on $\mathcal{L}$ via a direct link, unless the second qubit is right before or after the first qubit in the sequence $\mathcal{L}$. Formally put, for two qubits $i$ and $j$ on $\mathcal{L}$,  iff $(i,j)$ $\notin$ $G$ $\forall$ $i,j$ $\in$ $\mathcal{L}$,  $i<j\leq n$, $j\neq i+1$, then $\mathcal{L}\in\mathcal{C}_1$. Otherwise, $\mathcal{L}\in\mathcal{C}_2$ (see Fig.~\ref{fig:graph} for an example). Note that $\mathcal{C}_1\cap\mathcal{C}_2=\emptyset$, while $\mathcal{C}_1\cup\mathcal{C}_2$ constitutes the set of all possible simple paths between $a$ and $b$.  

\noindent\textbf{Distance between two nodes.} There can be a number of ways in which the distance between a pair of nodes in a graph can be quantified. However, the distance between two nodes $i$ and $j$ that is specific to a given path $\mathcal{L}$ of the form in Eq.~(\ref{eq:simple_path_main_text}) is unique. It is measured by the number of links that one has to traverse while going from $i$ to $j$ along $\mathcal{L}$, and can be represented by  $d_{(i,j)}^{\mathcal{L}}=j-i$, where we have assumed $j>i$ without any loss of generality.  Note that for a path $\mathcal{L}$, $d_{(1,n)}^{\mathcal{L}}=l=n-1$. 

All shortest paths between two given nodes $a$ and $b$ in a simple, connected, and undirected graph belongs to $\mathcal{C}_1$. This can be proved by assuming that there exists a shortest path (see Appendix~\ref{app:graphs} for a definition) $\mathcal{L}$ between two nodes $a\equiv 1$ and $b\equiv n$ that belongs to $\mathcal{C}_2$. However, by definition of $\mathcal{C}_2$, there exists at least one link $(i,j)$ such that $i,j$ $\in$ $\mathcal{L}$, $i<j\leq n$, and $j\neq i+1$. This implies that there exists a simple path between $a\equiv 1$ and $b\equiv n$, given by 
\begin{eqnarray}
\mathcal{L}^{(l^\prime)} = [1,2,\cdots,{i-1},i,j,{j+1},\cdots,n]\nonumber\\
\end{eqnarray}
of length $l^\prime=l-(j-i)+1$, which is $<l$ since $j>i+1$ by the definition of $\mathcal{C}_2$. Therefore, $\mathcal{L}$ is not a shortest path. However, note that  not all paths in $\mathcal{C}_1$ are shortest paths (see Fig.~\ref{fig:graph} for examples). 

We are now in a position to prove the following \textbf{Theorem}, which will be crucial for our purpose. 

\noindent\textbf{Theorem 1.} \emph{For a simple path $\mathcal{L}\in\mathcal{C}_1$ of the form given in Eq.~(\ref{eq:simple_path_main_text}) between a pair of nodes $a\equiv 1$ and $b\equiv n$  in a graph $G$, a sequence of LC operations on the nodes $\{2,\cdots,{n-1}\}$ always creates a link between the nodes $a$ and $b$, when the LC operations are performed on the nodes in the same order as they are in the sequence $\mathcal{L}$.}

\noindent\emph{Proof.} Let us consider the path $\mathcal{L}\in\mathcal{C}_1$, as given in Eq.~(\ref{eq:simple_path_main_text}), between the nodes $a$ and $b$ in a simple, connected, and undirected graph $G$. Let us denote the neighborhood of the node $2$ as $N_2$. The definition of $\mathcal{C}_1$ implies that only $1,3\in\mathcal{L}$ belongs to $N_2$, while $\{4,\cdots,n\}$ $\notin$ $N_2$.  Let us assume that the LC operation (see Appendix~\ref{app:graphs}) on node $2$ transforms the graph as $G$ $\rightarrow$ $G_2$ $=$ $\tau_2(G)$. The definition of LC operation implies creation of the  link $(1,3)$ thereby introducing $1$ in the neighborhood $N_3$ of $3$, while keeping the neighborhoods of the rest of the nodes $\{4,5,\cdots,n\}$ unchanged. Also, the LC operation $\tau_{2}$ does not affect the links $\{(i,i+1)$, $i=3,4,\cdots,n-1\}$. Therefore, in $G_{2}$, there exists a simple path of length $l^\prime=l-1$, given by $[a\equiv 1,3,4,\cdots,n\equiv b]$ $\in$ $\mathcal{C}_1$, between the pair of nodes $a$ and $b$. One can continue performing a total of $n-2$ LC operations successively on the nodes $2,3,\cdots,n-1$ in the same order as they are in the sequence $\mathcal{L}$, where during each individual LC operations on the node $i$, $i=2,3,\cdots,n-1$, the above arguments apply, and a link is created between the node $1$ and the node $i+1$. The last LC operation on the node $n-1$ creates a link between the nodes $1$ and $n$. Hence the proof. \hfill $\blacksquare$

\noindent\textbf{Corollary 1.1.} \emph{For a shortest path $\mathcal{L}$ of the form given in Eq.~(\ref{eq:simple_path_main_text})  between a pair of nodes $a\equiv 1$ and $b\equiv 2$ in a graph $G$, a sequence of LC operations on the nodes $\{2,\cdots,n-1\}$ always creates a link between the nodes $a$ and $b$, when the LC operations are performed on the nodes in the same order as they are in the sequence $\mathcal{L}$.}

\noindent\textbf{Corollary 1.2.} \emph{For a $\mathcal{C}_1$ path of length $l=n-1$, the number of LC operations required to create the link $(a,b)$ is $n_{\text{\scriptsize LC\normalsize}}=n-2$.}

\noindent\textbf{Theorem 1} suggests that the creation of a link between  $a$ and $b$ in a connected graph is ensured if one chooses only $\mathcal{C}_1$ paths between $a$ and $b$ and perform LC operations on the qubits in those paths. However, there usually exists a large number of $\mathcal{C}_2$ paths in a connected graph, and there is no intuition whatsoever for choosing a $\mathcal{C}_1$ path in order to obtain a \emph{tight} MLB of LE.  The challenge therefore is to develop an effective algorithm that can guarantee the creation of the link $(a,b)$ irrespective of whether the chosen path belongs to  $\mathcal{C}_1$ or $\mathcal{C}_2$, when LC operations only on the nodes belonging to the chosen path are performed.  Towards this goal, note that by virtue of the definition of $\mathcal{C}_2$, for a path $\mathcal{L}\in\mathcal{C}_2$, one can distil a number of $\mathcal{C}_1$ paths of different lengths  from the links constructing $\mathcal{L}$ and the links connecting the nodes in the $\mathcal{L}$, but not contributing to $\mathcal{L}$ (see Fig.~\ref{fig:graph} for examples). These $\mathcal{C}_1$ paths have support on the nodes $\in\mathcal{L}$ only, and therefore the condition of performing LC operations only on the nodes belonging to $\mathcal{L}$ is satisfied. \textbf{Theorem 1} ensures that each of these $\mathcal{C}_1$ paths can be chosen to create $(a,b)$ with certainty. Therefore, the challenge is to find a $\mathcal{C}_1$ path from the $\mathcal{C}_2$ path, which, in turn, leads to the creation of the link $(a,b)$ according to \textbf{Theorem 1}. It is easily understood how this is achieved via the $\mathcal{C}_2\rightarrow\mathcal{C}_1$ algorithm represented by the pseudocode in Appendix~\ref{app:c1c2}.

%\vspace{0.25cm}
%\begin{minipage}{8cm}
%\hrulefill
%\small 
%
%\noindent\textbf{$\mathcal{C}_2\rightarrow\mathcal{C}_1$ algorithm}
%
%\noindent\textbf{input} (1) graph $G$
%
%\noindent\hspace{0.9cm}(2) qubit-pair $\{a,b\}$ such that $(a,b)\notin G$
%
%\noindent\hspace{0.9cm}(3) a path $\mathcal{L}\in\mathcal{C}_2$ between $\{a,b\}$
%
%\noindent\textbf{initiate} \emph{list of nodes} $\mathcal{L}^\prime=[a]$
%
%\noindent\textbf{1.} \emph{determine} $N_a$: \emph{neighborhood of } $a\equiv1$
%
%\noindent\textbf{2.} \emph{determine} $N_a\cap \mathcal{L}$
%
%\noindent\textbf{3.} \emph{determine node} $i$ $\in$ $N_a\cap \mathcal{L}$ \emph{with maximum} $d^{\mathcal{L}}_{(a,i)}$
%
%\noindent\textbf{4.} \emph{append} $i$ \emph{to} $\mathcal{L}^\prime$
%
%\noindent\textbf{5. if} $i\neq b$ \textbf{then}
%
%\noindent\hspace{0.5cm}(i) $a=i$
%
%\noindent\hspace{0.5cm}(ii) \textbf{repeat 1-5}
%
%\noindent\textbf{6. else}
%
%\noindent\hspace{0.5cm}\textbf{ stop}
%
%\noindent\textbf{output} $\mathcal{L}^\prime \in\mathcal{C}_1$ connecting $a,b$
%
%\normalsize 
%\hrulefill
%\end{minipage}
%\vspace{0.25cm}

Note here that  to minimize effort, one has to use  the distilled $\mathcal{C}_1$ path having the shortest length, which may or may not be the shortest path between $a$ and $b$. However, there is no intuitive reason behind using the shortest path, or the shortest distilled path to create the link, since the value of the MLB computed via the graph-based method I  depends explicitly on the structure of the graph, and how the size of the neighborhood of  the qubit-pair $\{a,b\}$ in the transformed graph depends on the length of the chosen path between $a$ and $b$ in the original graph.   In a graph with high connectivity, there may exist other paths having length close to the shortest length which may provide a tighter value of the MLB compared to the same obtained from the shortest path (we elaborate more on this in Sec.~\ref{sec:apply}). The $\mathcal{C}_2\rightarrow\mathcal{C}_1$ algorithm focuses on finding any one of the distilled $\mathcal{C}_1$ paths that creates the link $(a,b)$. For example, in Fig.~\ref{fig:graph}, the $\mathcal{C}_2\rightarrow\mathcal{C}_1$ algorithm distils the $\mathcal{C}_1$ path $\mathcal{L}_2^{(3)}$ of length $3$, while the shortest distilled $\mathcal{C}_1$ path is $\mathcal{L}_2^{(2)}$ of length $2$, which is also one of the shortest paths between the qubits $1$ and $7$.

We now combine all these insights, and develop an LC-based graph-transformation algorithm that creates a link $(a,b)$ by performing LC operations on a subset of the set of nodes situated on a chosen path, where the subset of nodes is chosen by adapting the chosen path according to its category. We call this algorithm the \emph{Adaptive LC} (ALC) algorithm, which has been represented by the pseudocode in Appendix~\ref{app:alc}. The LC operation on a simple, connected, and undirected graph w.r.t. a chosen node, the distillation of a $\mathcal{C}_1$ path from a $\mathcal{C}_2$ path via the $\mathcal{C}_2\rightarrow \mathcal{C}_1$ algorithm, and the creation of a link between any two given nodes in a graph by using ALC algorithm have been realized in the form of the Python open-source package \textbf{ALCPack}~\cite{pal_github_2019}.

%\vspace{0.25cm}
%\begin{minipage}{8cm}
%\hrulefill
%\small 
%
%\noindent\textbf{Adaptive local complementation  algorithm}
%
%\noindent\textbf{input} (1) graph $G$
%
%\noindent\hspace{0.9cm}(2) qubit-pair $\{a,b\}$ such that $(a,b)\notin G$ and 
%
%\noindent\hspace{1.3cm} there exists a simple path $\mathcal{L}$ connecting $a$ and $b$ 
%
%\noindent\textbf{1.} \emph{determine category of} $\mathcal{L}$
%
%\noindent\textbf{2. if} $\mathcal{L}\in\mathcal{C}_2$ \textbf{then}
%
%\noindent\hspace{0.5cm}(i) \emph{apply $\mathcal{C}_2\rightarrow\mathcal{C}_1$ algorithm and determine the} 
%
%\noindent\hspace{0.9cm}shortest path $\mathcal{L}^\prime\in\mathcal{C_1}$
%
%\noindent\hspace{0.5cm}(ii) $\mathcal{L}=\mathcal{L}^\prime$
%
%  
%\noindent\textbf{3.} \emph{perform local complementation  operations on nodes} 
%
%\noindent\hspace{0.4cm}\emph{belonging to $\mathcal{L}$ in same sequence as $\mathcal{L}$}
%
%\noindent\textbf{output} \emph{transformed graph} $G^\prime$, $\{a,b\}$ $\mid$ $(a,b)\in G^\prime$ 
%
%\normalsize 
%\hrulefill
%\end{minipage}
%\vspace{0.25cm}

A word on the dependence of the run-time  of the adaptive LC algorithm on the system size $N$ is in order here. Since the algorithm takes into account the transformed graphs at each of its steps, it is difficult to determine an exact dependence of the run-time of the algorithm on system size. However, one can determine a bound on how the run-time scales with $N$. It is easy to see that the maximum size of the neighborhood $N_i$ of a node $i$ can be $\leq N-1$,  and it can host at most $\binom{N-1}{2}$ links if $G$ is connected and undirected. Therefore, the maximum number of links that can be created or deleted during an LC operation on a single node is   $\binom{N-1}{2}$. Since there can be at most $N-2$ nodes on a path $\mathcal{L}$ between $a$ and $b$, the total number of link operations during the ALC algorithm is $\leq (N-2)\binom{N-1}{2}\leq N^3$, which indicates a polynomial scaling with system size.

\section{Pseudo codes for different algorithms}
\label{app:pseudo_codes}
In this Appendix, we present pseudo-codes for the different algorithms introduced in this paper. 

\subsection{Graph-based method}
\label{app:gbm}

We now present the pseudo code representing the graph-based method to calculate the measurement-based lower bound for localizable entanglement. 

\noindent\hrulefill

\noindent\textbf{input} (1) $\ket{\psi_S}$ on system $S$ of size $N$, (2) qubit-pair $\{a,b\}$, and (3) noise map $\Lambda(.)$

\begin{enumerate}

\item\emph{obtain} $\mathcal{S}_G$ \emph{and} $U_{S\rightarrow G}$ \emph{corresponding to all}  $G\in\mathcal{S}_G$

\item \textbf{for} \emph{all} $G\in\mathcal{S}_G$, \textbf{do}
      \begin{enumerate}
      \item \textbf{if} \emph{link} $(a,b)\notin G$ \textbf{then}
            \begin{enumerate} 
            \item \textbf{for} \emph{all} $\mathcal{L}_{ab}\in\mathcal{S}_{\mathcal{L}_{ab}}$, \textbf{do}
                  \begin{enumerate}
                  \item\emph{perform the transformation} $G\rightarrow G^\prime$ \emph{via adaptive local complementations on} $\mathcal{L}_{ab}$
                  \item\emph{determine}  $U_{G\rightarrow G^\prime}$
                  \item\emph{determine noise transformation } $\Lambda\rightarrow \Lambda^\prime$
                  \item\emph{due to} $U_{S\rightarrow G^\prime}=U_{G\rightarrow G^\prime}U_{S\rightarrow G}$     
                  \item\emph{compute} $E^{\mathcal{P}}_{ab}(\rho^\prime)$ \emph{following}~\cite{amaro2018}
                  \item\emph{determine} $\mathcal{P}^\prime=U^{-1}_{S\rightarrow G^\prime}\mathcal{P}U_{S\rightarrow G^\prime}$
                  \end{enumerate}
            \end{enumerate}      
      \item \textbf{else} 
            \begin{enumerate} 
            \item\emph{determine} $\Lambda\rightarrow \Lambda^\prime$ \emph{due to} $U_{S\rightarrow G^\prime}=U_{S\rightarrow G}$     
            \item\emph{compute} $E^{\mathcal{P}}_{ab}(\rho^\prime)$ \emph{following}~\cite{amaro2018} 
            \item\emph{determine} $\mathcal{P}^\prime=U^{-1}_{S\rightarrow G}\mathcal{P}U_{S\rightarrow G}$
            \end{enumerate}
      \end{enumerate}
\end{enumerate}              
%\noindent\textbf{repeat} \textbf{1}-\textbf{3} $\forall$ $G\in\mathcal{S}_{G}$
\noindent\textbf{output} (1) $\max E^{\mathcal{P}}_{ab}(\rho^\prime)$, and (2) measurement setup $\mathcal{P}^\prime$

\noindent\hrulefill

\subsection{Modified graph-based method}
\label{app:mgbm}

The graph-based method will be  modified if the graph obtained from the stabilizer state already has a link between the chosen qubits. The modified  algorithm is as follows.

\noindent\hrulefill

\noindent\textbf{input} (1) $\ket{\psi_S}$ on system $S$ of size $N$, (2) qubit-pair $\{a,b\}$, and (3) noise map $\Lambda(.)$
\begin{enumerate}
\item \emph{obtain} $\mathcal{S}_G$ \emph{and} $U_{S\rightarrow G}$ \emph{corresponding to all}  $G\in\mathcal{S}_G$ \emph{such that} $(a,b)\in G$
\item \textbf{for} \emph{all} $G\in\mathcal{S}_G$, \textbf{do}
      \begin{enumerate}
      \item \emph{determine} $\Lambda\rightarrow \Lambda^\prime$ \emph{due to} $U_{S\rightarrow G}$     
      \item \emph{compute} $E^{\mathcal{P}}_{ab}(\rho^\prime)$ \emph{following}~\cite{amaro2018}
      \item \emph{determine} $\mathcal{P}^\prime=U^{-1}_{S\rightarrow G}\mathcal{P}U_{S\rightarrow G}$
      \end{enumerate}
\end{enumerate}
\noindent\textbf{output} (1)  $\max E^{\mathcal{P}}_{ab}(\rho^\prime)$, (2) measurement setup $\mathcal{P}^\prime$

\noindent\hrulefill

\subsection{Determination of the adjacency matrix}
\label{app:adj}

The following pseudo code is for obtaining the adjacency matrix of bicolorable graphs from the stabilizer states as per the  algebraic approach discussed in Appendix~\ref{app:algebraic_approach}. Note that similar code has been used to develop the \textbf{Stabgraph} package. 

\noindent \hrulefill

\noindent\textbf{input} Stabilizer state in the binary representation as given in Eq.~(\ref{eq:binary_stab})
\begin{enumerate}
\item \emph{obtain} $A^\prime$ \emph{via Gauss elimination on} $A$
\item \emph{determine $\mathcal{X}_{l,c}$ in bottom block via reordering rows}
\item \emph{apply same reordering in top block to get $A^\prime$ in} Eq.~(\ref{eq:binary_stab-ct})
\item \emph{identify $\mathcal{X}_{l,c}$,$\mathcal{X}_{l,t}$,$\mathcal{Z}_{l,c}$,$\mathcal{Z}_{l,t}$ from $A^\prime$} 
\item \emph{determine $B$ and $C$} (Eqs.~(\ref{eq:b}) and (\ref{eq:c}))
\item \emph{construct $\Gamma$} (Eq.~(\ref{eq:adjacency_gamma}))
\end{enumerate}
\noindent\textbf{output} adjacency matrix $\Gamma$ describing a graph state which is local unitary equivalent to the input stabilizer state.

\noindent\hrulefill

\subsection{Obtaining a \texorpdfstring{$\mathcal{C}_1$}{CI} path from a \texorpdfstring{$\mathcal{C}_2$}{C2} path}
\label{app:c1c2}

We now present the algorithm developed in Appendix~\ref{app:graph-based} to convert a category $\mathcal{C}_2$ path to a category $\mathcal{C}_1$ path.

\noindent\hrulefill

\noindent\textbf{input} (1) graph $G$, (2) qubit-pair $\{a,b\}$ such that $(a,b)\notin G$, and (3) a path $\mathcal{L}\in\mathcal{C}_2$ between $\{a,b\}$

\noindent\textbf{initiate} \emph{list of nodes} $\mathcal{L}^\prime=[a]$
\begin{enumerate}
\item \emph{determine} $N_a$: \emph{neighborhood of } $a\equiv1$
\item \emph{determine} $N_a\cap \mathcal{L}$
\item \emph{determine node} $i$ $\in$ $N_a\cap \mathcal{L}$ \emph{with maximum} $d^{\mathcal{L}}_{(a,i)}$
\item \emph{append} $i$ \emph{to} $\mathcal{L}^\prime$
\item \textbf{if} $i\neq b$ \textbf{then}
      \begin{enumerate}
      \item $a=i$
      \item \textbf{repeat 1-5}
      \end{enumerate}
\item \textbf{else}
      \begin{itemize}
      \item[]\textbf{ stop}
      \end{itemize}
\end{enumerate}
\noindent\textbf{output} $\mathcal{L}^\prime \in\mathcal{C}_1$ connecting $a,b$

\noindent\hrulefill

\subsection{Adaptive local complementation}
\label{app:alc}

We can now adapt the pseudo codes discussed in Appendix~\ref{app:c1c2} and obtain an \emph{adaptive  local complementation} algorithm, as follows. This also is the underlying algorithm for the \textbf{ALCPack}.

\noindent\hrulefill 

\noindent\textbf{input} (1) graph $G$, and (2) qubit-pair $\{a,b\}$ such that $(a,b)\notin G$ and there exists a simple path $\mathcal{L}$ connecting $a$ and $b$ 
\begin{enumerate}
\item \emph{determine category of} $\mathcal{L}$
\item \textbf{if} $\mathcal{L}\in\mathcal{C}_2$ \textbf{then}
      \begin{enumerate}
      \item \emph{apply $\mathcal{C}_2\rightarrow\mathcal{C}_1$ algorithm and determine the shortest path $\mathcal{L}^\prime\in\mathcal{C_1}$}
      \item $\mathcal{L}=\mathcal{L}^\prime$
      \end{enumerate}
\item \emph{perform local complementation  operations on nodes belonging to $\mathcal{L}$ in same sequence as $\mathcal{L}$}
\end{enumerate}
\noindent\textbf{output} \emph{transformed graph} $G^\prime$, $\{a,b\}$ $\mid$ $(a,b)\in G^\prime$ 

\noindent\hrulefill

\section{Graphs from color codes: Illustration}
\label{app:ct}
In this appendix, we demonstrate the procedure for obtaining a graph state from a topological color code  with an example. In the case of a four-qubit plaquette $P$ in a topological color code,  constituted of four qubits $c$, $t_1$, $t_2$, and $t_3$, the first step is to initialize the four qubits in the product state $\ket{+_c0_{t_1}0_{t_2}0_{t_3}}$, and then to apply CNOT gates $U^x_{(c,t_i)}$ successively over the qubit-pairs $\{c,t_1\}$, $\{c,t_2\}$, and $\{c,t_3\}$, always using qubit $c$ as the control qubit. This leads to the four-qubit GHZ state 
\begin{eqnarray}
\ket{\psi_{P}}&=&C^x_{(c,t_1)}C^x_{(c,t_2)}C^x_{(c,t_3)}\ket{+_c0_{t_1}0_{t_2}0_{t_3}},\nonumber\\
&=& \frac{1}{\sqrt{2}}(\ket{0000}+\ket{1111}),\nonumber \\ 
\end{eqnarray}  
over the plaquette $P$. Here, the CNOT operator $U^x_{(c,t_i)}$ is given by 
\begin{eqnarray}
U^x_{(c,t_i)}=\frac{1}{2}\left[(I_c+Z_c)I_{t_i}+(I_c-Z_c)X_{t_i}\right],
\end{eqnarray}
for a qubit-pair $\{c,t_i\}$, where $c$ denotes the control and $t_i$ denotes a target qubit.  The state $\ket{\psi_{P}}$ can  be further transformed to a graph state $\ket{\psi_{G_{P}}}=\frac{1}{\sqrt{2}}(\ket{0+++}+\ket{1---})$ corresponding to a simple, connected, and undirected  star-shaped  graph $G_{P}$ with the control qubit $c$ as the central qubit via local unitary transformations $H_{t_1}H_{t_2}H_{t_3}$ on the target qubits~\cite{hein2006}, where $H$ represents a Hadamard operator (see Fig.~\ref{fig:circuit}).  Note here that the convention of creating the graph state $\ket{\psi_{G_{P}}}$ involves initializing all the qubits in $\ket{+}$, and subsequently applying a controlled phase gate $U^z_{(c,t_i)}$ on all $(c,t_i)$ pairs of  nodes~\cite{hein2006}, where the controlled phase gate,
\begin{eqnarray}
U^z_{(c,t_i)}=\frac{1}{2}\left[(I_c+Z_c)I_{t_i}+(I_c-Z_c)Z_{t_i}\right],
\label{eq:czgate}
\end{eqnarray}  
is connected to $U^x_{(c,t_i)}$ via a Hadamard operation on the target qubit: $U^z_{(c,t_i)}=H_{t_i}U^x_{(c,t_i)}H_{t_i}$. In terms of the stabilizer operators, the plaquette stabilizer $S^x_P=X_cX_{t_1}X_{t_2}X_{t_3}$ is transformed to the  graph-state generator $X_{c}Z_{t_1}Z_{t_2}Z_{t_3}$ via application of Hadamard operations on the target qubits.

\section{Binary picture}
\label{app:binary}

Here we provide a few essential details on the binary picture of the stabilizer formalism~\cite{hein2006}. 

\noindent\textbf{Pauli operators.} In the binary picture, the single-qubit identity and Pauli operators are represented by the  column vectors
\begin{eqnarray}
	I&=&\left( \begin{array}{c}
	0\\ \hline 0\end{array}\right),\, X=\left( \begin{array}{c}
	0\\ \hline 1\end{array}\right),\, Y=\left( \begin{array}{c}
	1\\ \hline 1\end{array}\right),\, Z=\left( \begin{array}{c}
	1\\ \hline 0\end{array}\right),\nonumber \\ 
	\label{eq:binary_pauli}
\end{eqnarray}
and the products of these operators are mapped to the sum of the columns modulo $2$. Note that in the binary picture, the global phases are disregarded. 

\noindent\textbf{Clifford operations.} Single-qubit Clifford operators are $ 2\times 2 $ binary matrices that multiply from the left to the Pauli operators given in Eq.~(\ref{eq:binary_pauli}). Given that these operators do not transform Pauli matrices into the identity, their representative binary matrix is full-rank. As an examples, the Hadamard gate $ H $ and a $ \pi/2 $ $ z $-rotation $ U_Z $ are represented by
\begin{eqnarray}\label{single-qubit_unitaries}
	H=\left(\begin{array}{c|c}
	0 & 1 \\ \hline 1 & 0
	\end{array}\right),\quad  	U_Z=\left(\begin{array}{c|c}
	1 & 1 \\ \hline 0 & 1
	\end{array}\right)	.
\end{eqnarray}
One can check that $ H $ interchanges $ X $ and $ Z $, while $ U_Z $ changes  $ X $ to $ Y $.

\noindent\textbf{Stabilizer state.} A stabilizer state of $N$ qubits is represented by a $ 2N\times N $ binary matrix $ A $ formed by two $ N\times N $ blocks, 
\begin{equation}
	A=\left( \begin{array}{c}
	\mathcal{Z}\\ \hline \mathcal{X}\end{array}\right),
\end{equation}
where the $ j $-th stabilizer is represented by the $ j $-th column, and the qubit $ i $ is represented by the rows $ \mathcal{Z}_i $ and $ \mathcal{X}_i $ in such a way that the column of two elements, written as 
\begin{equation}
	\left( \begin{array}{c}
	\mathcal{Z}_{ij}\\ \hline \mathcal{X}_{ij}\end{array}\right),
\end{equation}
represents the operator from the set $ \{I,X,Y,Z\} $ that is applied to the qubit $i$ due to stabilizer $ j $. For example, the two-qubit maximally entangled state defined by the stabilizers $ S_1=X_1X_2 $ and $ S_2=Z_1Z_2 $ is represented by
\begin{equation}
A=\left( \begin{array}{cc}
0 & 1 \\ 0 & 1 \\\hline 1 & 0 \\ 1 & 0\end{array}\right).
\end{equation}
Another important example is a graph state defined on an underlying graph represented by the $ N\times N $ adjacency matrix $ \Gamma $. The stabilizer description of the graph state is given in Appendix~\ref{app:graphs}. Given that the neighborhood $ \mathcal{N}_j  $ of qubit $j$ includes the nodes corresponding to the non-vanishing elements of the $ j $-th column of $ \Gamma $, the representation of a graph state is completely defined by the adjacency matrix $ \Gamma $ as
\begin{equation}
		A=\left( \begin{array}{c}
	\Gamma\\ \hline I\end{array}\right)	,
\end{equation}
where $ I $ is the $ N\times N $ identity matrix. In a stabilizer state, all stabilizers are independent, implying that no product of them exists which equals to the identity operator. Since in the binary picture the product of Pauli operators is mapped to a sum of columns, all the columns of $ A $ must be linearly independent, or equivalently, $ \text{rank}(A)=N $. Moreover, an invertible recombination of the stabilizers operators leads to the same stabilizer operator. In the binary picture this recombination is represented by an invertible $ N\times N $ binary matrix $ R $ that multiplies $ A $ from the right. Finally, the commutation of the stabilizer operators defining a stabilizer state is guaranteed by the relation $ A^{\text{T}}DA=0 $, where $ D $ is the $ 2N\times 2N $ matrix:
\begin{equation}
	D=\left(\begin{array}{c|c}
	0 & I \\ \hline I & 0
	\end{array}\right)	.
\end{equation}

\noindent\textbf{Clifford operation on stabilizer state.} Clifford operations performed on multiple qubits of the stabilizer state are represented by $ 2N\times 2N $ binary matrices $ Q $ that multiply $ A $ from the left. Clifford operations can not map a stabilizer operator or a recombination of them into the identity operator, implying that multiplication of  $ Q $ to $ AR $ for all $ R $ can not result in vanishing a column. Consequently, $ Q $ is a full-rank matrix. For local operations like $ H $ or $ U_Z $ in Eq.~(\ref{single-qubit_unitaries}), $ Q $ is the tensor product of the single-qubit matrices. For example, 
\begin{equation}
	U_Z = \left(\begin{array}{c|c}
	I & \Lambda \\ \hline 0 & I
	\end{array}\right)
\end{equation}
represents a $ \pi/2 $ $ z $-rotation applied on every qubits $ i $ for which $ \Lambda_{ii}=1 $, $ \Lambda $ being a $ N\times N $ diagonal matrix.

\section{Uncorrelated single-qubit Pauli noise}
\label{app:noise}

We consider an operator-sum representation~\cite{holevo2012,nielsen2010} of the uncorrelated single-qubit Pauli noise applied to the stabilizer state $\ket{\psi_S}$, given by $\ket{\psi_S}\rightarrow\rho_S=\Lambda(\ket{\psi_S}\bra{\psi_S})$, where 
\begin{eqnarray}
\rho_S =  \sum_{\alpha=0}^{4^N-1}q_\alpha J_\alpha\ket{\psi_S}\bra{\psi_S} J_\alpha.
\label{eq:evolve}
\end{eqnarray}
Here, $\{\sqrt{q_\alpha}J_\alpha\}$ are the Kraus operators satisfying  the completeness condition $\sum_{\alpha}J_\alpha^\dagger J_\alpha=I$, and $\{q_\alpha\}$ is a probability distribution, representing the strength of the noise.  The individual operators, $J_{\alpha}$ can be written as 
\begin{eqnarray}
J_\alpha=\bigotimes_{i=1}^N\sigma_{\alpha_i},
\label{eq:kraus_pauli}
\end{eqnarray}
whereas
\begin{eqnarray}
q_\alpha=\prod_{i=1}^Nq_{\alpha_i},
\label{eq:kraus_prob}
\end{eqnarray}
with $\alpha_i\in\{0,1,2,3\}$, $\sum_{\alpha_i=0}^3q_{\alpha_i}=1$, and $\sigma_{0}=I_i$, $\sigma_{1}=X_i$, $\sigma_{2}=Y_i$, and $\sigma_{3}=Z_i$.   Here we interpret the  index $\alpha$ as the multi-index $\alpha\equiv\alpha_1\alpha_2\cdots\alpha_N$, where the value of $\alpha$ is the decimal representation of the base $4$ string $\alpha_1\alpha_2\cdots\alpha_N$. Examples of uncorrelated Pauli noise include phase-flip ($q_{0}=1-\frac{q}{2}$, $q_{3}=\frac{q}{2}$, $q_{1}=q_{2}=0$) , bit-flip ($q_{0}=1-\frac{q}{2}$, $q_{1}=\frac{q}{2}$, $q_{2}=q_{3}=0$), bit-phase-flip ($q_{0}=1-\frac{q}{2}$, $q_{2}=\frac{q}{2}$, $q_{1}=q_{3}=0$), and depolarizing ($q_{0}=1-\frac{3q}{4}$, $q_{1}=q_2=q_3=\frac{q}{4}$) noises.

\section{Stabilizer expectation values under noise}
\label{app:stab_expt}

For the purpose of demonstration, we choose the phase-flip noise, where due to the specific values of the probabilities $\{q_{\alpha_i}\}$, only $2^N$ terms in the sum in Eq.~(\ref{eq:evolve}) survive. We rewrite Eq.~(\ref{eq:evolve}) for the PF noise as 
\begin{eqnarray}
\rho_S &=&  \sum_{\alpha=0}^{2^N-1}q_\alpha J_\alpha\rho_L J_\alpha,
\label{eq:evolve_pf}
\end{eqnarray}
where the index $\alpha$ has similar definition as before, only with $\alpha_i\in\{0,1\}$, $\sum_{\alpha_i=0}^1q_{\alpha_i}=1$, and $\sigma_{0}=I_i$, $\sigma_{1}=Z_i$. From the form of the Kraus operators and the fact that $S^z\ket{\psi_S}=\ket{\psi_S}$, it is easy to see that $\omega_z=1$. However, in the case of $S^x$, 
\begin{eqnarray}
\omega_x &=& \sum_{\alpha=0}^{2^N-1}q_\alpha \text{Tr}\left[S^x J_\alpha\ket{\psi_S}\bra{\psi_S} J_\alpha\right]. 
\label{eq:wlb_function}
\end{eqnarray} 
Writing $J_{\alpha}=J_{\alpha_{R_x}}\otimes J_{\alpha_{\overline{R}_x}}$ and $q_\alpha=q_{\alpha_{R_x}} q_{\alpha_{\overline{R}_x}}$, where $R_x$ is the set of $n_x$ qubits constructing the support of $S^x$ such that $R_x\cup\overline{R}_x$ constructs the entire system, and $R_x\cap\overline{R}_x=\emptyset$, with $\alpha_{R_x}$ ($\alpha_{\overline{R}_x}$) defined similarly as $\alpha$ for qubits in $R_x$ ($\overline{R}_x$), Eq.~(\ref{eq:wlb_function}) becomes 
\begin{eqnarray}
\omega_x &=& \sum_{\alpha=0}^{2^{n_x}-1}q_{\alpha_{R_x}} \text{Tr}\left[S^x J_{\alpha_{R_x}}\ket{\psi_S}\bra{\psi_S}J_{\alpha_{R_x}}\right],
\label{eq:wlb_function_2}
\end{eqnarray}  
since the probabilities $q_{\alpha_{\overline{R}_x}}$ sum up to 1. The form of $S^x$  suggests that the value of the  trace in the parenthesis in Eq.~(\ref{eq:wlb_function_2}) is $+1 (-1)$ if the number of qubits in $R_x$ for which $\alpha_i=1$ is even (odd). Using the explicit forms of $q_{\alpha_i}$ in terms of $q$ for the PF noise (see Appendix~\ref{app:noise}), Eq.~(\ref{eq:wlb_function_2}) yields 
\begin{eqnarray}
\omega_x &=& \sum_{m=0}^n (-1)^m{n\choose m}\left(\frac{q}{2}\right)^{m}\left(1-\frac{q}{2}\right)^{n-m},
\end{eqnarray}
which, upon  using the identities
\begin{eqnarray}
\sum_{m\text{ even}} {n\choose m}\left(\frac{q}{2}\right)^{m}\left(1-\frac{q}{2}\right)^{n-m}&=&\frac{1}{2}\left[1+(1-q)^{n_x}\right],\nonumber \\
\sum_{m\text{ odd}} {n\choose m}\left(\frac{q}{2}\right)^{m}\left(1-\frac{q}{2}\right)^{n-m}&=&\frac{1}{2}\left[1-(1-q)^{n_x}\right],\nonumber \\
\end{eqnarray} 
leads to 
\begin{eqnarray}
\omega_x=(1-q)^{n_x}.
\end{eqnarray}
Lastly, using the commutation of $S^x$ and $S^z$, it is easy to show that $\omega_{xz}=\omega_x$. 

We point out here that the calculation for the bit-flip and the depolarizing channels would be exactly similar to the same for the phase-flip channel. However, in the case of the depolarizing channel, all $4^N$ terms in the operator-sum representation have to be considered.

\bibliography{le_cc_v2_lib_2}{}

\begin{thebibliography}{120}%
\makeatletter
\providecommand \@ifxundefined [1]{%
 \@ifx{#1\undefined}
}%
\providecommand \@ifnum [1]{%
 \ifnum #1\expandafter \@firstoftwo
 \else \expandafter \@secondoftwo
 \fi
}%
\providecommand \@ifx [1]{%
 \ifx #1\expandafter \@firstoftwo
 \else \expandafter \@secondoftwo
 \fi
}%
\providecommand \natexlab [1]{#1}%
\providecommand \enquote  [1]{``#1''}%
\providecommand \bibnamefont  [1]{#1}%
\providecommand \bibfnamefont [1]{#1}%
\providecommand \citenamefont [1]{#1}%
\providecommand \href@noop [0]{\@secondoftwo}%
\providecommand \href [0]{\begingroup \@sanitize@url \@href}%
\providecommand \@href[1]{\@@startlink{#1}\@@href}%
\providecommand \@@href[1]{\endgroup#1\@@endlink}%
\providecommand \@sanitize@url [0]{\catcode `\\12\catcode `\$12\catcode
  `\&12\catcode `\#12\catcode `\^12\catcode `\_12\catcode `\%12\relax}%
\providecommand \@@startlink[1]{}%
\providecommand \@@endlink[0]{}%
\providecommand \url  [0]{\begingroup\@sanitize@url \@url }%
\providecommand \@url [1]{\endgroup\@href {#1}{\urlprefix }}%
\providecommand \urlprefix  [0]{URL }%
\providecommand \Eprint [0]{\href }%
\providecommand \doibase [0]{http://dx.doi.org/}%
\providecommand \selectlanguage [0]{\@gobble}%
\providecommand \bibinfo  [0]{\@secondoftwo}%
\providecommand \bibfield  [0]{\@secondoftwo}%
\providecommand \translation [1]{[#1]}%
\providecommand \BibitemOpen [0]{}%
\providecommand \bibitemStop [0]{}%
\providecommand \bibitemNoStop [0]{.\EOS\space}%
\providecommand \EOS [0]{\spacefactor3000\relax}%
\providecommand \BibitemShut  [1]{\csname bibitem#1\endcsname}%
\let\auto@bib@innerbib\@empty
%</preamble>
\bibitem [{\citenamefont {Bennett}\ and\ \citenamefont
  {DiVincenzo}(2000)}]{bennett2000}%
  \BibitemOpen
  \bibfield  {author} {\bibinfo {author} {\bibfnamefont {C.~H.}\ \bibnamefont
  {Bennett}}\ and\ \bibinfo {author} {\bibfnamefont {D.~P.}\ \bibnamefont
  {DiVincenzo}},\ }\href {\doibase 10.1038/35005001} {\bibfield  {journal}
  {\bibinfo  {journal} {Nature}\ }\textbf {\bibinfo {volume} {404}},\ \bibinfo
  {pages} {247} (\bibinfo {year} {2000})}\BibitemShut {NoStop}%
\bibitem [{\citenamefont {Horodecki}\ \emph {et~al.}(2009)\citenamefont
  {Horodecki}, \citenamefont {Horodecki}, \citenamefont {Horodecki},\ and\
  \citenamefont {Horodecki}}]{horodecki2009}%
  \BibitemOpen
  \bibfield  {author} {\bibinfo {author} {\bibfnamefont {R.}~\bibnamefont
  {Horodecki}}, \bibinfo {author} {\bibfnamefont {P.}~\bibnamefont
  {Horodecki}}, \bibinfo {author} {\bibfnamefont {M.}~\bibnamefont
  {Horodecki}}, \ and\ \bibinfo {author} {\bibfnamefont {K.}~\bibnamefont
  {Horodecki}},\ }\href {\doibase 10.1103/RevModPhys.81.865} {\bibfield
  {journal} {\bibinfo  {journal} {Rev. Mod. Phys.}\ }\textbf {\bibinfo {volume}
  {81}},\ \bibinfo {pages} {865} (\bibinfo {year} {2009})}\BibitemShut
  {NoStop}%
\bibitem [{\citenamefont {Bennett}\ \emph {et~al.}(1993)\citenamefont
  {Bennett}, \citenamefont {Brassard}, \citenamefont {Cr\'epeau}, \citenamefont
  {Jozsa}, \citenamefont {Peres},\ and\ \citenamefont
  {Wootters}}]{bennett1993}%
  \BibitemOpen
  \bibfield  {author} {\bibinfo {author} {\bibfnamefont {C.~H.}\ \bibnamefont
  {Bennett}}, \bibinfo {author} {\bibfnamefont {G.}~\bibnamefont {Brassard}},
  \bibinfo {author} {\bibfnamefont {C.}~\bibnamefont {Cr\'epeau}}, \bibinfo
  {author} {\bibfnamefont {R.}~\bibnamefont {Jozsa}}, \bibinfo {author}
  {\bibfnamefont {A.}~\bibnamefont {Peres}}, \ and\ \bibinfo {author}
  {\bibfnamefont {W.~K.}\ \bibnamefont {Wootters}},\ }\href {\doibase
  10.1103/PhysRevLett.70.1895} {\bibfield  {journal} {\bibinfo  {journal}
  {Phys. Rev. Lett.}\ }\textbf {\bibinfo {volume} {70}},\ \bibinfo {pages}
  {1895} (\bibinfo {year} {1993})}\BibitemShut {NoStop}%
\bibitem [{\citenamefont {Bouwmeester}\ \emph {et~al.}(1997)\citenamefont
  {Bouwmeester}, \citenamefont {Pan}, \citenamefont {Mattle}, \citenamefont
  {Eibl}, \citenamefont {Weinfurter},\ and\ \citenamefont
  {Zeilinger}}]{bouwmeester1997}%
  \BibitemOpen
  \bibfield  {author} {\bibinfo {author} {\bibfnamefont {D.}~\bibnamefont
  {Bouwmeester}}, \bibinfo {author} {\bibfnamefont {J.-W.}\ \bibnamefont
  {Pan}}, \bibinfo {author} {\bibfnamefont {K.}~\bibnamefont {Mattle}},
  \bibinfo {author} {\bibfnamefont {M.}~\bibnamefont {Eibl}}, \bibinfo {author}
  {\bibfnamefont {H.}~\bibnamefont {Weinfurter}}, \ and\ \bibinfo {author}
  {\bibfnamefont {A.}~\bibnamefont {Zeilinger}},\ }\href {\doibase
  10.1038/37539} {\bibfield  {journal} {\bibinfo  {journal} {Nature}\ }\textbf
  {\bibinfo {volume} {390}},\ \bibinfo {pages} {575} (\bibinfo {year}
  {1997})}\BibitemShut {NoStop}%
\bibitem [{\citenamefont {Bennett}\ and\ \citenamefont
  {Wiesner}(1992)}]{bennett1992}%
  \BibitemOpen
  \bibfield  {author} {\bibinfo {author} {\bibfnamefont {C.~H.}\ \bibnamefont
  {Bennett}}\ and\ \bibinfo {author} {\bibfnamefont {S.~J.}\ \bibnamefont
  {Wiesner}},\ }\href {\doibase 10.1103/PhysRevLett.69.2881} {\bibfield
  {journal} {\bibinfo  {journal} {Phys. Rev. Lett.}\ }\textbf {\bibinfo
  {volume} {69}},\ \bibinfo {pages} {2881} (\bibinfo {year}
  {1992})}\BibitemShut {NoStop}%
\bibitem [{\citenamefont {Mattle}\ \emph {et~al.}(1996)\citenamefont {Mattle},
  \citenamefont {Weinfurter}, \citenamefont {Kwiat},\ and\ \citenamefont
  {Zeilinger}}]{mattle1996}%
  \BibitemOpen
  \bibfield  {author} {\bibinfo {author} {\bibfnamefont {K.}~\bibnamefont
  {Mattle}}, \bibinfo {author} {\bibfnamefont {H.}~\bibnamefont {Weinfurter}},
  \bibinfo {author} {\bibfnamefont {P.~G.}\ \bibnamefont {Kwiat}}, \ and\
  \bibinfo {author} {\bibfnamefont {A.}~\bibnamefont {Zeilinger}},\ }\href
  {\doibase 10.1103/PhysRevLett.76.4656} {\bibfield  {journal} {\bibinfo
  {journal} {Phys. Rev. Lett.}\ }\textbf {\bibinfo {volume} {76}},\ \bibinfo
  {pages} {4656} (\bibinfo {year} {1996})}\BibitemShut {NoStop}%
\bibitem [{\citenamefont {Sen(De)}\ and\ \citenamefont
  {Sen}(2011)}]{sende2010}%
  \BibitemOpen
  \bibfield  {author} {\bibinfo {author} {\bibfnamefont {A.}~\bibnamefont
  {Sen(De)}}\ and\ \bibinfo {author} {\bibfnamefont {U.}~\bibnamefont {Sen}},\
  }\href {https://arxiv.org/abs/1105.2412} {\bibfield  {journal} {\bibinfo
  {journal} {Phys. News}\ }\textbf {\bibinfo {volume} {40}},\ \bibinfo {pages}
  {17} (\bibinfo {year} {2011})},\ \Eprint
  {http://arxiv.org/abs/arXiv:1105.2412} {arXiv:1105.2412} \BibitemShut
  {NoStop}%
\bibitem [{\citenamefont {Ekert}(1991)}]{ekert1991}%
  \BibitemOpen
  \bibfield  {author} {\bibinfo {author} {\bibfnamefont {A.~K.}\ \bibnamefont
  {Ekert}},\ }\href {\doibase 10.1103/PhysRevLett.67.661} {\bibfield  {journal}
  {\bibinfo  {journal} {Phys. Rev. Lett.}\ }\textbf {\bibinfo {volume} {67}},\
  \bibinfo {pages} {661} (\bibinfo {year} {1991})}\BibitemShut {NoStop}%
\bibitem [{\citenamefont {Jennewein}\ \emph {et~al.}(2000)\citenamefont
  {Jennewein}, \citenamefont {Simon}, \citenamefont {Weihs}, \citenamefont
  {Weinfurter},\ and\ \citenamefont {Zeilinger}}]{jennewein2000}%
  \BibitemOpen
  \bibfield  {author} {\bibinfo {author} {\bibfnamefont {T.}~\bibnamefont
  {Jennewein}}, \bibinfo {author} {\bibfnamefont {C.}~\bibnamefont {Simon}},
  \bibinfo {author} {\bibfnamefont {G.}~\bibnamefont {Weihs}}, \bibinfo
  {author} {\bibfnamefont {H.}~\bibnamefont {Weinfurter}}, \ and\ \bibinfo
  {author} {\bibfnamefont {A.}~\bibnamefont {Zeilinger}},\ }\href {\doibase
  10.1103/PhysRevLett.84.4729} {\bibfield  {journal} {\bibinfo  {journal}
  {Phys. Rev. Lett.}\ }\textbf {\bibinfo {volume} {84}},\ \bibinfo {pages}
  {4729} (\bibinfo {year} {2000})}\BibitemShut {NoStop}%
\bibitem [{\citenamefont {Raussendorf}\ and\ \citenamefont
  {Briegel}(2001)}]{raussendorf2001}%
  \BibitemOpen
  \bibfield  {author} {\bibinfo {author} {\bibfnamefont {R.}~\bibnamefont
  {Raussendorf}}\ and\ \bibinfo {author} {\bibfnamefont {H.~J.}\ \bibnamefont
  {Briegel}},\ }\href {\doibase 10.1103/PhysRevLett.86.5188} {\bibfield
  {journal} {\bibinfo  {journal} {Phys. Rev. Lett.}\ }\textbf {\bibinfo
  {volume} {86}},\ \bibinfo {pages} {5188} (\bibinfo {year}
  {2001})}\BibitemShut {NoStop}%
\bibitem [{\citenamefont {Raussendorf}\ \emph {et~al.}(2003)\citenamefont
  {Raussendorf}, \citenamefont {Browne},\ and\ \citenamefont
  {Briegel}}]{raussendorf2003}%
  \BibitemOpen
  \bibfield  {author} {\bibinfo {author} {\bibfnamefont {R.}~\bibnamefont
  {Raussendorf}}, \bibinfo {author} {\bibfnamefont {D.~E.}\ \bibnamefont
  {Browne}}, \ and\ \bibinfo {author} {\bibfnamefont {H.~J.}\ \bibnamefont
  {Briegel}},\ }\href {\doibase 10.1103/PhysRevA.68.022312} {\bibfield
  {journal} {\bibinfo  {journal} {Phys. Rev. A}\ }\textbf {\bibinfo {volume}
  {68}},\ \bibinfo {pages} {022312} (\bibinfo {year} {2003})}\BibitemShut
  {NoStop}%
\bibitem [{\citenamefont {Briegel}\ \emph {et~al.}(2009)\citenamefont
  {Briegel}, \citenamefont {Browne}, \citenamefont {D{\"u}r}, \citenamefont
  {Raussendorf},\ and\ \citenamefont {Van~den Nest}}]{briegel2009}%
  \BibitemOpen
  \bibfield  {author} {\bibinfo {author} {\bibfnamefont {H.~J.}\ \bibnamefont
  {Briegel}}, \bibinfo {author} {\bibfnamefont {D.~E.}\ \bibnamefont {Browne}},
  \bibinfo {author} {\bibfnamefont {W.}~\bibnamefont {D{\"u}r}}, \bibinfo
  {author} {\bibfnamefont {R.}~\bibnamefont {Raussendorf}}, \ and\ \bibinfo
  {author} {\bibfnamefont {M.}~\bibnamefont {Van~den Nest}},\ }\href
  {https://doi.org/10.1038/nphys1157} {\bibfield  {journal} {\bibinfo
  {journal} {Nat. Phys.}\ }\textbf {\bibinfo {volume} {5}},\ \bibinfo {pages}
  {19} (\bibinfo {year} {2009})}\BibitemShut {NoStop}%
\bibitem [{\citenamefont {Pollmann}\ \emph {et~al.}(2010)\citenamefont
  {Pollmann}, \citenamefont {Turner}, \citenamefont {Berg},\ and\ \citenamefont
  {Oshikawa}}]{pollmann2010}%
  \BibitemOpen
  \bibfield  {author} {\bibinfo {author} {\bibfnamefont {F.}~\bibnamefont
  {Pollmann}}, \bibinfo {author} {\bibfnamefont {A.~M.}\ \bibnamefont
  {Turner}}, \bibinfo {author} {\bibfnamefont {E.}~\bibnamefont {Berg}}, \ and\
  \bibinfo {author} {\bibfnamefont {M.}~\bibnamefont {Oshikawa}},\ }\href
  {\doibase 10.1103/PhysRevB.81.064439} {\bibfield  {journal} {\bibinfo
  {journal} {Phys. Rev. B}\ }\textbf {\bibinfo {volume} {81}},\ \bibinfo
  {pages} {064439} (\bibinfo {year} {2010})}\BibitemShut {NoStop}%
\bibitem [{\citenamefont {Chen}\ \emph {et~al.}(2010)\citenamefont {Chen},
  \citenamefont {Gu},\ and\ \citenamefont {Wen}}]{chen2010}%
  \BibitemOpen
  \bibfield  {author} {\bibinfo {author} {\bibfnamefont {X.}~\bibnamefont
  {Chen}}, \bibinfo {author} {\bibfnamefont {Z.-C.}\ \bibnamefont {Gu}}, \ and\
  \bibinfo {author} {\bibfnamefont {X.-G.}\ \bibnamefont {Wen}},\ }\href
  {\doibase 10.1103/PhysRevB.82.155138} {\bibfield  {journal} {\bibinfo
  {journal} {Phys. Rev. B}\ }\textbf {\bibinfo {volume} {82}},\ \bibinfo
  {pages} {155138} (\bibinfo {year} {2010})}\BibitemShut {NoStop}%
\bibitem [{\citenamefont {Jiang}\ \emph {et~al.}(2012)\citenamefont {Jiang},
  \citenamefont {Wang},\ and\ \citenamefont {Balents}}]{jiang2012}%
  \BibitemOpen
  \bibfield  {author} {\bibinfo {author} {\bibfnamefont {H.-C.}\ \bibnamefont
  {Jiang}}, \bibinfo {author} {\bibfnamefont {Z.}~\bibnamefont {Wang}}, \ and\
  \bibinfo {author} {\bibfnamefont {L.}~\bibnamefont {Balents}},\ }\href
  {https://doi.org/10.1038/nphys2465} {\bibfield  {journal} {\bibinfo
  {journal} {Nat. Phys.}\ }\textbf {\bibinfo {volume} {8}},\ \bibinfo {pages}
  {902} (\bibinfo {year} {2012})}\BibitemShut {NoStop}%
\bibitem [{\citenamefont {Osterloh}\ \emph {et~al.}(2002)\citenamefont
  {Osterloh}, \citenamefont {Amico}, \citenamefont {Falci},\ and\ \citenamefont
  {Fazio}}]{osterloh2002}%
  \BibitemOpen
  \bibfield  {author} {\bibinfo {author} {\bibfnamefont {A.}~\bibnamefont
  {Osterloh}}, \bibinfo {author} {\bibfnamefont {L.}~\bibnamefont {Amico}},
  \bibinfo {author} {\bibfnamefont {G.}~\bibnamefont {Falci}}, \ and\ \bibinfo
  {author} {\bibfnamefont {R.}~\bibnamefont {Fazio}},\ }\href {\doibase
  10.1038/416608a} {\bibfield  {journal} {\bibinfo  {journal} {Nature}\
  }\textbf {\bibinfo {volume} {416}},\ \bibinfo {pages} {608} (\bibinfo {year}
  {2002})}\BibitemShut {NoStop}%
\bibitem [{\citenamefont {Osborne}\ and\ \citenamefont
  {Nielsen}(2002)}]{osborne2002}%
  \BibitemOpen
  \bibfield  {author} {\bibinfo {author} {\bibfnamefont {T.~J.}\ \bibnamefont
  {Osborne}}\ and\ \bibinfo {author} {\bibfnamefont {M.~A.}\ \bibnamefont
  {Nielsen}},\ }\href {\doibase 10.1103/PhysRevA.66.032110} {\bibfield
  {journal} {\bibinfo  {journal} {Phys. Rev. A}\ }\textbf {\bibinfo {volume}
  {66}},\ \bibinfo {pages} {032110} (\bibinfo {year} {2002})}\BibitemShut
  {NoStop}%
\bibitem [{\citenamefont {Amico}\ \emph {et~al.}(2008)\citenamefont {Amico},
  \citenamefont {Fazio}, \citenamefont {Osterloh},\ and\ \citenamefont
  {Vedral}}]{amico2008}%
  \BibitemOpen
  \bibfield  {author} {\bibinfo {author} {\bibfnamefont {L.}~\bibnamefont
  {Amico}}, \bibinfo {author} {\bibfnamefont {R.}~\bibnamefont {Fazio}},
  \bibinfo {author} {\bibfnamefont {A.}~\bibnamefont {Osterloh}}, \ and\
  \bibinfo {author} {\bibfnamefont {V.}~\bibnamefont {Vedral}},\ }\href
  {\doibase 10.1103/RevModPhys.80.517} {\bibfield  {journal} {\bibinfo
  {journal} {Rev. Mod. Phys.}\ }\textbf {\bibinfo {volume} {80}},\ \bibinfo
  {pages} {517} (\bibinfo {year} {2008})}\BibitemShut {NoStop}%
\bibitem [{\citenamefont {Chiara}\ and\ \citenamefont
  {Sanpera}(2018)}]{chiara2017}%
  \BibitemOpen
  \bibfield  {author} {\bibinfo {author} {\bibfnamefont {G.~D.}\ \bibnamefont
  {Chiara}}\ and\ \bibinfo {author} {\bibfnamefont {A.}~\bibnamefont
  {Sanpera}},\ }\href {\doibase 10.1088/1361-6633/aabf61} {\bibfield  {journal}
  {\bibinfo  {journal} {Rep. Prog. Phys.}\ }\textbf {\bibinfo {volume} {81}},\
  \bibinfo {pages} {074002} (\bibinfo {year} {2018})}\BibitemShut {NoStop}%
\bibitem [{\citenamefont {Sarovar}\ \emph {et~al.}(2010)\citenamefont
  {Sarovar}, \citenamefont {Ishizaki}, \citenamefont {Fleming},\ and\
  \citenamefont {Whaley}}]{sarovar2010}%
  \BibitemOpen
  \bibfield  {author} {\bibinfo {author} {\bibfnamefont {M.}~\bibnamefont
  {Sarovar}}, \bibinfo {author} {\bibfnamefont {A.}~\bibnamefont {Ishizaki}},
  \bibinfo {author} {\bibfnamefont {G.~R.}\ \bibnamefont {Fleming}}, \ and\
  \bibinfo {author} {\bibfnamefont {K.~B.}\ \bibnamefont {Whaley}},\ }\href
  {https://doi.org/10.1038/nphys1652} {\bibfield  {journal} {\bibinfo
  {journal} {Nat. Phys.}\ }\textbf {\bibinfo {volume} {6}},\ \bibinfo {pages}
  {462} (\bibinfo {year} {2010})}\BibitemShut {NoStop}%
\bibitem [{\citenamefont {Zhu}\ \emph {et~al.}(2012)\citenamefont {Zhu},
  \citenamefont {Kais}, \citenamefont {Aspuru-Guzik}, \citenamefont
  {Rodriques}, \citenamefont {Brock},\ and\ \citenamefont {Love}}]{zhu2012}%
  \BibitemOpen
  \bibfield  {author} {\bibinfo {author} {\bibfnamefont {J.}~\bibnamefont
  {Zhu}}, \bibinfo {author} {\bibfnamefont {S.}~\bibnamefont {Kais}}, \bibinfo
  {author} {\bibfnamefont {A.}~\bibnamefont {Aspuru-Guzik}}, \bibinfo {author}
  {\bibfnamefont {S.}~\bibnamefont {Rodriques}}, \bibinfo {author}
  {\bibfnamefont {B.}~\bibnamefont {Brock}}, \ and\ \bibinfo {author}
  {\bibfnamefont {P.~J.}\ \bibnamefont {Love}},\ }\href {\doibase
  10.1063/1.4742333} {\bibfield  {journal} {\bibinfo  {journal} {J. Chem.
  Phys.}\ }\textbf {\bibinfo {volume} {137}},\ \bibinfo {pages} {074112}
  (\bibinfo {year} {2012})}\BibitemShut {NoStop}%
\bibitem [{\citenamefont {Lambert}\ \emph {et~al.}(2013)\citenamefont
  {Lambert}, \citenamefont {Chen}, \citenamefont {Cheng}, \citenamefont {Li},
  \citenamefont {Chen},\ and\ \citenamefont {Nori}}]{lambert2013}%
  \BibitemOpen
  \bibfield  {author} {\bibinfo {author} {\bibfnamefont {N.}~\bibnamefont
  {Lambert}}, \bibinfo {author} {\bibfnamefont {Y.-N.}\ \bibnamefont {Chen}},
  \bibinfo {author} {\bibfnamefont {Y.}~\bibnamefont {Cheng}}, \bibinfo
  {author} {\bibfnamefont {C.-M.}\ \bibnamefont {Li}}, \bibinfo {author}
  {\bibfnamefont {G.}~\bibnamefont {Chen}}, \ and\ \bibinfo {author}
  {\bibfnamefont {F.}~\bibnamefont {Nori}},\ }\href {\doibase
  10.1038/nphys2474} {\bibfield  {journal} {\bibinfo  {journal} {Nat. Phys.}\
  }\textbf {\bibinfo {volume} {9}},\ \bibinfo {pages} {10} (\bibinfo {year}
  {2013})}\BibitemShut {NoStop}%
\bibitem [{\citenamefont {Chanda}\ \emph {et~al.}(2014)\citenamefont {Chanda},
  \citenamefont {Mishra}, \citenamefont {Sen(De)},\ and\ \citenamefont
  {Sen}}]{chanda2014}%
  \BibitemOpen
  \bibfield  {author} {\bibinfo {author} {\bibfnamefont {T.}~\bibnamefont
  {Chanda}}, \bibinfo {author} {\bibfnamefont {U.}~\bibnamefont {Mishra}},
  \bibinfo {author} {\bibfnamefont {A.}~\bibnamefont {Sen(De)}}, \ and\
  \bibinfo {author} {\bibfnamefont {U.}~\bibnamefont {Sen}},\ }\href
  {https://arxiv.org/abs/1412.6519} {\bibfield  {journal} {\bibinfo  {journal}
  {arXiv:1412.6519}\ } (\bibinfo {year} {2014})}\BibitemShut {NoStop}%
\bibitem [{\citenamefont {Cai}\ \emph {et~al.}(2010)\citenamefont {Cai},
  \citenamefont {Guerreschi},\ and\ \citenamefont {Briegel}}]{cai2010}%
  \BibitemOpen
  \bibfield  {author} {\bibinfo {author} {\bibfnamefont {J.}~\bibnamefont
  {Cai}}, \bibinfo {author} {\bibfnamefont {G.~G.}\ \bibnamefont {Guerreschi}},
  \ and\ \bibinfo {author} {\bibfnamefont {H.~J.}\ \bibnamefont {Briegel}},\
  }\href {\doibase 10.1103/PhysRevLett.104.220502} {\bibfield  {journal}
  {\bibinfo  {journal} {Phys. Rev. Lett.}\ }\textbf {\bibinfo {volume} {104}},\
  \bibinfo {pages} {220502} (\bibinfo {year} {2010})}\BibitemShut {NoStop}%
\bibitem [{\citenamefont {Hubeny}(2015)}]{hubeny2015}%
  \BibitemOpen
  \bibfield  {author} {\bibinfo {author} {\bibfnamefont {V.~E.}\ \bibnamefont
  {Hubeny}},\ }\href {\doibase 10.1088/0264-9381/32/12/124010} {\bibfield
  {journal} {\bibinfo  {journal} {Classical Quant. Grav.}\ }\textbf {\bibinfo
  {volume} {32}},\ \bibinfo {pages} {124010} (\bibinfo {year}
  {2015})}\BibitemShut {NoStop}%
\bibitem [{\citenamefont {Pastawski}\ \emph {et~al.}(2015)\citenamefont
  {Pastawski}, \citenamefont {Yoshida}, \citenamefont {Harlow},\ and\
  \citenamefont {Preskill}}]{pastawski2015}%
  \BibitemOpen
  \bibfield  {author} {\bibinfo {author} {\bibfnamefont {F.}~\bibnamefont
  {Pastawski}}, \bibinfo {author} {\bibfnamefont {B.}~\bibnamefont {Yoshida}},
  \bibinfo {author} {\bibfnamefont {D.}~\bibnamefont {Harlow}}, \ and\ \bibinfo
  {author} {\bibfnamefont {J.}~\bibnamefont {Preskill}},\ }\href {\doibase
  10.1007/JHEP06(2015)149} {\bibfield  {journal} {\bibinfo  {journal} {J. High.
  Energy Phys.}\ }\textbf {\bibinfo {volume} {2015}},\ \bibinfo {pages} {149}
  (\bibinfo {year} {2015})}\BibitemShut {NoStop}%
\bibitem [{\citenamefont {Almheiri}\ \emph {et~al.}(2015)\citenamefont
  {Almheiri}, \citenamefont {Dong},\ and\ \citenamefont
  {Harlow}}]{almheiri2015}%
  \BibitemOpen
  \bibfield  {author} {\bibinfo {author} {\bibfnamefont {A.}~\bibnamefont
  {Almheiri}}, \bibinfo {author} {\bibfnamefont {X.}~\bibnamefont {Dong}}, \
  and\ \bibinfo {author} {\bibfnamefont {D.}~\bibnamefont {Harlow}},\ }\href
  {\doibase 10.1007/JHEP04(2015)163} {\bibfield  {journal} {\bibinfo  {journal}
  {J. High Energy Phys.}\ }\textbf {\bibinfo {volume} {4}},\ \bibinfo {pages}
  {163} (\bibinfo {year} {2015})}\BibitemShut {NoStop}%
\bibitem [{\citenamefont {{Jahn}}\ \emph {et~al.}(2017)\citenamefont {{Jahn}},
  \citenamefont {{Gluza}}, \citenamefont {{Pastawski}},\ and\ \citenamefont
  {{Eisert}}}]{jahn2017}%
  \BibitemOpen
  \bibfield  {author} {\bibinfo {author} {\bibfnamefont {A.}~\bibnamefont
  {{Jahn}}}, \bibinfo {author} {\bibfnamefont {M.}~\bibnamefont {{Gluza}}},
  \bibinfo {author} {\bibfnamefont {F.}~\bibnamefont {{Pastawski}}}, \ and\
  \bibinfo {author} {\bibfnamefont {J.}~\bibnamefont {{Eisert}}},\ }\href
  {https://arxiv.org/abs/1711.03109} {\bibfield  {journal} {\bibinfo  {journal}
  {arXiv:1711.03109}\ } (\bibinfo {year} {2017})}\BibitemShut {NoStop}%
\bibitem [{\citenamefont {Leibfried}\ \emph {et~al.}(2003)\citenamefont
  {Leibfried}, \citenamefont {Blatt}, \citenamefont {Monroe},\ and\
  \citenamefont {Wineland}}]{leibfried2003}%
  \BibitemOpen
  \bibfield  {author} {\bibinfo {author} {\bibfnamefont {D.}~\bibnamefont
  {Leibfried}}, \bibinfo {author} {\bibfnamefont {R.}~\bibnamefont {Blatt}},
  \bibinfo {author} {\bibfnamefont {C.}~\bibnamefont {Monroe}}, \ and\ \bibinfo
  {author} {\bibfnamefont {D.}~\bibnamefont {Wineland}},\ }\href {\doibase
  10.1103/RevModPhys.75.281} {\bibfield  {journal} {\bibinfo  {journal} {Rev.
  Mod. Phys.}\ }\textbf {\bibinfo {volume} {75}},\ \bibinfo {pages} {281}
  (\bibinfo {year} {2003})}\BibitemShut {NoStop}%
\bibitem [{\citenamefont {Leibfried}\ \emph {et~al.}(2005)\citenamefont
  {Leibfried}, \citenamefont {Knill}, \citenamefont {Seidelin}, \citenamefont
  {Britton}, \citenamefont {Blakestad}, \citenamefont {Chiaverini},
  \citenamefont {Hume}, \citenamefont {Itano}, \citenamefont {Jost},
  \citenamefont {Langer}, \citenamefont {Ozeri}, \citenamefont {Reichle},\ and\
  \citenamefont {Wineland}}]{leibfried2005}%
  \BibitemOpen
  \bibfield  {author} {\bibinfo {author} {\bibfnamefont {D.}~\bibnamefont
  {Leibfried}}, \bibinfo {author} {\bibfnamefont {E.}~\bibnamefont {Knill}},
  \bibinfo {author} {\bibfnamefont {S.}~\bibnamefont {Seidelin}}, \bibinfo
  {author} {\bibfnamefont {J.}~\bibnamefont {Britton}}, \bibinfo {author}
  {\bibfnamefont {R.~B.}\ \bibnamefont {Blakestad}}, \bibinfo {author}
  {\bibfnamefont {J.}~\bibnamefont {Chiaverini}}, \bibinfo {author}
  {\bibfnamefont {D.~B.}\ \bibnamefont {Hume}}, \bibinfo {author}
  {\bibfnamefont {W.~M.}\ \bibnamefont {Itano}}, \bibinfo {author}
  {\bibfnamefont {J.~D.}\ \bibnamefont {Jost}}, \bibinfo {author}
  {\bibfnamefont {C.}~\bibnamefont {Langer}}, \bibinfo {author} {\bibfnamefont
  {R.}~\bibnamefont {Ozeri}}, \bibinfo {author} {\bibfnamefont
  {R.}~\bibnamefont {Reichle}}, \ and\ \bibinfo {author} {\bibfnamefont
  {D.~J.}\ \bibnamefont {Wineland}},\ }\href {\doibase 10.1038/nature04251}
  {\bibfield  {journal} {\bibinfo  {journal} {Nature}\ }\textbf {\bibinfo
  {volume} {438}},\ \bibinfo {pages} {639} (\bibinfo {year}
  {2005})}\BibitemShut {NoStop}%
\bibitem [{\citenamefont {Brown}\ \emph {et~al.}(2016)\citenamefont {Brown},
  \citenamefont {Kim},\ and\ \citenamefont {Monroe}}]{brown2016}%
  \BibitemOpen
  \bibfield  {author} {\bibinfo {author} {\bibfnamefont {K.~R.}\ \bibnamefont
  {Brown}}, \bibinfo {author} {\bibfnamefont {J.}~\bibnamefont {Kim}}, \ and\
  \bibinfo {author} {\bibfnamefont {C.}~\bibnamefont {Monroe}},\ }\href
  {https://doi.org/10.1038/npjqi.2016.34} {\bibfield  {journal} {\bibinfo
  {journal} {Nature Phys. J. Quant. Inf.}\ }\textbf {\bibinfo {volume} {2}},\
  \bibinfo {pages} {16034 EP } (\bibinfo {year} {2016})}\BibitemShut {NoStop}%
\bibitem [{\citenamefont {Raimond}\ \emph {et~al.}(2001)\citenamefont
  {Raimond}, \citenamefont {Brune},\ and\ \citenamefont
  {Haroche}}]{raimond2001}%
  \BibitemOpen
  \bibfield  {author} {\bibinfo {author} {\bibfnamefont {J.~M.}\ \bibnamefont
  {Raimond}}, \bibinfo {author} {\bibfnamefont {M.}~\bibnamefont {Brune}}, \
  and\ \bibinfo {author} {\bibfnamefont {S.}~\bibnamefont {Haroche}},\ }\href
  {\doibase 10.1103/RevModPhys.73.565} {\bibfield  {journal} {\bibinfo
  {journal} {Rev. Mod. Phys.}\ }\textbf {\bibinfo {volume} {73}},\ \bibinfo
  {pages} {565} (\bibinfo {year} {2001})}\BibitemShut {NoStop}%
\bibitem [{\citenamefont {Prevedel}\ \emph {et~al.}(2009)\citenamefont
  {Prevedel}, \citenamefont {Cronenberg}, \citenamefont {Tame}, \citenamefont
  {Paternostro}, \citenamefont {Walther}, \citenamefont {Kim},\ and\
  \citenamefont {Zeilinger}}]{prevedel2009}%
  \BibitemOpen
  \bibfield  {author} {\bibinfo {author} {\bibfnamefont {R.}~\bibnamefont
  {Prevedel}}, \bibinfo {author} {\bibfnamefont {G.}~\bibnamefont
  {Cronenberg}}, \bibinfo {author} {\bibfnamefont {M.~S.}\ \bibnamefont
  {Tame}}, \bibinfo {author} {\bibfnamefont {M.}~\bibnamefont {Paternostro}},
  \bibinfo {author} {\bibfnamefont {P.}~\bibnamefont {Walther}}, \bibinfo
  {author} {\bibfnamefont {M.~S.}\ \bibnamefont {Kim}}, \ and\ \bibinfo
  {author} {\bibfnamefont {A.}~\bibnamefont {Zeilinger}},\ }\href {\doibase
  10.1103/PhysRevLett.103.020503} {\bibfield  {journal} {\bibinfo  {journal}
  {Phys. Rev. Lett.}\ }\textbf {\bibinfo {volume} {103}},\ \bibinfo {pages}
  {020503} (\bibinfo {year} {2009})}\BibitemShut {NoStop}%
\bibitem [{\citenamefont {Barz}(2015)}]{barz2015}%
  \BibitemOpen
  \bibfield  {author} {\bibinfo {author} {\bibfnamefont {S.}~\bibnamefont
  {Barz}},\ }\href {http://stacks.iop.org/0953-4075/48/i=8/a=083001} {\bibfield
   {journal} {\bibinfo  {journal} {J. Phys. B}\ }\textbf {\bibinfo {volume}
  {48}},\ \bibinfo {pages} {083001} (\bibinfo {year} {2015})}\BibitemShut
  {NoStop}%
\bibitem [{\citenamefont {Negrevergne}\ \emph {et~al.}(2006)\citenamefont
  {Negrevergne}, \citenamefont {Mahesh}, \citenamefont {Ryan}, \citenamefont
  {Ditty}, \citenamefont {Cyr-Racine}, \citenamefont {Power}, \citenamefont
  {Boulant}, \citenamefont {Havel}, \citenamefont {Cory},\ and\ \citenamefont
  {Laflamme}}]{negrevergne2006}%
  \BibitemOpen
  \bibfield  {author} {\bibinfo {author} {\bibfnamefont {C.}~\bibnamefont
  {Negrevergne}}, \bibinfo {author} {\bibfnamefont {T.~S.}\ \bibnamefont
  {Mahesh}}, \bibinfo {author} {\bibfnamefont {C.~A.}\ \bibnamefont {Ryan}},
  \bibinfo {author} {\bibfnamefont {M.}~\bibnamefont {Ditty}}, \bibinfo
  {author} {\bibfnamefont {F.}~\bibnamefont {Cyr-Racine}}, \bibinfo {author}
  {\bibfnamefont {W.}~\bibnamefont {Power}}, \bibinfo {author} {\bibfnamefont
  {N.}~\bibnamefont {Boulant}}, \bibinfo {author} {\bibfnamefont
  {T.}~\bibnamefont {Havel}}, \bibinfo {author} {\bibfnamefont {D.~G.}\
  \bibnamefont {Cory}}, \ and\ \bibinfo {author} {\bibfnamefont
  {R.}~\bibnamefont {Laflamme}},\ }\href {\doibase
  10.1103/PhysRevLett.96.170501} {\bibfield  {journal} {\bibinfo  {journal}
  {Phys. Rev. Lett.}\ }\textbf {\bibinfo {volume} {96}},\ \bibinfo {pages}
  {170501} (\bibinfo {year} {2006})}\BibitemShut {NoStop}%
\bibitem [{\citenamefont {Clarke}\ and\ \citenamefont
  {Wilhelm}(2008)}]{clarke2008}%
  \BibitemOpen
  \bibfield  {author} {\bibinfo {author} {\bibfnamefont {J.}~\bibnamefont
  {Clarke}}\ and\ \bibinfo {author} {\bibfnamefont {F.~K.}\ \bibnamefont
  {Wilhelm}},\ }\href {\doibase 10.1038/nature07128} {\bibfield  {journal}
  {\bibinfo  {journal} {Nature}\ }\textbf {\bibinfo {volume} {453}},\ \bibinfo
  {pages} {1031} (\bibinfo {year} {2008})}\BibitemShut {NoStop}%
\bibitem [{\citenamefont {Berkley}\ \emph {et~al.}(2003)\citenamefont
  {Berkley}, \citenamefont {Xu}, \citenamefont {Ramos}, \citenamefont {Gubrud},
  \citenamefont {Strauch}, \citenamefont {Johnson}, \citenamefont {Anderson},
  \citenamefont {Dragt}, \citenamefont {Lobb},\ and\ \citenamefont
  {Wellstood}}]{berkley2003}%
  \BibitemOpen
  \bibfield  {author} {\bibinfo {author} {\bibfnamefont {A.~J.}\ \bibnamefont
  {Berkley}}, \bibinfo {author} {\bibfnamefont {H.}~\bibnamefont {Xu}},
  \bibinfo {author} {\bibfnamefont {R.~C.}\ \bibnamefont {Ramos}}, \bibinfo
  {author} {\bibfnamefont {M.~A.}\ \bibnamefont {Gubrud}}, \bibinfo {author}
  {\bibfnamefont {F.~W.}\ \bibnamefont {Strauch}}, \bibinfo {author}
  {\bibfnamefont {P.~R.}\ \bibnamefont {Johnson}}, \bibinfo {author}
  {\bibfnamefont {J.~R.}\ \bibnamefont {Anderson}}, \bibinfo {author}
  {\bibfnamefont {A.~J.}\ \bibnamefont {Dragt}}, \bibinfo {author}
  {\bibfnamefont {C.~J.}\ \bibnamefont {Lobb}}, \ and\ \bibinfo {author}
  {\bibfnamefont {F.~C.}\ \bibnamefont {Wellstood}},\ }\href {\doibase
  10.1126/science.1084528} {\bibfield  {journal} {\bibinfo  {journal}
  {Science}\ }\textbf {\bibinfo {volume} {300}},\ \bibinfo {pages} {1548}
  (\bibinfo {year} {2003})}\BibitemShut {NoStop}%
\bibitem [{\citenamefont {Hald}\ \emph {et~al.}(1999)\citenamefont {Hald},
  \citenamefont {S\o{}rensen}, \citenamefont {Schori},\ and\ \citenamefont
  {Polzik}}]{hald1999}%
  \BibitemOpen
  \bibfield  {author} {\bibinfo {author} {\bibfnamefont {J.}~\bibnamefont
  {Hald}}, \bibinfo {author} {\bibfnamefont {J.~L.}\ \bibnamefont
  {S\o{}rensen}}, \bibinfo {author} {\bibfnamefont {C.}~\bibnamefont {Schori}},
  \ and\ \bibinfo {author} {\bibfnamefont {E.~S.}\ \bibnamefont {Polzik}},\
  }\href {\doibase 10.1103/PhysRevLett.83.1319} {\bibfield  {journal} {\bibinfo
   {journal} {Phys. Rev. Lett.}\ }\textbf {\bibinfo {volume} {83}},\ \bibinfo
  {pages} {1319} (\bibinfo {year} {1999})}\BibitemShut {NoStop}%
\bibitem [{\citenamefont {Mandel}\ \emph {et~al.}(2003)\citenamefont {Mandel},
  \citenamefont {Greiner}, \citenamefont {Widera}, \citenamefont {Rom},
  \citenamefont {Hansch},\ and\ \citenamefont {Bloch}}]{mandel2003}%
  \BibitemOpen
  \bibfield  {author} {\bibinfo {author} {\bibfnamefont {O.}~\bibnamefont
  {Mandel}}, \bibinfo {author} {\bibfnamefont {M.}~\bibnamefont {Greiner}},
  \bibinfo {author} {\bibfnamefont {A.}~\bibnamefont {Widera}}, \bibinfo
  {author} {\bibfnamefont {T.}~\bibnamefont {Rom}}, \bibinfo {author}
  {\bibfnamefont {T.~W.}\ \bibnamefont {Hansch}}, \ and\ \bibinfo {author}
  {\bibfnamefont {I.}~\bibnamefont {Bloch}},\ }\href {\doibase
  10.1038/nature02008} {\bibfield  {journal} {\bibinfo  {journal} {Nature}\
  }\textbf {\bibinfo {volume} {425}},\ \bibinfo {pages} {937} (\bibinfo {year}
  {2003})}\BibitemShut {NoStop}%
\bibitem [{\citenamefont {Bloch}(2005)}]{bloch2005}%
  \BibitemOpen
  \bibfield  {author} {\bibinfo {author} {\bibfnamefont {I.}~\bibnamefont
  {Bloch}},\ }\href {\doibase 10.1088/0953-4075/38/9/013} {\bibfield  {journal}
  {\bibinfo  {journal} {J. Phys. B: At. Mol. Opt. Phys.}\ }\textbf {\bibinfo
  {volume} {38}},\ \bibinfo {pages} {S629} (\bibinfo {year}
  {2005})}\BibitemShut {NoStop}%
\bibitem [{\citenamefont {Bloch}\ \emph {et~al.}(2008)\citenamefont {Bloch},
  \citenamefont {Dalibard},\ and\ \citenamefont {Zwerger}}]{bloch2008}%
  \BibitemOpen
  \bibfield  {author} {\bibinfo {author} {\bibfnamefont {I.}~\bibnamefont
  {Bloch}}, \bibinfo {author} {\bibfnamefont {J.}~\bibnamefont {Dalibard}}, \
  and\ \bibinfo {author} {\bibfnamefont {W.}~\bibnamefont {Zwerger}},\ }\href
  {\doibase 10.1103/RevModPhys.80.885} {\bibfield  {journal} {\bibinfo
  {journal} {Rev. Mod. Phys.}\ }\textbf {\bibinfo {volume} {80}},\ \bibinfo
  {pages} {885} (\bibinfo {year} {2008})}\BibitemShut {NoStop}%
\bibitem [{\citenamefont {Bernien}\ \emph {et~al.}(2013)\citenamefont
  {Bernien}, \citenamefont {Hensen}, \citenamefont {Pfaff}, \citenamefont
  {Koolstra}, \citenamefont {Blok}, \citenamefont {Robledo}, \citenamefont
  {Taminiau}, \citenamefont {Markham}, \citenamefont {Twitchen}, \citenamefont
  {Childress},\ and\ \citenamefont {Hanson}}]{bernien2013}%
  \BibitemOpen
  \bibfield  {author} {\bibinfo {author} {\bibfnamefont {H.}~\bibnamefont
  {Bernien}}, \bibinfo {author} {\bibfnamefont {B.}~\bibnamefont {Hensen}},
  \bibinfo {author} {\bibfnamefont {W.}~\bibnamefont {Pfaff}}, \bibinfo
  {author} {\bibfnamefont {G.}~\bibnamefont {Koolstra}}, \bibinfo {author}
  {\bibfnamefont {M.~S.}\ \bibnamefont {Blok}}, \bibinfo {author}
  {\bibfnamefont {L.}~\bibnamefont {Robledo}}, \bibinfo {author} {\bibfnamefont
  {T.~H.}\ \bibnamefont {Taminiau}}, \bibinfo {author} {\bibfnamefont
  {M.}~\bibnamefont {Markham}}, \bibinfo {author} {\bibfnamefont {D.~J.}\
  \bibnamefont {Twitchen}}, \bibinfo {author} {\bibfnamefont {L.}~\bibnamefont
  {Childress}}, \ and\ \bibinfo {author} {\bibfnamefont {R.}~\bibnamefont
  {Hanson}},\ }\href {https://doi.org/10.1038/nature12016} {\bibfield
  {journal} {\bibinfo  {journal} {Nature}\ }\textbf {\bibinfo {volume} {497}},\
  \bibinfo {pages} {86} (\bibinfo {year} {2013})}\BibitemShut {NoStop}%
\bibitem [{\citenamefont {Togan}\ \emph {et~al.}(2010)\citenamefont {Togan},
  \citenamefont {Chu}, \citenamefont {Trifonov}, \citenamefont {Jiang},
  \citenamefont {Maze}, \citenamefont {Childress}, \citenamefont {Dutt},
  \citenamefont {S{\o}rensen}, \citenamefont {Hemmer}, \citenamefont {Zibrov},\
  and\ \citenamefont {Lukin}}]{togan2010}%
  \BibitemOpen
  \bibfield  {author} {\bibinfo {author} {\bibfnamefont {E.}~\bibnamefont
  {Togan}}, \bibinfo {author} {\bibfnamefont {Y.}~\bibnamefont {Chu}}, \bibinfo
  {author} {\bibfnamefont {A.~S.}\ \bibnamefont {Trifonov}}, \bibinfo {author}
  {\bibfnamefont {L.}~\bibnamefont {Jiang}}, \bibinfo {author} {\bibfnamefont
  {J.}~\bibnamefont {Maze}}, \bibinfo {author} {\bibfnamefont {L.}~\bibnamefont
  {Childress}}, \bibinfo {author} {\bibfnamefont {M.~V.~G.}\ \bibnamefont
  {Dutt}}, \bibinfo {author} {\bibfnamefont {A.~S.}\ \bibnamefont
  {S{\o}rensen}}, \bibinfo {author} {\bibfnamefont {P.~R.}\ \bibnamefont
  {Hemmer}}, \bibinfo {author} {\bibfnamefont {A.~S.}\ \bibnamefont {Zibrov}},
  \ and\ \bibinfo {author} {\bibfnamefont {M.~D.}\ \bibnamefont {Lukin}},\
  }\href {https://doi.org/10.1038/nature09256} {\bibfield  {journal} {\bibinfo
  {journal} {Nature}\ }\textbf {\bibinfo {volume} {466}},\ \bibinfo {pages}
  {730} (\bibinfo {year} {2010})}\BibitemShut {NoStop}%
\bibitem [{\citenamefont {Marinkovi\ifmmode~\acute{c}\else \'{c}\fi{}}\ \emph
  {et~al.}(2018)\citenamefont {Marinkovi\ifmmode~\acute{c}\else \'{c}\fi{}},
  \citenamefont {Wallucks}, \citenamefont {Riedinger}, \citenamefont {Hong},
  \citenamefont {Aspelmeyer},\ and\ \citenamefont
  {Gr\"oblacher}}]{marinkovic2018}%
  \BibitemOpen
  \bibfield  {author} {\bibinfo {author} {\bibfnamefont {I.}~\bibnamefont
  {Marinkovi\ifmmode~\acute{c}\else \'{c}\fi{}}}, \bibinfo {author}
  {\bibfnamefont {A.}~\bibnamefont {Wallucks}}, \bibinfo {author}
  {\bibfnamefont {R.}~\bibnamefont {Riedinger}}, \bibinfo {author}
  {\bibfnamefont {S.}~\bibnamefont {Hong}}, \bibinfo {author} {\bibfnamefont
  {M.}~\bibnamefont {Aspelmeyer}}, \ and\ \bibinfo {author} {\bibfnamefont
  {S.}~\bibnamefont {Gr\"oblacher}},\ }\href {\doibase
  10.1103/PhysRevLett.121.220404} {\bibfield  {journal} {\bibinfo  {journal}
  {Phys. Rev. Lett.}\ }\textbf {\bibinfo {volume} {121}},\ \bibinfo {pages}
  {220404} (\bibinfo {year} {2018})}\BibitemShut {NoStop}%
\bibitem [{\citenamefont {DiVincenzo}(2000)}]{divincenzo2000}%
  \BibitemOpen
  \bibfield  {author} {\bibinfo {author} {\bibfnamefont {D.~P.}\ \bibnamefont
  {DiVincenzo}},\ }\href {\doibase
  10.1002/1521-3978(200009)48:9/11<771::AID-PROP771>3.0.CO;2-E} {\bibfield
  {journal} {\bibinfo  {journal} {Fortschr. Phys.}\ }\textbf {\bibinfo {volume}
  {48}},\ \bibinfo {pages} {771} (\bibinfo {year} {2000})}\BibitemShut
  {NoStop}%
\bibitem [{\citenamefont {Gottesman}(2010)}]{gottesman2010}%
  \BibitemOpen
  \bibfield  {author} {\bibinfo {author} {\bibfnamefont {D.}~\bibnamefont
  {Gottesman}},\ }in\ \href {https://arxiv.org/abs/0904.2557} {\emph {\bibinfo
  {booktitle} {Quantum information science and its contributions to
  mathematics, Proceedings of Symposia in Applied Mathematics}}},\
  Vol.~\bibinfo {volume} {68}\ (\bibinfo {year} {2010})\ pp.\ \bibinfo {pages}
  {13--58}\BibitemShut {NoStop}%
\bibitem [{\citenamefont {Benhelm}\ \emph {et~al.}(2008)\citenamefont
  {Benhelm}, \citenamefont {Kirchmair}, \citenamefont {Roos},\ and\
  \citenamefont {Blatt}}]{benhelm2008}%
  \BibitemOpen
  \bibfield  {author} {\bibinfo {author} {\bibfnamefont {J.}~\bibnamefont
  {Benhelm}}, \bibinfo {author} {\bibfnamefont {G.}~\bibnamefont {Kirchmair}},
  \bibinfo {author} {\bibfnamefont {C.~F.}\ \bibnamefont {Roos}}, \ and\
  \bibinfo {author} {\bibfnamefont {R.}~\bibnamefont {Blatt}},\ }\href
  {https://doi.org/10.1038/nphys961} {\bibfield  {journal} {\bibinfo  {journal}
  {Nat. Phys.}\ }\textbf {\bibinfo {volume} {4}},\ \bibinfo {pages} {463}
  (\bibinfo {year} {2008})}\BibitemShut {NoStop}%
\bibitem [{\citenamefont {Chow}\ \emph {et~al.}(2012)\citenamefont {Chow},
  \citenamefont {Gambetta}, \citenamefont {C\'orcoles}, \citenamefont {Merkel},
  \citenamefont {Smolin}, \citenamefont {Rigetti}, \citenamefont {Poletto},
  \citenamefont {Keefe}, \citenamefont {Rothwell}, \citenamefont {Rozen},
  \citenamefont {Ketchen},\ and\ \citenamefont {Steffen}}]{chow2012}%
  \BibitemOpen
  \bibfield  {author} {\bibinfo {author} {\bibfnamefont {J.~M.}\ \bibnamefont
  {Chow}}, \bibinfo {author} {\bibfnamefont {J.~M.}\ \bibnamefont {Gambetta}},
  \bibinfo {author} {\bibfnamefont {A.~D.}\ \bibnamefont {C\'orcoles}},
  \bibinfo {author} {\bibfnamefont {S.~T.}\ \bibnamefont {Merkel}}, \bibinfo
  {author} {\bibfnamefont {J.~A.}\ \bibnamefont {Smolin}}, \bibinfo {author}
  {\bibfnamefont {C.}~\bibnamefont {Rigetti}}, \bibinfo {author} {\bibfnamefont
  {S.}~\bibnamefont {Poletto}}, \bibinfo {author} {\bibfnamefont {G.~A.}\
  \bibnamefont {Keefe}}, \bibinfo {author} {\bibfnamefont {M.~B.}\ \bibnamefont
  {Rothwell}}, \bibinfo {author} {\bibfnamefont {J.~R.}\ \bibnamefont {Rozen}},
  \bibinfo {author} {\bibfnamefont {M.~B.}\ \bibnamefont {Ketchen}}, \ and\
  \bibinfo {author} {\bibfnamefont {M.}~\bibnamefont {Steffen}},\ }\href
  {\doibase 10.1103/PhysRevLett.109.060501} {\bibfield  {journal} {\bibinfo
  {journal} {Phys. Rev. Lett.}\ }\textbf {\bibinfo {volume} {109}},\ \bibinfo
  {pages} {060501} (\bibinfo {year} {2012})}\BibitemShut {NoStop}%
\bibitem [{\citenamefont {Barends}\ \emph {et~al.}(2014)\citenamefont
  {Barends}, \citenamefont {Kelly}, \citenamefont {Megrant}, \citenamefont
  {Veitia}, \citenamefont {Sank}, \citenamefont {Jeffrey}, \citenamefont
  {White}, \citenamefont {Mutus}, \citenamefont {Fowler}, \citenamefont
  {Campbell}, \citenamefont {Chen}, \citenamefont {Chen}, \citenamefont
  {Chiaro}, \citenamefont {Dunsworth}, \citenamefont {Neill}, \citenamefont
  {O'Malley}, \citenamefont {Roushan}, \citenamefont {Vainsencher},
  \citenamefont {Wenner}, \citenamefont {Korotkov}, \citenamefont {Cleland},\
  and\ \citenamefont {Martinis}}]{barends2014}%
  \BibitemOpen
  \bibfield  {author} {\bibinfo {author} {\bibfnamefont {R.}~\bibnamefont
  {Barends}}, \bibinfo {author} {\bibfnamefont {J.}~\bibnamefont {Kelly}},
  \bibinfo {author} {\bibfnamefont {A.}~\bibnamefont {Megrant}}, \bibinfo
  {author} {\bibfnamefont {A.}~\bibnamefont {Veitia}}, \bibinfo {author}
  {\bibfnamefont {D.}~\bibnamefont {Sank}}, \bibinfo {author} {\bibfnamefont
  {E.}~\bibnamefont {Jeffrey}}, \bibinfo {author} {\bibfnamefont {T.~C.}\
  \bibnamefont {White}}, \bibinfo {author} {\bibfnamefont {J.}~\bibnamefont
  {Mutus}}, \bibinfo {author} {\bibfnamefont {A.~G.}\ \bibnamefont {Fowler}},
  \bibinfo {author} {\bibfnamefont {B.}~\bibnamefont {Campbell}}, \bibinfo
  {author} {\bibfnamefont {Y.}~\bibnamefont {Chen}}, \bibinfo {author}
  {\bibfnamefont {Z.}~\bibnamefont {Chen}}, \bibinfo {author} {\bibfnamefont
  {B.}~\bibnamefont {Chiaro}}, \bibinfo {author} {\bibfnamefont
  {A.}~\bibnamefont {Dunsworth}}, \bibinfo {author} {\bibfnamefont
  {C.}~\bibnamefont {Neill}}, \bibinfo {author} {\bibfnamefont
  {P.}~\bibnamefont {O'Malley}}, \bibinfo {author} {\bibfnamefont
  {P.}~\bibnamefont {Roushan}}, \bibinfo {author} {\bibfnamefont
  {A.}~\bibnamefont {Vainsencher}}, \bibinfo {author} {\bibfnamefont
  {J.}~\bibnamefont {Wenner}}, \bibinfo {author} {\bibfnamefont {A.~N.}\
  \bibnamefont {Korotkov}}, \bibinfo {author} {\bibfnamefont {A.~N.}\
  \bibnamefont {Cleland}}, \ and\ \bibinfo {author} {\bibfnamefont {J.~M.}\
  \bibnamefont {Martinis}},\ }\href {https://doi.org/10.1038/nature13171}
  {\bibfield  {journal} {\bibinfo  {journal} {Nature}\ }\textbf {\bibinfo
  {volume} {508}},\ \bibinfo {pages} {500} (\bibinfo {year}
  {2014})}\BibitemShut {NoStop}%
\bibitem [{\citenamefont {Preskill}(2018)}]{preskill2018}%
  \BibitemOpen
  \bibfield  {author} {\bibinfo {author} {\bibfnamefont {J.}~\bibnamefont
  {Preskill}},\ }\href {\doibase 10.22331/q-2018-08-06-79} {\bibfield
  {journal} {\bibinfo  {journal} {{Quantum}}\ }\textbf {\bibinfo {volume}
  {2}},\ \bibinfo {pages} {79} (\bibinfo {year} {2018})}\BibitemShut {NoStop}%
\bibitem [{\citenamefont {Paler}\ \emph {et~al.}(2018)\citenamefont {Paler},
  \citenamefont {Zulehner},\ and\ \citenamefont {Wille}}]{paler2018}%
  \BibitemOpen
  \bibfield  {author} {\bibinfo {author} {\bibfnamefont {A.}~\bibnamefont
  {Paler}}, \bibinfo {author} {\bibfnamefont {A.}~\bibnamefont {Zulehner}}, \
  and\ \bibinfo {author} {\bibfnamefont {R.}~\bibnamefont {Wille}},\ }\href
  {https://arxiv.org/abs/1806.07241} {\bibfield  {journal} {\bibinfo  {journal}
  {arXiv:1806.07241}\ } (\bibinfo {year} {2018})}\BibitemShut {NoStop}%
\bibitem [{\citenamefont {{Nash}}\ \emph {et~al.}(2019)\citenamefont {{Nash}},
  \citenamefont {{Gheorghiu}},\ and\ \citenamefont {{Mosca}}}]{nash2019}%
  \BibitemOpen
  \bibfield  {author} {\bibinfo {author} {\bibfnamefont {B.}~\bibnamefont
  {{Nash}}}, \bibinfo {author} {\bibfnamefont {V.}~\bibnamefont {{Gheorghiu}}},
  \ and\ \bibinfo {author} {\bibfnamefont {M.}~\bibnamefont {{Mosca}}},\ }\href
  {https://arxiv.org/abs/1904.01972} {\bibfield  {journal} {\bibinfo  {journal}
  {arXiv:1904.01972}\ } (\bibinfo {year} {2019})}\BibitemShut {NoStop}%
\bibitem [{\citenamefont {Preskill}(2012)}]{preskill2012}%
  \BibitemOpen
  \bibfield  {author} {\bibinfo {author} {\bibfnamefont {J.}~\bibnamefont
  {Preskill}},\ }\href {https://arxiv.org/abs/1203.5813} {\bibfield  {journal}
  {\bibinfo  {journal} {arXiv:1203.5813}\ } (\bibinfo {year}
  {2012})}\BibitemShut {NoStop}%
\bibitem [{\citenamefont {Fujii}(2015)}]{fujii2015}%
  \BibitemOpen
  \bibfield  {author} {\bibinfo {author} {\bibfnamefont {K.}~\bibnamefont
  {Fujii}},\ }\href {https://arxiv.org/abs/1504.01444} {\bibfield  {journal}
  {\bibinfo  {journal} {arXiv:1504.01444}\ } (\bibinfo {year}
  {2015})}\BibitemShut {NoStop}%
\bibitem [{\citenamefont {Ibort}\ \emph {et~al.}(2009)\citenamefont {Ibort},
  \citenamefont {Man'ko}, \citenamefont {Marmo}, \citenamefont {Simoni},\ and\
  \citenamefont {Ventriglia}}]{ibort2009}%
  \BibitemOpen
  \bibfield  {author} {\bibinfo {author} {\bibfnamefont {A.}~\bibnamefont
  {Ibort}}, \bibinfo {author} {\bibfnamefont {V.~I.}\ \bibnamefont {Man'ko}},
  \bibinfo {author} {\bibfnamefont {G.}~\bibnamefont {Marmo}}, \bibinfo
  {author} {\bibfnamefont {A.}~\bibnamefont {Simoni}}, \ and\ \bibinfo {author}
  {\bibfnamefont {F.}~\bibnamefont {Ventriglia}},\ }\href {\doibase
  10.1088/0031-8949/79/06/065013} {\bibfield  {journal} {\bibinfo  {journal}
  {Physica Scripta}\ }\textbf {\bibinfo {volume} {79}},\ \bibinfo {pages}
  {065013} (\bibinfo {year} {2009})}\BibitemShut {NoStop}%
\bibitem [{\citenamefont {Lvovsky}\ and\ \citenamefont
  {Raymer}(2009)}]{lvovsky2009}%
  \BibitemOpen
  \bibfield  {author} {\bibinfo {author} {\bibfnamefont {A.~I.}\ \bibnamefont
  {Lvovsky}}\ and\ \bibinfo {author} {\bibfnamefont {M.~G.}\ \bibnamefont
  {Raymer}},\ }\href {\doibase 10.1103/RevModPhys.81.299} {\bibfield  {journal}
  {\bibinfo  {journal} {Rev. Mod. Phys.}\ }\textbf {\bibinfo {volume} {81}},\
  \bibinfo {pages} {299} (\bibinfo {year} {2009})}\BibitemShut {NoStop}%
\bibitem [{\citenamefont {Huang}(2014)}]{huang2014}%
  \BibitemOpen
  \bibfield  {author} {\bibinfo {author} {\bibfnamefont {Y.}~\bibnamefont
  {Huang}},\ }\href {\doibase 10.1088/1367-2630/16/3/033027} {\bibfield
  {journal} {\bibinfo  {journal} {New Journal of Physics}\ }\textbf {\bibinfo
  {volume} {16}},\ \bibinfo {pages} {033027} (\bibinfo {year}
  {2014})}\BibitemShut {NoStop}%
\bibitem [{\citenamefont {Nayak}\ \emph {et~al.}(2008)\citenamefont {Nayak},
  \citenamefont {Simon}, \citenamefont {Stern}, \citenamefont {Freedman},\ and\
  \citenamefont {Das~Sarma}}]{nayak2008}%
  \BibitemOpen
  \bibfield  {author} {\bibinfo {author} {\bibfnamefont {C.}~\bibnamefont
  {Nayak}}, \bibinfo {author} {\bibfnamefont {S.~H.}\ \bibnamefont {Simon}},
  \bibinfo {author} {\bibfnamefont {A.}~\bibnamefont {Stern}}, \bibinfo
  {author} {\bibfnamefont {M.}~\bibnamefont {Freedman}}, \ and\ \bibinfo
  {author} {\bibfnamefont {S.}~\bibnamefont {Das~Sarma}},\ }\href {\doibase
  10.1103/RevModPhys.80.1083} {\bibfield  {journal} {\bibinfo  {journal} {Rev.
  Mod. Phys.}\ }\textbf {\bibinfo {volume} {80}},\ \bibinfo {pages} {1083}
  (\bibinfo {year} {2008})}\BibitemShut {NoStop}%
\bibitem [{\citenamefont {Pachos}\ and\ \citenamefont
  {Simon}(2014)}]{pachos2014}%
  \BibitemOpen
  \bibfield  {author} {\bibinfo {author} {\bibfnamefont {J.~K.}\ \bibnamefont
  {Pachos}}\ and\ \bibinfo {author} {\bibfnamefont {S.~H.}\ \bibnamefont
  {Simon}},\ }\href {\doibase 10.1088/1367-2630/16/6/065003} {\bibfield
  {journal} {\bibinfo  {journal} {New J. Phys.}\ }\textbf {\bibinfo {volume}
  {16}},\ \bibinfo {pages} {065003} (\bibinfo {year} {2014})}\BibitemShut
  {NoStop}%
\bibitem [{\citenamefont {Lahtinen}\ and\ \citenamefont
  {Pachos}(2017)}]{lahtinen2017}%
  \BibitemOpen
  \bibfield  {author} {\bibinfo {author} {\bibfnamefont {V.}~\bibnamefont
  {Lahtinen}}\ and\ \bibinfo {author} {\bibfnamefont {J.~K.}\ \bibnamefont
  {Pachos}},\ }\href {\doibase 10.21468/SciPostPhys.3.3.021} {\bibfield
  {journal} {\bibinfo  {journal} {SciPost Phys.}\ }\textbf {\bibinfo {volume}
  {3}},\ \bibinfo {pages} {021} (\bibinfo {year} {2017})}\BibitemShut {NoStop}%
\bibitem [{\citenamefont {Dennis}\ \emph {et~al.}(2002)\citenamefont {Dennis},
  \citenamefont {Kitaev}, \citenamefont {Landahl},\ and\ \citenamefont
  {Preskill}}]{kitaev2001}%
  \BibitemOpen
  \bibfield  {author} {\bibinfo {author} {\bibfnamefont {E.}~\bibnamefont
  {Dennis}}, \bibinfo {author} {\bibfnamefont {A.}~\bibnamefont {Kitaev}},
  \bibinfo {author} {\bibfnamefont {A.}~\bibnamefont {Landahl}}, \ and\
  \bibinfo {author} {\bibfnamefont {J.}~\bibnamefont {Preskill}},\ }\href
  {\doibase 10.1063/1.1499754} {\bibfield  {journal} {\bibinfo  {journal} {J.
  Math. Phys.}\ }\textbf {\bibinfo {volume} {43}},\ \bibinfo {pages} {4452}
  (\bibinfo {year} {2002})}\BibitemShut {NoStop}%
\bibitem [{\citenamefont {Kitaev}(2006)}]{kitaev2006}%
  \BibitemOpen
  \bibfield  {author} {\bibinfo {author} {\bibfnamefont {A.}~\bibnamefont
  {Kitaev}},\ }\href {\doibase https://doi.org/10.1016/j.aop.2005.10.005}
  {\bibfield  {journal} {\bibinfo  {journal} {Ann. Phys.}\ }\textbf {\bibinfo
  {volume} {321}},\ \bibinfo {pages} {2 } (\bibinfo {year} {2006})}\BibitemShut
  {NoStop}%
\bibitem [{\citenamefont {Bombin}\ and\ \citenamefont
  {Martin-Delgado}(2006)}]{bombin2006}%
  \BibitemOpen
  \bibfield  {author} {\bibinfo {author} {\bibfnamefont {H.}~\bibnamefont
  {Bombin}}\ and\ \bibinfo {author} {\bibfnamefont {M.~A.}\ \bibnamefont
  {Martin-Delgado}},\ }\href {\doibase 10.1103/PhysRevLett.97.180501}
  {\bibfield  {journal} {\bibinfo  {journal} {Phys. Rev. Lett.}\ }\textbf
  {\bibinfo {volume} {97}},\ \bibinfo {pages} {180501} (\bibinfo {year}
  {2006})}\BibitemShut {NoStop}%
\bibitem [{\citenamefont {Bombin}\ and\ \citenamefont
  {Martin-Delgado}(2007)}]{bombin2007}%
  \BibitemOpen
  \bibfield  {author} {\bibinfo {author} {\bibfnamefont {H.}~\bibnamefont
  {Bombin}}\ and\ \bibinfo {author} {\bibfnamefont {M.~A.}\ \bibnamefont
  {Martin-Delgado}},\ }\href {\doibase 10.1103/PhysRevLett.98.160502}
  {\bibfield  {journal} {\bibinfo  {journal} {Phys. Rev. Lett.}\ }\textbf
  {\bibinfo {volume} {98}},\ \bibinfo {pages} {160502} (\bibinfo {year}
  {2007})}\BibitemShut {NoStop}%
\bibitem [{\citenamefont {Dusuel}\ \emph {et~al.}(2011)\citenamefont {Dusuel},
  \citenamefont {Kamfor}, \citenamefont {Or\'us}, \citenamefont {Schmidt},\
  and\ \citenamefont {Vidal}}]{dusuel2011}%
  \BibitemOpen
  \bibfield  {author} {\bibinfo {author} {\bibfnamefont {S.}~\bibnamefont
  {Dusuel}}, \bibinfo {author} {\bibfnamefont {M.}~\bibnamefont {Kamfor}},
  \bibinfo {author} {\bibfnamefont {R.}~\bibnamefont {Or\'us}}, \bibinfo
  {author} {\bibfnamefont {K.~P.}\ \bibnamefont {Schmidt}}, \ and\ \bibinfo
  {author} {\bibfnamefont {J.}~\bibnamefont {Vidal}},\ }\href {\doibase
  10.1103/PhysRevLett.106.107203} {\bibfield  {journal} {\bibinfo  {journal}
  {Phys. Rev. Lett.}\ }\textbf {\bibinfo {volume} {106}},\ \bibinfo {pages}
  {107203} (\bibinfo {year} {2011})}\BibitemShut {NoStop}%
\bibitem [{\citenamefont {Jahromi}\ \emph {et~al.}(2013)\citenamefont
  {Jahromi}, \citenamefont {Kargarian}, \citenamefont {Masoudi},\ and\
  \citenamefont {Schmidt}}]{jahromi2013}%
  \BibitemOpen
  \bibfield  {author} {\bibinfo {author} {\bibfnamefont {S.~S.}\ \bibnamefont
  {Jahromi}}, \bibinfo {author} {\bibfnamefont {M.}~\bibnamefont {Kargarian}},
  \bibinfo {author} {\bibfnamefont {S.~F.}\ \bibnamefont {Masoudi}}, \ and\
  \bibinfo {author} {\bibfnamefont {K.~P.}\ \bibnamefont {Schmidt}},\ }\href
  {\doibase 10.1103/PhysRevB.87.094413} {\bibfield  {journal} {\bibinfo
  {journal} {Phys. Rev. B}\ }\textbf {\bibinfo {volume} {87}},\ \bibinfo
  {pages} {094413} (\bibinfo {year} {2013})}\BibitemShut {NoStop}%
\bibitem [{\citenamefont {Zarei}(2015)}]{zarei2015}%
  \BibitemOpen
  \bibfield  {author} {\bibinfo {author} {\bibfnamefont {M.~H.}\ \bibnamefont
  {Zarei}},\ }\href {\doibase 10.1103/PhysRevA.91.022319} {\bibfield  {journal}
  {\bibinfo  {journal} {Phys. Rev. A}\ }\textbf {\bibinfo {volume} {91}},\
  \bibinfo {pages} {022319} (\bibinfo {year} {2015})}\BibitemShut {NoStop}%
\bibitem [{\citenamefont {Jamadagni}\ \emph {et~al.}(2018)\citenamefont
  {Jamadagni}, \citenamefont {Weimer},\ and\ \citenamefont
  {Bhattacharyya}}]{jamadagni2018}%
  \BibitemOpen
  \bibfield  {author} {\bibinfo {author} {\bibfnamefont {A.}~\bibnamefont
  {Jamadagni}}, \bibinfo {author} {\bibfnamefont {H.}~\bibnamefont {Weimer}}, \
  and\ \bibinfo {author} {\bibfnamefont {A.}~\bibnamefont {Bhattacharyya}},\
  }\href {\doibase 10.1103/PhysRevB.98.235147} {\bibfield  {journal} {\bibinfo
  {journal} {Phys. Rev. B}\ }\textbf {\bibinfo {volume} {98}},\ \bibinfo
  {pages} {235147} (\bibinfo {year} {2018})}\BibitemShut {NoStop}%
\bibitem [{\citenamefont {Stace}\ \emph {et~al.}(2009)\citenamefont {Stace},
  \citenamefont {Barrett},\ and\ \citenamefont {Doherty}}]{stace2009}%
  \BibitemOpen
  \bibfield  {author} {\bibinfo {author} {\bibfnamefont {T.~M.}\ \bibnamefont
  {Stace}}, \bibinfo {author} {\bibfnamefont {S.~D.}\ \bibnamefont {Barrett}},
  \ and\ \bibinfo {author} {\bibfnamefont {A.~C.}\ \bibnamefont {Doherty}},\
  }\href {\doibase 10.1103/PhysRevLett.102.200501} {\bibfield  {journal}
  {\bibinfo  {journal} {Phys. Rev. Lett.}\ }\textbf {\bibinfo {volume} {102}},\
  \bibinfo {pages} {200501} (\bibinfo {year} {2009})}\BibitemShut {NoStop}%
\bibitem [{\citenamefont {Stace}\ and\ \citenamefont
  {Barrett}(2010)}]{stace2010}%
  \BibitemOpen
  \bibfield  {author} {\bibinfo {author} {\bibfnamefont {T.~M.}\ \bibnamefont
  {Stace}}\ and\ \bibinfo {author} {\bibfnamefont {S.~D.}\ \bibnamefont
  {Barrett}},\ }\href {\doibase 10.1103/PhysRevA.81.022317} {\bibfield
  {journal} {\bibinfo  {journal} {Phys. Rev. A}\ }\textbf {\bibinfo {volume}
  {81}},\ \bibinfo {pages} {022317} (\bibinfo {year} {2010})}\BibitemShut
  {NoStop}%
\bibitem [{\citenamefont {Vodola}\ \emph {et~al.}(2018)\citenamefont {Vodola},
  \citenamefont {Amaro}, \citenamefont {Martin-Delgado},\ and\ \citenamefont
  {M\"uller}}]{vodola2018}%
  \BibitemOpen
  \bibfield  {author} {\bibinfo {author} {\bibfnamefont {D.}~\bibnamefont
  {Vodola}}, \bibinfo {author} {\bibfnamefont {D.}~\bibnamefont {Amaro}},
  \bibinfo {author} {\bibfnamefont {M.~A.}\ \bibnamefont {Martin-Delgado}}, \
  and\ \bibinfo {author} {\bibfnamefont {M.}~\bibnamefont {M\"uller}},\ }\href
  {\doibase 10.1103/PhysRevLett.121.060501} {\bibfield  {journal} {\bibinfo
  {journal} {Phys. Rev. Lett.}\ }\textbf {\bibinfo {volume} {121}},\ \bibinfo
  {pages} {060501} (\bibinfo {year} {2018})}\BibitemShut {NoStop}%
\bibitem [{\citenamefont {Katzgraber}\ \emph {et~al.}(2009)\citenamefont
  {Katzgraber}, \citenamefont {Bombin},\ and\ \citenamefont
  {Martin-Delgado}}]{katzgraber2009}%
  \BibitemOpen
  \bibfield  {author} {\bibinfo {author} {\bibfnamefont {H.~G.}\ \bibnamefont
  {Katzgraber}}, \bibinfo {author} {\bibfnamefont {H.}~\bibnamefont {Bombin}},
  \ and\ \bibinfo {author} {\bibfnamefont {M.~A.}\ \bibnamefont
  {Martin-Delgado}},\ }\href {\doibase 10.1103/PhysRevLett.103.090501}
  {\bibfield  {journal} {\bibinfo  {journal} {Phys. Rev. Lett.}\ }\textbf
  {\bibinfo {volume} {103}},\ \bibinfo {pages} {090501} (\bibinfo {year}
  {2009})}\BibitemShut {NoStop}%
\bibitem [{\citenamefont {Nigg}\ \emph {et~al.}(2014)\citenamefont {Nigg},
  \citenamefont {M{\"u}ller}, \citenamefont {Martinez}, \citenamefont
  {Schindler}, \citenamefont {Hennrich}, \citenamefont {Monz}, \citenamefont
  {Martin-Delgado},\ and\ \citenamefont {Blatt}}]{nigg2014}%
  \BibitemOpen
  \bibfield  {author} {\bibinfo {author} {\bibfnamefont {D.}~\bibnamefont
  {Nigg}}, \bibinfo {author} {\bibfnamefont {M.}~\bibnamefont {M{\"u}ller}},
  \bibinfo {author} {\bibfnamefont {E.~A.}\ \bibnamefont {Martinez}}, \bibinfo
  {author} {\bibfnamefont {P.}~\bibnamefont {Schindler}}, \bibinfo {author}
  {\bibfnamefont {M.}~\bibnamefont {Hennrich}}, \bibinfo {author}
  {\bibfnamefont {T.}~\bibnamefont {Monz}}, \bibinfo {author} {\bibfnamefont
  {M.~A.}\ \bibnamefont {Martin-Delgado}}, \ and\ \bibinfo {author}
  {\bibfnamefont {R.}~\bibnamefont {Blatt}},\ }\href {\doibase
  10.1126/science.1253742} {\bibfield  {journal} {\bibinfo  {journal}
  {Science}\ }\textbf {\bibinfo {volume} {345}},\ \bibinfo {pages} {302}
  (\bibinfo {year} {2014})}\BibitemShut {NoStop}%
\bibitem [{\citenamefont {Linke}\ \emph {et~al.}(2017)\citenamefont {Linke},
  \citenamefont {Gutierrez}, \citenamefont {Landsman}, \citenamefont {Figgatt},
  \citenamefont {Debnath}, \citenamefont {Brown},\ and\ \citenamefont
  {Monroe}}]{linke2017}%
  \BibitemOpen
  \bibfield  {author} {\bibinfo {author} {\bibfnamefont {N.~M.}\ \bibnamefont
  {Linke}}, \bibinfo {author} {\bibfnamefont {M.}~\bibnamefont {Gutierrez}},
  \bibinfo {author} {\bibfnamefont {K.~A.}\ \bibnamefont {Landsman}}, \bibinfo
  {author} {\bibfnamefont {C.}~\bibnamefont {Figgatt}}, \bibinfo {author}
  {\bibfnamefont {S.}~\bibnamefont {Debnath}}, \bibinfo {author} {\bibfnamefont
  {K.~R.}\ \bibnamefont {Brown}}, \ and\ \bibinfo {author} {\bibfnamefont
  {C.}~\bibnamefont {Monroe}},\ }\href@noop {} {\bibfield  {journal} {\bibinfo
  {journal} {Sci. Adv.}\ }\textbf {\bibinfo {volume} {3}},\ \bibinfo {pages}
  {10} (\bibinfo {year} {2017})}\BibitemShut {NoStop}%
\bibitem [{\citenamefont {Wright}\ \emph {et~al.}(2019)\citenamefont {Wright},
  \citenamefont {Beck}, \citenamefont {Debnath}, \citenamefont {Amini},
  \citenamefont {Nam}, \citenamefont {Grzesiak}, \citenamefont {Chen},
  \citenamefont {Pisenti}, \citenamefont {Chmielewski}, \citenamefont
  {Collins}, \citenamefont {Hudek}, \citenamefont {Mizrahi}, \citenamefont
  {Wong-Campos}, \citenamefont {Allen}, \citenamefont {Apisdorf}, \citenamefont
  {Solomon}, \citenamefont {Williams}, \citenamefont {Ducore}, \citenamefont
  {Blinov}, \citenamefont {Kreikemeier}, \citenamefont {Chaplin}, \citenamefont
  {Keesan}, \citenamefont {Monroe},\ and\ \citenamefont {Kim}}]{wright2019}%
  \BibitemOpen
  \bibfield  {author} {\bibinfo {author} {\bibfnamefont {K.}~\bibnamefont
  {Wright}}, \bibinfo {author} {\bibfnamefont {K.~M.}\ \bibnamefont {Beck}},
  \bibinfo {author} {\bibfnamefont {S.}~\bibnamefont {Debnath}}, \bibinfo
  {author} {\bibfnamefont {J.~M.}\ \bibnamefont {Amini}}, \bibinfo {author}
  {\bibfnamefont {Y.}~\bibnamefont {Nam}}, \bibinfo {author} {\bibfnamefont
  {N.}~\bibnamefont {Grzesiak}}, \bibinfo {author} {\bibfnamefont {J.-S.}\
  \bibnamefont {Chen}}, \bibinfo {author} {\bibfnamefont {N.~C.}\ \bibnamefont
  {Pisenti}}, \bibinfo {author} {\bibfnamefont {M.}~\bibnamefont
  {Chmielewski}}, \bibinfo {author} {\bibfnamefont {C.}~\bibnamefont
  {Collins}}, \bibinfo {author} {\bibfnamefont {K.~M.}\ \bibnamefont {Hudek}},
  \bibinfo {author} {\bibfnamefont {J.}~\bibnamefont {Mizrahi}}, \bibinfo
  {author} {\bibfnamefont {J.~D.}\ \bibnamefont {Wong-Campos}}, \bibinfo
  {author} {\bibfnamefont {S.}~\bibnamefont {Allen}}, \bibinfo {author}
  {\bibfnamefont {J.}~\bibnamefont {Apisdorf}}, \bibinfo {author}
  {\bibfnamefont {P.}~\bibnamefont {Solomon}}, \bibinfo {author} {\bibfnamefont
  {M.}~\bibnamefont {Williams}}, \bibinfo {author} {\bibfnamefont {A.~M.}\
  \bibnamefont {Ducore}}, \bibinfo {author} {\bibfnamefont {A.}~\bibnamefont
  {Blinov}}, \bibinfo {author} {\bibfnamefont {S.~M.}\ \bibnamefont
  {Kreikemeier}}, \bibinfo {author} {\bibfnamefont {V.}~\bibnamefont
  {Chaplin}}, \bibinfo {author} {\bibfnamefont {M.}~\bibnamefont {Keesan}},
  \bibinfo {author} {\bibfnamefont {C.}~\bibnamefont {Monroe}}, \ and\ \bibinfo
  {author} {\bibfnamefont {J.}~\bibnamefont {Kim}},\ }\href
  {https://arxiv.org/abs/1903.08181} {\bibfield  {journal} {\bibinfo  {journal}
  {arXiv:1903.08181}\ } (\bibinfo {year} {2019})}\BibitemShut {NoStop}%
\bibitem [{\citenamefont {Kelly}\ \emph {et~al.}(2015)\citenamefont {Kelly},
  \citenamefont {Barends}, \citenamefont {Fowler}, \citenamefont {Megrant},
  \citenamefont {Jeffrey}, \citenamefont {White}, \citenamefont {Sank},
  \citenamefont {Mutus}, \citenamefont {Campbell}, \citenamefont {Chen},
  \citenamefont {Chen}, \citenamefont {Chiaro}, \citenamefont {Dunsworth},
  \citenamefont {Hoi}, \citenamefont {Neill}, \citenamefont {O'Malley},
  \citenamefont {Quintana}, \citenamefont {Roushan}, \citenamefont
  {Vainsencher}, \citenamefont {Wenner}, \citenamefont {Cleland},\ and\
  \citenamefont {Martinis}}]{kelly2015}%
  \BibitemOpen
  \bibfield  {author} {\bibinfo {author} {\bibfnamefont {J.}~\bibnamefont
  {Kelly}}, \bibinfo {author} {\bibfnamefont {R.}~\bibnamefont {Barends}},
  \bibinfo {author} {\bibfnamefont {A.~G.}\ \bibnamefont {Fowler}}, \bibinfo
  {author} {\bibfnamefont {A.}~\bibnamefont {Megrant}}, \bibinfo {author}
  {\bibfnamefont {E.}~\bibnamefont {Jeffrey}}, \bibinfo {author} {\bibfnamefont
  {T.~C.}\ \bibnamefont {White}}, \bibinfo {author} {\bibfnamefont
  {D.}~\bibnamefont {Sank}}, \bibinfo {author} {\bibfnamefont {J.~Y.}\
  \bibnamefont {Mutus}}, \bibinfo {author} {\bibfnamefont {B.}~\bibnamefont
  {Campbell}}, \bibinfo {author} {\bibfnamefont {Y.}~\bibnamefont {Chen}},
  \bibinfo {author} {\bibfnamefont {Z.}~\bibnamefont {Chen}}, \bibinfo {author}
  {\bibfnamefont {B.}~\bibnamefont {Chiaro}}, \bibinfo {author} {\bibfnamefont
  {A.}~\bibnamefont {Dunsworth}}, \bibinfo {author} {\bibfnamefont {I.-C.}\
  \bibnamefont {Hoi}}, \bibinfo {author} {\bibfnamefont {C.}~\bibnamefont
  {Neill}}, \bibinfo {author} {\bibfnamefont {P.~J.~J.}\ \bibnamefont
  {O'Malley}}, \bibinfo {author} {\bibfnamefont {C.}~\bibnamefont {Quintana}},
  \bibinfo {author} {\bibfnamefont {P.}~\bibnamefont {Roushan}}, \bibinfo
  {author} {\bibfnamefont {A.}~\bibnamefont {Vainsencher}}, \bibinfo {author}
  {\bibfnamefont {J.}~\bibnamefont {Wenner}}, \bibinfo {author} {\bibfnamefont
  {A.~N.}\ \bibnamefont {Cleland}}, \ and\ \bibinfo {author} {\bibfnamefont
  {J.~M.}\ \bibnamefont {Martinis}},\ }\href
  {https://doi.org/10.1038/nature14270} {\bibfield  {journal} {\bibinfo
  {journal} {Nature}\ }\textbf {\bibinfo {volume} {519}},\ \bibinfo {pages}
  {66} (\bibinfo {year} {2015})}\BibitemShut {NoStop}%
\bibitem [{\citenamefont {Gambetta}\ \emph {et~al.}(2017)\citenamefont
  {Gambetta}, \citenamefont {Chow},\ and\ \citenamefont
  {Steffen}}]{gambetta2017}%
  \BibitemOpen
  \bibfield  {author} {\bibinfo {author} {\bibfnamefont {J.~M.}\ \bibnamefont
  {Gambetta}}, \bibinfo {author} {\bibfnamefont {J.~M.}\ \bibnamefont {Chow}},
  \ and\ \bibinfo {author} {\bibfnamefont {M.}~\bibnamefont {Steffen}},\ }\href
  {\doibase 10.1038/s41534-016-0004-0} {\bibfield  {journal} {\bibinfo
  {journal} {Nature Phys. J. Quant. Inf.}\ }\textbf {\bibinfo {volume} {3}},\
  \bibinfo {pages} {2} (\bibinfo {year} {2017})}\BibitemShut {NoStop}%
\bibitem [{\citenamefont {Castelnovo}\ and\ \citenamefont
  {Chamon}(2007)}]{castelnovo2007}%
  \BibitemOpen
  \bibfield  {author} {\bibinfo {author} {\bibfnamefont {C.}~\bibnamefont
  {Castelnovo}}\ and\ \bibinfo {author} {\bibfnamefont {C.}~\bibnamefont
  {Chamon}},\ }\href {\doibase 10.1103/PhysRevB.76.184442} {\bibfield
  {journal} {\bibinfo  {journal} {Phys. Rev. B}\ }\textbf {\bibinfo {volume}
  {76}},\ \bibinfo {pages} {184442} (\bibinfo {year} {2007})}\BibitemShut
  {NoStop}%
\bibitem [{\citenamefont {Castelnovo}\ and\ \citenamefont
  {Chamon}(2008)}]{castelnovo2008}%
  \BibitemOpen
  \bibfield  {author} {\bibinfo {author} {\bibfnamefont {C.}~\bibnamefont
  {Castelnovo}}\ and\ \bibinfo {author} {\bibfnamefont {C.}~\bibnamefont
  {Chamon}},\ }\href {\doibase 10.1103/PhysRevB.78.155120} {\bibfield
  {journal} {\bibinfo  {journal} {Phys. Rev. B}\ }\textbf {\bibinfo {volume}
  {78}},\ \bibinfo {pages} {155120} (\bibinfo {year} {2008})}\BibitemShut
  {NoStop}%
\bibitem [{\citenamefont {Schmitz}\ \emph {et~al.}(2019)\citenamefont
  {Schmitz}, \citenamefont {Huang},\ and\ \citenamefont {Prem}}]{schmitz2019}%
  \BibitemOpen
  \bibfield  {author} {\bibinfo {author} {\bibfnamefont {A.~T.}\ \bibnamefont
  {Schmitz}}, \bibinfo {author} {\bibfnamefont {S.-J.}\ \bibnamefont {Huang}},
  \ and\ \bibinfo {author} {\bibfnamefont {A.}~\bibnamefont {Prem}},\ }\href
  {\doibase 10.1103/PhysRevB.99.205109} {\bibfield  {journal} {\bibinfo
  {journal} {Phys. Rev. B}\ }\textbf {\bibinfo {volume} {99}},\ \bibinfo
  {pages} {205109} (\bibinfo {year} {2019})}\BibitemShut {NoStop}%
\bibitem [{\citenamefont {DiVincenzo}\ \emph {et~al.}(1998)\citenamefont
  {DiVincenzo}, \citenamefont {Fuchs}, \citenamefont {Mabuchi}, \citenamefont
  {Smolin}, \citenamefont {Thapliyal},\ and\ \citenamefont
  {Uhlmann}}]{divincenzo1998}%
  \BibitemOpen
  \bibfield  {author} {\bibinfo {author} {\bibfnamefont {D.~P.}\ \bibnamefont
  {DiVincenzo}}, \bibinfo {author} {\bibfnamefont {C.~A.}\ \bibnamefont
  {Fuchs}}, \bibinfo {author} {\bibfnamefont {H.}~\bibnamefont {Mabuchi}},
  \bibinfo {author} {\bibfnamefont {J.~A.}\ \bibnamefont {Smolin}}, \bibinfo
  {author} {\bibfnamefont {A.}~\bibnamefont {Thapliyal}}, \ and\ \bibinfo
  {author} {\bibfnamefont {A.}~\bibnamefont {Uhlmann}},\ }\href
  {https://arxiv.org/abs/quant-ph/9803033} {\bibfield  {journal} {\bibinfo
  {journal} {arXiv:quant-ph/9803033}\ } (\bibinfo {year} {1998})}\BibitemShut
  {NoStop}%
\bibitem [{\citenamefont {Verstraete}\ \emph
  {et~al.}(2004{\natexlab{a}})\citenamefont {Verstraete}, \citenamefont
  {Popp},\ and\ \citenamefont {Cirac}}]{verstraete2004}%
  \BibitemOpen
  \bibfield  {author} {\bibinfo {author} {\bibfnamefont {F.}~\bibnamefont
  {Verstraete}}, \bibinfo {author} {\bibfnamefont {M.}~\bibnamefont {Popp}}, \
  and\ \bibinfo {author} {\bibfnamefont {J.~I.}\ \bibnamefont {Cirac}},\ }\href
  {\doibase 10.1103/PhysRevLett.92.027901} {\bibfield  {journal} {\bibinfo
  {journal} {Phys. Rev. Lett.}\ }\textbf {\bibinfo {volume} {92}},\ \bibinfo
  {pages} {027901} (\bibinfo {year} {2004}{\natexlab{a}})}\BibitemShut
  {NoStop}%
\bibitem [{\citenamefont {Popp}\ \emph {et~al.}(2005)\citenamefont {Popp},
  \citenamefont {Verstraete}, \citenamefont {Mart\'{\i}n-Delgado},\ and\
  \citenamefont {Cirac}}]{popp2005}%
  \BibitemOpen
  \bibfield  {author} {\bibinfo {author} {\bibfnamefont {M.}~\bibnamefont
  {Popp}}, \bibinfo {author} {\bibfnamefont {F.}~\bibnamefont {Verstraete}},
  \bibinfo {author} {\bibfnamefont {M.~A.}\ \bibnamefont
  {Mart\'{\i}n-Delgado}}, \ and\ \bibinfo {author} {\bibfnamefont {J.~I.}\
  \bibnamefont {Cirac}},\ }\href {\doibase 10.1103/PhysRevA.71.042306}
  {\bibfield  {journal} {\bibinfo  {journal} {Phys. Rev. A}\ }\textbf {\bibinfo
  {volume} {71}},\ \bibinfo {pages} {042306} (\bibinfo {year}
  {2005})}\BibitemShut {NoStop}%
\bibitem [{\citenamefont {Sadhukhan}\ \emph {et~al.}(2017)\citenamefont
  {Sadhukhan}, \citenamefont {Roy}, \citenamefont {Pal}, \citenamefont
  {Rakshit}, \citenamefont {Sen(De)},\ and\ \citenamefont
  {Sen}}]{sadhukhan2017}%
  \BibitemOpen
  \bibfield  {author} {\bibinfo {author} {\bibfnamefont {D.}~\bibnamefont
  {Sadhukhan}}, \bibinfo {author} {\bibfnamefont {S.~S.}\ \bibnamefont {Roy}},
  \bibinfo {author} {\bibfnamefont {A.~K.}\ \bibnamefont {Pal}}, \bibinfo
  {author} {\bibfnamefont {D.}~\bibnamefont {Rakshit}}, \bibinfo {author}
  {\bibfnamefont {A.}~\bibnamefont {Sen(De)}}, \ and\ \bibinfo {author}
  {\bibfnamefont {U.}~\bibnamefont {Sen}},\ }\href {\doibase
  10.1103/PhysRevA.95.022301} {\bibfield  {journal} {\bibinfo  {journal} {Phys.
  Rev. A}\ }\textbf {\bibinfo {volume} {95}},\ \bibinfo {pages} {022301}
  (\bibinfo {year} {2017})}\BibitemShut {NoStop}%
\bibitem [{\citenamefont {Kitaev}\ and\ \citenamefont
  {Preskill}(2006)}]{kitaev2006a}%
  \BibitemOpen
  \bibfield  {author} {\bibinfo {author} {\bibfnamefont {A.}~\bibnamefont
  {Kitaev}}\ and\ \bibinfo {author} {\bibfnamefont {J.}~\bibnamefont
  {Preskill}},\ }\href {\doibase 10.1103/PhysRevLett.96.110404} {\bibfield
  {journal} {\bibinfo  {journal} {Phys. Rev. Lett.}\ }\textbf {\bibinfo
  {volume} {96}},\ \bibinfo {pages} {110404} (\bibinfo {year}
  {2006})}\BibitemShut {NoStop}%
\bibitem [{\citenamefont {Levin}\ and\ \citenamefont {Wen}(2006)}]{levin2006}%
  \BibitemOpen
  \bibfield  {author} {\bibinfo {author} {\bibfnamefont {M.}~\bibnamefont
  {Levin}}\ and\ \bibinfo {author} {\bibfnamefont {X.-G.}\ \bibnamefont
  {Wen}},\ }\href {\doibase 10.1103/PhysRevLett.96.110405} {\bibfield
  {journal} {\bibinfo  {journal} {Phys. Rev. Lett.}\ }\textbf {\bibinfo
  {volume} {96}},\ \bibinfo {pages} {110405} (\bibinfo {year}
  {2006})}\BibitemShut {NoStop}%
\bibitem [{\citenamefont {Linke}\ \emph {et~al.}(2018)\citenamefont {Linke},
  \citenamefont {Johri}, \citenamefont {Figgatt}, \citenamefont {Landsman},
  \citenamefont {Matsuura},\ and\ \citenamefont {Monroe}}]{linke2018}%
  \BibitemOpen
  \bibfield  {author} {\bibinfo {author} {\bibfnamefont {N.~M.}\ \bibnamefont
  {Linke}}, \bibinfo {author} {\bibfnamefont {S.}~\bibnamefont {Johri}},
  \bibinfo {author} {\bibfnamefont {C.}~\bibnamefont {Figgatt}}, \bibinfo
  {author} {\bibfnamefont {K.~A.}\ \bibnamefont {Landsman}}, \bibinfo {author}
  {\bibfnamefont {A.~Y.}\ \bibnamefont {Matsuura}}, \ and\ \bibinfo {author}
  {\bibfnamefont {C.}~\bibnamefont {Monroe}},\ }\href {\doibase
  10.1103/PhysRevA.98.052334} {\bibfield  {journal} {\bibinfo  {journal} {Phys.
  Rev. A}\ }\textbf {\bibinfo {volume} {98}},\ \bibinfo {pages} {052334}
  (\bibinfo {year} {2018})}\BibitemShut {NoStop}%
\bibitem [{\citenamefont {Brydges}\ \emph {et~al.}(2019)\citenamefont
  {Brydges}, \citenamefont {Elben}, \citenamefont {Jurcevic}, \citenamefont
  {Vermersch}, \citenamefont {Maier}, \citenamefont {Lanyon}, \citenamefont
  {Zoller}, \citenamefont {Blatt},\ and\ \citenamefont {Roos}}]{brydges2019}%
  \BibitemOpen
  \bibfield  {author} {\bibinfo {author} {\bibfnamefont {T.}~\bibnamefont
  {Brydges}}, \bibinfo {author} {\bibfnamefont {A.}~\bibnamefont {Elben}},
  \bibinfo {author} {\bibfnamefont {P.}~\bibnamefont {Jurcevic}}, \bibinfo
  {author} {\bibfnamefont {B.}~\bibnamefont {Vermersch}}, \bibinfo {author}
  {\bibfnamefont {C.}~\bibnamefont {Maier}}, \bibinfo {author} {\bibfnamefont
  {B.~P.}\ \bibnamefont {Lanyon}}, \bibinfo {author} {\bibfnamefont
  {P.}~\bibnamefont {Zoller}}, \bibinfo {author} {\bibfnamefont
  {R.}~\bibnamefont {Blatt}}, \ and\ \bibinfo {author} {\bibfnamefont {C.~F.}\
  \bibnamefont {Roos}},\ }\href {\doibase 10.1126/science.aau4963} {\bibfield
  {journal} {\bibinfo  {journal} {Science}\ }\textbf {\bibinfo {volume}
  {364}},\ \bibinfo {pages} {260} (\bibinfo {year} {2019})}\BibitemShut
  {NoStop}%
\bibitem [{\citenamefont {Kitaev}(2003)}]{kitaev2003}%
  \BibitemOpen
  \bibfield  {author} {\bibinfo {author} {\bibfnamefont {A.}~\bibnamefont
  {Kitaev}},\ }\href {\doibase 10.1016/S0003-4916(02)00018-0} {\bibfield
  {journal} {\bibinfo  {journal} {Ann. Phys.}\ }\textbf {\bibinfo {volume}
  {303}},\ \bibinfo {pages} {2 } (\bibinfo {year} {2003})}\BibitemShut
  {NoStop}%
\bibitem [{\citenamefont {Kargarian}(2008)}]{kargarian2008}%
  \BibitemOpen
  \bibfield  {author} {\bibinfo {author} {\bibfnamefont {M.}~\bibnamefont
  {Kargarian}},\ }\href {\doibase 10.1103/PhysRevA.78.062312} {\bibfield
  {journal} {\bibinfo  {journal} {Phys. Rev. A}\ }\textbf {\bibinfo {volume}
  {78}},\ \bibinfo {pages} {062312} (\bibinfo {year} {2008})}\BibitemShut
  {NoStop}%
\bibitem [{\citenamefont {Van~den Nest}\ \emph {et~al.}(2004)\citenamefont
  {Van~den Nest}, \citenamefont {Dehaene},\ and\ \citenamefont
  {De~Moor}}]{van-den-nest2004}%
  \BibitemOpen
  \bibfield  {author} {\bibinfo {author} {\bibfnamefont {M.}~\bibnamefont
  {Van~den Nest}}, \bibinfo {author} {\bibfnamefont {J.}~\bibnamefont
  {Dehaene}}, \ and\ \bibinfo {author} {\bibfnamefont {B.}~\bibnamefont
  {De~Moor}},\ }\href {\doibase 10.1103/PhysRevA.69.022316} {\bibfield
  {journal} {\bibinfo  {journal} {Phys. Rev. A}\ }\textbf {\bibinfo {volume}
  {69}},\ \bibinfo {pages} {022316} (\bibinfo {year} {2004})}\BibitemShut
  {NoStop}%
\bibitem [{\citenamefont {Hein}\ \emph {et~al.}(2006)\citenamefont {Hein},
  \citenamefont {D\"{u}r}, \citenamefont {Eisert}, \citenamefont {Raussendorf},
  \citenamefont {Van~den Nest},\ and\ \citenamefont {J.~Briegel}}]{hein2006}%
  \BibitemOpen
  \bibfield  {author} {\bibinfo {author} {\bibfnamefont {M.}~\bibnamefont
  {Hein}}, \bibinfo {author} {\bibfnamefont {W.}~\bibnamefont {D\"{u}r}},
  \bibinfo {author} {\bibfnamefont {J.}~\bibnamefont {Eisert}}, \bibinfo
  {author} {\bibfnamefont {R.}~\bibnamefont {Raussendorf}}, \bibinfo {author}
  {\bibfnamefont {M.}~\bibnamefont {Van~den Nest}}, \ and\ \bibinfo {author}
  {\bibfnamefont {H.}~\bibnamefont {J.~Briegel}},\ }\href
  {https://arxiv.org/abs/quant-ph/0602096} {\bibfield  {journal} {\bibinfo
  {journal} {arXiv:quant-ph/0602096}\ } (\bibinfo {year} {2006})}\BibitemShut
  {NoStop}%
\bibitem [{\citenamefont {Amaro}\ \emph {et~al.}(2018)\citenamefont {Amaro},
  \citenamefont {MÃŒller},\ and\ \citenamefont {Pal}}]{amaro2018}%
  \BibitemOpen
  \bibfield  {author} {\bibinfo {author} {\bibfnamefont {D.}~\bibnamefont
  {Amaro}}, \bibinfo {author} {\bibfnamefont {M.}~\bibnamefont {MÃŒller}}, \
  and\ \bibinfo {author} {\bibfnamefont {A.~K.}\ \bibnamefont {Pal}},\ }\href
  {\doibase 10.1088/1367-2630/aac485} {\bibfield  {journal} {\bibinfo
  {journal} {New J. Phys.}\ }\textbf {\bibinfo {volume} {20}},\ \bibinfo
  {pages} {063017} (\bibinfo {year} {2018})}\BibitemShut {NoStop}%
\bibitem [{\citenamefont {Verstraete}\ \emph
  {et~al.}(2004{\natexlab{b}})\citenamefont {Verstraete}, \citenamefont
  {Mart\'{\i}n-Delgado},\ and\ \citenamefont {Cirac}}]{verstraete2004a}%
  \BibitemOpen
  \bibfield  {author} {\bibinfo {author} {\bibfnamefont {F.}~\bibnamefont
  {Verstraete}}, \bibinfo {author} {\bibfnamefont {M.~A.}\ \bibnamefont
  {Mart\'{\i}n-Delgado}}, \ and\ \bibinfo {author} {\bibfnamefont {J.~I.}\
  \bibnamefont {Cirac}},\ }\href {\doibase 10.1103/PhysRevLett.92.087201}
  {\bibfield  {journal} {\bibinfo  {journal} {Phys. Rev. Lett.}\ }\textbf
  {\bibinfo {volume} {92}},\ \bibinfo {pages} {087201} (\bibinfo {year}
  {2004}{\natexlab{b}})}\BibitemShut {NoStop}%
\bibitem [{\citenamefont {Jin}\ and\ \citenamefont {Korepin}(2004)}]{jin2004}%
  \BibitemOpen
  \bibfield  {author} {\bibinfo {author} {\bibfnamefont {B.-Q.}\ \bibnamefont
  {Jin}}\ and\ \bibinfo {author} {\bibfnamefont {V.~E.}\ \bibnamefont
  {Korepin}},\ }\href {\doibase 10.1103/PhysRevA.69.062314} {\bibfield
  {journal} {\bibinfo  {journal} {Phys. Rev. A}\ }\textbf {\bibinfo {volume}
  {69}},\ \bibinfo {pages} {062314} (\bibinfo {year} {2004})}\BibitemShut
  {NoStop}%
\bibitem [{\citenamefont {Skr\o{}vseth}\ and\ \citenamefont
  {Bartlett}(2009)}]{skrovseth2009}%
  \BibitemOpen
  \bibfield  {author} {\bibinfo {author} {\bibfnamefont {S.~O.}\ \bibnamefont
  {Skr\o{}vseth}}\ and\ \bibinfo {author} {\bibfnamefont {S.~D.}\ \bibnamefont
  {Bartlett}},\ }\href {\doibase 10.1103/PhysRevA.80.022316} {\bibfield
  {journal} {\bibinfo  {journal} {Phys. Rev. A}\ }\textbf {\bibinfo {volume}
  {80}},\ \bibinfo {pages} {022316} (\bibinfo {year} {2009})}\BibitemShut
  {NoStop}%
\bibitem [{\citenamefont {Smacchia}\ \emph {et~al.}(2011)\citenamefont
  {Smacchia}, \citenamefont {Amico}, \citenamefont {Facchi}, \citenamefont
  {Fazio}, \citenamefont {Florio}, \citenamefont {Pascazio},\ and\
  \citenamefont {Vedral}}]{smacchia2011}%
  \BibitemOpen
  \bibfield  {author} {\bibinfo {author} {\bibfnamefont {P.}~\bibnamefont
  {Smacchia}}, \bibinfo {author} {\bibfnamefont {L.}~\bibnamefont {Amico}},
  \bibinfo {author} {\bibfnamefont {P.}~\bibnamefont {Facchi}}, \bibinfo
  {author} {\bibfnamefont {R.}~\bibnamefont {Fazio}}, \bibinfo {author}
  {\bibfnamefont {G.}~\bibnamefont {Florio}}, \bibinfo {author} {\bibfnamefont
  {S.}~\bibnamefont {Pascazio}}, \ and\ \bibinfo {author} {\bibfnamefont
  {V.}~\bibnamefont {Vedral}},\ }\href {\doibase 10.1103/PhysRevA.84.022304}
  {\bibfield  {journal} {\bibinfo  {journal} {Phys. Rev. A}\ }\textbf {\bibinfo
  {volume} {84}},\ \bibinfo {pages} {022304} (\bibinfo {year}
  {2011})}\BibitemShut {NoStop}%
\bibitem [{\citenamefont {Montes}\ and\ \citenamefont
  {Hamma}(2012)}]{montes2012}%
  \BibitemOpen
  \bibfield  {author} {\bibinfo {author} {\bibfnamefont {S.}~\bibnamefont
  {Montes}}\ and\ \bibinfo {author} {\bibfnamefont {A.}~\bibnamefont {Hamma}},\
  }\href {\doibase 10.1103/PhysRevE.86.021101} {\bibfield  {journal} {\bibinfo
  {journal} {Phys. Rev. E}\ }\textbf {\bibinfo {volume} {86}},\ \bibinfo
  {pages} {021101} (\bibinfo {year} {2012})}\BibitemShut {NoStop}%
\bibitem [{\citenamefont {Ac{\'i}n}\ \emph {et~al.}(2007)\citenamefont
  {Ac{\'i}n}, \citenamefont {Cirac},\ and\ \citenamefont
  {Lewenstein}}]{acin2007}%
  \BibitemOpen
  \bibfield  {author} {\bibinfo {author} {\bibfnamefont {A.}~\bibnamefont
  {Ac{\'i}n}}, \bibinfo {author} {\bibfnamefont {J.~I.}\ \bibnamefont {Cirac}},
  \ and\ \bibinfo {author} {\bibfnamefont {M.}~\bibnamefont {Lewenstein}},\
  }\href {https://doi.org/10.1038/nphys549} {\bibfield  {journal} {\bibinfo
  {journal} {Nat. Phys.}\ }\textbf {\bibinfo {volume} {3}},\ \bibinfo {pages}
  {256} (\bibinfo {year} {2007})}\BibitemShut {NoStop}%
\bibitem [{\citenamefont {Terhal}(2002)}]{terhal2002}%
  \BibitemOpen
  \bibfield  {author} {\bibinfo {author} {\bibfnamefont {B.~M.}\ \bibnamefont
  {Terhal}},\ }\href {\doibase 10.1016/S0304-3975(02)00139-1} {\bibfield
  {journal} {\bibinfo  {journal} {Theor. Comput. Sci.}\ }\textbf {\bibinfo
  {volume} {287}},\ \bibinfo {pages} {313} (\bibinfo {year}
  {2002})}\BibitemShut {NoStop}%
\bibitem [{\citenamefont {G\"uhne}\ \emph {et~al.}(2002)\citenamefont
  {G\"uhne}, \citenamefont {Hyllus}, \citenamefont {Bru\ss{}}, \citenamefont
  {Ekert}, \citenamefont {Lewenstein}, \citenamefont {Macchiavello},\ and\
  \citenamefont {Sanpera}}]{guhne2002}%
  \BibitemOpen
  \bibfield  {author} {\bibinfo {author} {\bibfnamefont {O.}~\bibnamefont
  {G\"uhne}}, \bibinfo {author} {\bibfnamefont {P.}~\bibnamefont {Hyllus}},
  \bibinfo {author} {\bibfnamefont {D.}~\bibnamefont {Bru\ss{}}}, \bibinfo
  {author} {\bibfnamefont {A.}~\bibnamefont {Ekert}}, \bibinfo {author}
  {\bibfnamefont {M.}~\bibnamefont {Lewenstein}}, \bibinfo {author}
  {\bibfnamefont {C.}~\bibnamefont {Macchiavello}}, \ and\ \bibinfo {author}
  {\bibfnamefont {A.}~\bibnamefont {Sanpera}},\ }\href {\doibase
  10.1103/PhysRevA.66.062305} {\bibfield  {journal} {\bibinfo  {journal} {Phys.
  Rev. A}\ }\textbf {\bibinfo {volume} {66}},\ \bibinfo {pages} {062305}
  (\bibinfo {year} {2002})}\BibitemShut {NoStop}%
\bibitem [{\citenamefont {Bourennane}\ \emph {et~al.}(2004)\citenamefont
  {Bourennane}, \citenamefont {Eibl}, \citenamefont {Kurtsiefer}, \citenamefont
  {Gaertner}, \citenamefont {Weinfurter}, \citenamefont {G\"uhne},
  \citenamefont {Hyllus}, \citenamefont {Bru\ss{}}, \citenamefont
  {Lewenstein},\ and\ \citenamefont {Sanpera}}]{bourennane2004}%
  \BibitemOpen
  \bibfield  {author} {\bibinfo {author} {\bibfnamefont {M.}~\bibnamefont
  {Bourennane}}, \bibinfo {author} {\bibfnamefont {M.}~\bibnamefont {Eibl}},
  \bibinfo {author} {\bibfnamefont {C.}~\bibnamefont {Kurtsiefer}}, \bibinfo
  {author} {\bibfnamefont {S.}~\bibnamefont {Gaertner}}, \bibinfo {author}
  {\bibfnamefont {H.}~\bibnamefont {Weinfurter}}, \bibinfo {author}
  {\bibfnamefont {O.}~\bibnamefont {G\"uhne}}, \bibinfo {author} {\bibfnamefont
  {P.}~\bibnamefont {Hyllus}}, \bibinfo {author} {\bibfnamefont
  {D.}~\bibnamefont {Bru\ss{}}}, \bibinfo {author} {\bibfnamefont
  {M.}~\bibnamefont {Lewenstein}}, \ and\ \bibinfo {author} {\bibfnamefont
  {A.}~\bibnamefont {Sanpera}},\ }\href {\doibase
  10.1103/PhysRevLett.92.087902} {\bibfield  {journal} {\bibinfo  {journal}
  {Phys. Rev. Lett.}\ }\textbf {\bibinfo {volume} {92}},\ \bibinfo {pages}
  {087902} (\bibinfo {year} {2004})}\BibitemShut {NoStop}%
\bibitem [{\citenamefont {G{\"u}hne}\ and\ \citenamefont
  {T{\'o}th}(2009)}]{guhne2009}%
  \BibitemOpen
  \bibfield  {author} {\bibinfo {author} {\bibfnamefont {O.}~\bibnamefont
  {G{\"u}hne}}\ and\ \bibinfo {author} {\bibfnamefont {G.}~\bibnamefont
  {T{\'o}th}},\ }\href
  {http://www.sciencedirect.com/science/article/pii/S0370157309000623}
  {\bibfield  {journal} {\bibinfo  {journal} {Phys. Rep.}\ }\textbf {\bibinfo
  {volume} {474}},\ \bibinfo {pages} {1} (\bibinfo {year} {2009})}\BibitemShut
  {NoStop}%
\bibitem [{\citenamefont {T\'oth}\ and\ \citenamefont
  {G\"uhne}(2005)}]{guhne2005}%
  \BibitemOpen
  \bibfield  {author} {\bibinfo {author} {\bibfnamefont {G.}~\bibnamefont
  {T\'oth}}\ and\ \bibinfo {author} {\bibfnamefont {O.}~\bibnamefont
  {G\"uhne}},\ }\href {\doibase 10.1103/PhysRevA.72.022340} {\bibfield
  {journal} {\bibinfo  {journal} {Phys. Rev. A}\ }\textbf {\bibinfo {volume}
  {72}},\ \bibinfo {pages} {022340} (\bibinfo {year} {2005})}\BibitemShut
  {NoStop}%
\bibitem [{\citenamefont {Alba}\ \emph {et~al.}(2010)\citenamefont {Alba},
  \citenamefont {T\'oth},\ and\ \citenamefont {Garc\'{\i}a-Ripoll}}]{alba2010}%
  \BibitemOpen
  \bibfield  {author} {\bibinfo {author} {\bibfnamefont {E.}~\bibnamefont
  {Alba}}, \bibinfo {author} {\bibfnamefont {G.}~\bibnamefont {T\'oth}}, \ and\
  \bibinfo {author} {\bibfnamefont {J.~J.}\ \bibnamefont
  {Garc\'{\i}a-Ripoll}},\ }\href {\doibase 10.1103/PhysRevA.82.062321}
  {\bibfield  {journal} {\bibinfo  {journal} {Phys. Rev. A}\ }\textbf {\bibinfo
  {volume} {82}},\ \bibinfo {pages} {062321} (\bibinfo {year}
  {2010})}\BibitemShut {NoStop}%
\bibitem [{\citenamefont {Friis}\ \emph {et~al.}(2018)\citenamefont {Friis},
  \citenamefont {Marty}, \citenamefont {Maier}, \citenamefont {Hempel},
  \citenamefont {Holz\"apfel}, \citenamefont {Jurcevic}, \citenamefont
  {Plenio}, \citenamefont {Huber}, \citenamefont {Roos}, \citenamefont
  {Blatt},\ and\ \citenamefont {Lanyon}}]{friis2018}%
  \BibitemOpen
  \bibfield  {author} {\bibinfo {author} {\bibfnamefont {N.}~\bibnamefont
  {Friis}}, \bibinfo {author} {\bibfnamefont {O.}~\bibnamefont {Marty}},
  \bibinfo {author} {\bibfnamefont {C.}~\bibnamefont {Maier}}, \bibinfo
  {author} {\bibfnamefont {C.}~\bibnamefont {Hempel}}, \bibinfo {author}
  {\bibfnamefont {M.}~\bibnamefont {Holz\"apfel}}, \bibinfo {author}
  {\bibfnamefont {P.}~\bibnamefont {Jurcevic}}, \bibinfo {author}
  {\bibfnamefont {M.~B.}\ \bibnamefont {Plenio}}, \bibinfo {author}
  {\bibfnamefont {M.}~\bibnamefont {Huber}}, \bibinfo {author} {\bibfnamefont
  {C.}~\bibnamefont {Roos}}, \bibinfo {author} {\bibfnamefont {R.}~\bibnamefont
  {Blatt}}, \ and\ \bibinfo {author} {\bibfnamefont {B.}~\bibnamefont
  {Lanyon}},\ }\href {\doibase 10.1103/PhysRevX.8.021012} {\bibfield  {journal}
  {\bibinfo  {journal} {Phys. Rev. X}\ }\textbf {\bibinfo {volume} {8}},\
  \bibinfo {pages} {021012} (\bibinfo {year} {2018})}\BibitemShut {NoStop}%
\bibitem [{\citenamefont {Amaro}\ and\ \citenamefont
  {M\"{u}ller}(2019)}]{amaro2019a}%
  \BibitemOpen
  \bibfield  {author} {\bibinfo {author} {\bibfnamefont {D.}~\bibnamefont
  {Amaro}}\ and\ \bibinfo {author} {\bibfnamefont {M.}~\bibnamefont
  {M\"{u}ller}},\ }\href {https://arxiv.org/abs/1911.01144} {\bibfield
  {journal} {\bibinfo  {journal} {arXiv:1911.01144}\ } (\bibinfo {year}
  {2019})}\BibitemShut {NoStop}%
\bibitem [{\citenamefont {Lang}\ and\ \citenamefont
  {B\"uchler}(2012)}]{lang2012}%
  \BibitemOpen
  \bibfield  {author} {\bibinfo {author} {\bibfnamefont {N.}~\bibnamefont
  {Lang}}\ and\ \bibinfo {author} {\bibfnamefont {H.~P.}\ \bibnamefont
  {B\"uchler}},\ }\href {\doibase 10.1103/PhysRevA.86.022336} {\bibfield
  {journal} {\bibinfo  {journal} {Phys. Rev. A}\ }\textbf {\bibinfo {volume}
  {86}},\ \bibinfo {pages} {022336} (\bibinfo {year} {2012})}\BibitemShut
  {NoStop}%
\bibitem [{\citenamefont {Brand\~ao}(2005)}]{brandao2005}%
  \BibitemOpen
  \bibfield  {author} {\bibinfo {author} {\bibfnamefont {F.~G. S.~L.}\
  \bibnamefont {Brand\~ao}},\ }\href {\doibase 10.1103/PhysRevA.72.022310}
  {\bibfield  {journal} {\bibinfo  {journal} {Phys. Rev. A}\ }\textbf {\bibinfo
  {volume} {72}},\ \bibinfo {pages} {022310} (\bibinfo {year}
  {2005})}\BibitemShut {NoStop}%
\bibitem [{\citenamefont {Brand\~ao}\ and\ \citenamefont
  {Viana}(2006)}]{brandao2006}%
  \BibitemOpen
  \bibfield  {author} {\bibinfo {author} {\bibfnamefont {F.~G. S.~L.}\
  \bibnamefont {Brand\~ao}}\ and\ \bibinfo {author} {\bibfnamefont {R.~O.}\
  \bibnamefont {Viana}},\ }\href {\doibase 10.1142/S0219749906001803}
  {\bibfield  {journal} {\bibinfo  {journal} {Int. J. Quant. Inf.}\ }\textbf
  {\bibinfo {volume} {04}},\ \bibinfo {pages} {331} (\bibinfo {year}
  {2006})}\BibitemShut {NoStop}%
\bibitem [{\citenamefont {Eisert}\ \emph {et~al.}(2007)\citenamefont {Eisert},
  \citenamefont {Brand\~ao},\ and\ \citenamefont {Audenaert}}]{eisert2007}%
  \BibitemOpen
  \bibfield  {author} {\bibinfo {author} {\bibfnamefont {J.}~\bibnamefont
  {Eisert}}, \bibinfo {author} {\bibfnamefont {F.~G. S.~L.}\ \bibnamefont
  {Brand\~ao}}, \ and\ \bibinfo {author} {\bibfnamefont {K.~M.~R.}\
  \bibnamefont {Audenaert}},\ }\href
  {http://stacks.iop.org/1367-2630/9/i=3/a=046} {\bibfield  {journal} {\bibinfo
   {journal} {New J. Phys.}\ }\textbf {\bibinfo {volume} {9}},\ \bibinfo
  {pages} {46} (\bibinfo {year} {2007})}\BibitemShut {NoStop}%
\bibitem [{\citenamefont {G\"uhne}\ \emph {et~al.}(2007)\citenamefont
  {G\"uhne}, \citenamefont {Reimpell},\ and\ \citenamefont
  {Werner}}]{guhne2007}%
  \BibitemOpen
  \bibfield  {author} {\bibinfo {author} {\bibfnamefont {O.}~\bibnamefont
  {G\"uhne}}, \bibinfo {author} {\bibfnamefont {M.}~\bibnamefont {Reimpell}}, \
  and\ \bibinfo {author} {\bibfnamefont {R.~F.}\ \bibnamefont {Werner}},\
  }\href {\doibase 10.1103/PhysRevLett.98.110502} {\bibfield  {journal}
  {\bibinfo  {journal} {Phys. Rev. Lett.}\ }\textbf {\bibinfo {volume} {98}},\
  \bibinfo {pages} {110502} (\bibinfo {year} {2007})}\BibitemShut {NoStop}%
\bibitem [{\citenamefont {G\"uhne}\ \emph {et~al.}(2008)\citenamefont
  {G\"uhne}, \citenamefont {Reimpell},\ and\ \citenamefont
  {Werner}}]{guhne2008}%
  \BibitemOpen
  \bibfield  {author} {\bibinfo {author} {\bibfnamefont {O.}~\bibnamefont
  {G\"uhne}}, \bibinfo {author} {\bibfnamefont {M.}~\bibnamefont {Reimpell}}, \
  and\ \bibinfo {author} {\bibfnamefont {R.~F.}\ \bibnamefont {Werner}},\
  }\href {\doibase 10.1103/PhysRevA.77.052317} {\bibfield  {journal} {\bibinfo
  {journal} {Phys. Rev. A}\ }\textbf {\bibinfo {volume} {77}},\ \bibinfo
  {pages} {052317} (\bibinfo {year} {2008})}\BibitemShut {NoStop}%
\bibitem [{\citenamefont {Amaro}(2019)}]{amaro_github_2019}%
  \BibitemOpen
  \bibfield  {author} {\bibinfo {author} {\bibfnamefont {D.}~\bibnamefont
  {Amaro}},\ }\href@noop {} {\enquote {\bibinfo {title} {stabgraph},}\
  }\bibinfo {howpublished} {availale at
  \url{https://github.com/davamaro/stabgraph}} (\bibinfo {year}
  {2019})\BibitemShut {NoStop}%
\bibitem [{\citenamefont {Amaro}\ \emph {et~al.}(2019)\citenamefont {Amaro},
  \citenamefont {M\"{u}ller},\ and\ \citenamefont {Pal}}]{pal_github_2019}%
  \BibitemOpen
  \bibfield  {author} {\bibinfo {author} {\bibfnamefont {D.}~\bibnamefont
  {Amaro}}, \bibinfo {author} {\bibfnamefont {M.}~\bibnamefont {M\"{u}ller}}, \
  and\ \bibinfo {author} {\bibfnamefont {A.~K.}\ \bibnamefont {Pal}},\
  }\href@noop {} {\enquote {\bibinfo {title} {alcpack},}\ }\bibinfo
  {howpublished} {available at \url{https://github.com/amitkpal/alcpack}}
  (\bibinfo {year} {2019})\BibitemShut {NoStop}%
\bibitem [{\citenamefont {Peres}(1996)}]{peres1996}%
  \BibitemOpen
  \bibfield  {author} {\bibinfo {author} {\bibfnamefont {A.}~\bibnamefont
  {Peres}},\ }\href {\doibase 10.1103/PhysRevLett.77.1413} {\bibfield
  {journal} {\bibinfo  {journal} {Phys. Rev. Lett.}\ }\textbf {\bibinfo
  {volume} {77}},\ \bibinfo {pages} {1413} (\bibinfo {year}
  {1996})}\BibitemShut {NoStop}%
\bibitem [{\citenamefont {Horodecki}\ \emph {et~al.}(1996)\citenamefont
  {Horodecki}, \citenamefont {Horodecki},\ and\ \citenamefont
  {Horodecki}}]{horodecki1996}%
  \BibitemOpen
  \bibfield  {author} {\bibinfo {author} {\bibfnamefont {M.}~\bibnamefont
  {Horodecki}}, \bibinfo {author} {\bibfnamefont {P.}~\bibnamefont
  {Horodecki}}, \ and\ \bibinfo {author} {\bibfnamefont {R.}~\bibnamefont
  {Horodecki}},\ }\href {\doibase
  https://doi.org/10.1016/S0375-9601(96)00706-2} {\bibfield  {journal}
  {\bibinfo  {journal} {Phys. Lett. A}\ }\textbf {\bibinfo {volume} {223}},\
  \bibinfo {pages} {1 } (\bibinfo {year} {1996})}\BibitemShut {NoStop}%
\bibitem [{\citenamefont {Greenberger}\ \emph {et~al.}(1989)\citenamefont
  {Greenberger}, \citenamefont {Horne},\ and\ \citenamefont
  {Zeilinger}}]{greenberger1989}%
  \BibitemOpen
  \bibfield  {author} {\bibinfo {author} {\bibfnamefont {D.~M.}\ \bibnamefont
  {Greenberger}}, \bibinfo {author} {\bibfnamefont {M.~A.}\ \bibnamefont
  {Horne}}, \ and\ \bibinfo {author} {\bibfnamefont {A.}~\bibnamefont
  {Zeilinger}},\ }\href
  {http://inis.iaea.org/search/search.aspx?orig_q=RN:22064349} {\emph {\bibinfo
  {title} {Bell's theorem, quantum theory and conceptions of the universe}}}\
  (\bibinfo  {publisher} {Kluwer},\ \bibinfo {address} {Netherlands},\ \bibinfo
  {year} {1989})\BibitemShut {NoStop}%
\bibitem [{\citenamefont {Nielsen}\ and\ \citenamefont
  {Chuang}(2010)}]{nielsen2010}%
  \BibitemOpen
  \bibfield  {author} {\bibinfo {author} {\bibfnamefont {M.~A.}\ \bibnamefont
  {Nielsen}}\ and\ \bibinfo {author} {\bibfnamefont {I.~L.}\ \bibnamefont
  {Chuang}},\ }\href@noop {} {\emph {\bibinfo {title} {Quantum Computation and
  Quantum Information}}}\ (\bibinfo  {publisher} {Cambridge University Press},\
  \bibinfo {year} {2010})\BibitemShut {NoStop}%
\bibitem [{\citenamefont {Holevo}\ and\ \citenamefont
  {Giovannetti}(2012)}]{holevo2012}%
  \BibitemOpen
  \bibfield  {author} {\bibinfo {author} {\bibfnamefont {A.~S.}\ \bibnamefont
  {Holevo}}\ and\ \bibinfo {author} {\bibfnamefont {V.}~\bibnamefont
  {Giovannetti}},\ }\href {\doibase 10.1088/0034-4885/75/4/046001} {\bibfield
  {journal} {\bibinfo  {journal} {Rep. Prog. Phys.}\ }\textbf {\bibinfo
  {volume} {75}},\ \bibinfo {pages} {046001} (\bibinfo {year}
  {2012})}\BibitemShut {NoStop}%
\end{thebibliography}%
\bibliographystyle{apsrev4-1}

\end{document}